\documentclass[%
twocolumn,
superscriptaddress,
 amsmath,amssymb,
prb,
]{revtex4-1}

\usepackage{graphicx}
\usepackage{dcolumn}
\usepackage{bm}
\usepackage{hyperref}
\usepackage{color}
\usepackage{braket}
\usepackage{subcaption}
\usepackage{mathtools}
\usepackage{bbold}
\usepackage{comment}

\begin{document}

\preprint{}

\title{Coupled Dynamical Boltzmann Transport Equations with Long-range Electron-phonon and Electron-electron Interactions in 2D Materials}

\author{Francesco Macheda}
\affiliation{Dipartimento di Scienze e Metodi dell’Ingegneria,
University of Modena and Reggio Emilia, Reggio Emilia, Italy}
\affiliation{Dipartimento di Fisica, Università di Roma La Sapienza, Roma, Italy }%
\author{Thibault Sohier}
\affiliation{Laboratoire Charles Coulomb (L2C), Université de Montpellier, CNRS, Montpellier, France}


\begin{abstract}
We study the interplay between long-range electron-phonon and electron-electron interactions in electrostatically doped two-dimensional semiconductors, including interlayer couplings in van der Waals heterostructures. We evaluate the effects of those interactions on transport properties by writing dynamically coupled Boltzmann equations for the electrons and for the electrodynamically active excitations. We develop a theory with a general validity, and apply it both to simplified parabolic models, and to the realistic BN-encapsulated graphene system which we present in an accompanying paper \onlinecite{accpaper}. We show that dynamical screening effects are of fundamental importance in order to correctly describe the electronic transport properties of two-dimensional materials, and in particular the scattering from polar phonons, whether those come from the semiconductor itself or the surrounding layers.  
\end{abstract}

\maketitle

\section{Introduction}
The electron-phonon interaction (e-ph) is the primary intrinsic mechanism that determines the electrical resistivity of a crystal at finite temperature\cite{ziman2001electrons}. It is well known that the e-ph interaction can be generally split into a long-range Coulomb mediated component in the Fr\"olich form \cite{Frohlich01071954}, and short-range terms that are mediated via local fields effects, i.e. electrostatic fields that average to zero over the unit cell of the crystal \cite{PhysRevB.13.694,PhysRevB.1.910}.

Long-range interactions are among the leading factors limiting mobility in two dimensional heterostructures, whether due to intrinsic optical phonons or remote ones from surrounding layers \cite{Ma2014a, PhysRevApplied.18.054062}. Fr\"olich-type coupling manifests in polar materials, and stems from the motion of effectively charged atoms generating long-range electric fields in the crystal. These have the ability to scatter free-carriers (electrons) and, accordingly, degrade their mobility. However, unlike local fields, long-range electric fields are also extremely sensitive to the presence of free-carriers in the system. In particular, the dynamical generation of electron-hole pairs leads to a screening of those fields. The dynamically screened Fr\"olich interaction is not the only long-range field detrimental to transport properties: free-carriers can also interact via the Coulomb mediated dynamical electron-electron (e-e) interactions. Thus, both phonons and dynamical electron-hole pairs influence electronic transport properties.

However, many of the past and recent efforts to compute mobilities, either from models \cite{PhysRevB.3.2534,10.1063/1.1659368} or from first principles \cite{Ponce2020,PhysRevB.94.201201,PhysRevB.97.045201,Ponce2021,Brunin2020a,Gopalan2022,PhysRevResearch.3.043022}, completely disregard dynamical screening effects and e-e interactions. Further, most work in the Bloch's ansatz \cite{ziman2001electrons} i.e. assuming that phonons remain in equilibrium in presence of an electric field. Some very recent developments \cite{Macheda2018,PhysRevB.102.245202,PhysRevB.101.075202,PhysRevLett.134.226301} partially or totally remove Bloch's ansatz in order to take into account out-of-equilibrium dynamics of phonon population, 
and some \cite{PhysRevB.51.14247,Hauber2017a,PhysRevApplied.18.054062,mocatti2025nonequilibriumphotocarrierphonondynamics,paul2025coulombdragdrivenelectronhole} investigate the role of e-e interactions or dynamical screening effects. 
Combining those ingredients within a general ab initio framework applicable to van der Waals heterostructures remains a challenge, as it requires a treatment of electron-phonon dynamics on the same footing.

We progress over these limitations and build two coupled dynamical Boltzmann equations, one for the electrons and one for the system's electrodynamic excitations, i.e. for phonons coupled with light and electron-hole pairs. We take into account the long-range e-ph and e-e interactions, fully dynamical screening effects, and the deviation of the excitations population from equilibrium. Crucially, we do not treat polar phonons as sharp spectral excitations, and devise a systematic way to include anharmonic dissipation, which is fundamental to determine the electronic conductivity. 
We built on and progress over the work of Hauber et al. \onlinecite{Hauber2017a} by developing a general \textit{ab initio} methodology that can be applied to realistic systems, including van der Waals heterostructures. We bring a crucial improvement to consistently account for dimensionality and dynamical screening effects on the dispersion and spectral shape of polar phonons.
Here, we report the general theory and decline it for i) a simple parabolic band materials, where we can easily explore the physical trends; and ii) realistic multiband systems embedded in van der Waals heterostructures via the Van der Waals Electrodynamics (VED) framework \cite{PhysRevB.110.115407}, where the calculation of realistic transport properties is of crucial technological interest. We then apply our method to a simplified parabolic band model for two dimensional h-BN and 2H-MoS$_2$, coupled to a single polar longitudinal-optical (LO) phonon, while we study the realistic BN-encapsulated graphene system in an accompanying paper \cite{accpaper} including both LO and out-of-plane optical (ZO) phonons. Generally, we find that both dynamical screening and e-e interactions strongly impact the mobility value in the doping range typical for transport experiments, giving rise to non trivial trends with respect to the carrier concentration. Finally, our findings show that an adequate treatment of the dynamical screening is needed to correctly quantify the coupling between electrons and phonons, with a potential impact on the quantitative determination of Raman scattering \cite{Piscanec2004,Venanzi2023,Gruneis2009}, excited carriers relaxation \cite{Kim2011,Marini2021,Harb2006,Bernardi2014,Betz2013,Tielrooij2018}, and superconductivity \cite{Bardeen1957,Schrieffer1999,Marsiglio2020}. 

The paper is outlined as follow: in Sec. \ref{sec:gen} we outline the problem; in Sec. \ref{sec:scat} we derive the general expression for the inverse carrier lifetime due to long-range interactions as a function of the dielectric functions; in Sec. \ref{sec:BTE} we present the coupled Boltzmann equations; in Sec. \ref{sec:resp} we present the form of the dielectric functions and of electrodynamical  excitations, highlighting their nature; in Sec. \ref{sec:localfields} we discuss the treatment of short-range interactions, or local field effect; in Sec. \ref{sec:compmeth} we describe our computational methodology; in Sec. \ref{sec:results} we present the results regarding the simple parabolic systems; finally, in Sec. \ref{sec:concl} we draw our conclusions. Most sections are divided in two subsections, separately discussing the parabolic band materials and the VED case.

\section{General overview of the problem}
\label{sec:gen}
In a closed, defect-free system composed of electrons and phonons, and in the presence of an external driving force, 
quasi-momentum is conserved in absence of Umklapp scattering \cite{peierls1955quantum}, i.e. in absence of quasi-momentum transfer from the quasiparticles of the crystal to the crystal itself, and current degradation cannot happen. 
This is the case if e-e and e-ph are of macroscopic nature, and the hydrodynamic regime is then established. This is a well known phenomena in the study of the thermal conductivity \cite{PhysRev.113.1046,coulter2025coupledelectronphononhydrodynamicsviscous} and electronic conductivity in the regimes where normal e-e collisions dominate \cite{doi:10.1126/science.aad0201}. Macroscopic e-ph interactions appear in crystal in the Fr\"olich form \cite{PhysRevB.13.694,PhysRevLett.115.176401}, which has non-null average over the unit cell of the crystal. In this case, Umklapp scattering is completely negligible. It is then fundamental to invoke another intrinsic mechanisms that indirectly degrade the electronic current, namely anharmonic scattering \cite{peierls1955quantum}. This acts as follows: the electric field drives out of equilibrium the electrons which, as a consequence,  scatter more frequently with phonons. The phonons are thus in turn driven out of equilibrium, and scatter more frequently with other phonons and electrons, dissipating quasi-momentum when the interaction is of the anharmonic kind. 

On the contrary, when short-range e-ph interactions dominate, the quasi-momentum transfer from electrons to the crystal is very large. This is because short-range interactions allow for Umklapp scattering and therefore are expected to degrade the electronic current. 

Both the above scattering types are present in a realistic material, and we will see in the next sections how to treat them. Nonetheless, in the above picture electrons contributing to the conductivity (free-carriers) are considered as test-charges. 
In the realistic case, many electrons contribute to transport, and they interact with each other with the Coulomb interaction, which lead to exchange and correlation effects driven by the screened Coulomb interaction $w$ \cite{PhysRev.139.A796,Reining_16}. We consider $w$ in the RPA approximation, i.e. in a mean field approach. Then, the role of the Coulomb interaction is double: it modifies the screening properties of the medium via the typical Lindhard response function \cite{Giuliani2005}, and it introduces the additional scattering channel between electrons, which does not dissipate quasi-momentum directly but can indirectly impact the transport properties of the system. We will see that both effects are indeed embedded in the dynamical dielectric response of the material. In particular, scattering is related to the imaginary part of the dielectric function, which for the free-carriers is connected to the generation of electron-hole pairs excitations that mediate e-e scattering.

The simplest conserving theory that can describe the above mechanisms altogether is the Boltzmann Transport Equation (BTE). Indeed, it has been shown \cite{PhysRevB.99.165107} that using the BTE for electrons is equivalent to consider all the ladder vertex corrections to the current-current response function, even in the case of inelastic scattering, in the semiclassical approximation for the electronic Green's function \cite{PhysRev.131.993,lihm2025beyondquasiparticlestransportvertexcorrection}. We extend the use of the BTE, inspired by the work of by Hauber et al. \cite{Hauber2017a}, to study the dynamical coupling between electrons and excitations (phonons and electron-hole pairs). To justify the form of the BTEs, we start from the relation between the electronic inverse lifetime $\tau^{-1,\rm{el}}$ and the electronic self-energy $\Sigma^{\rm{el}}$. Under precise approximations, the self-energy is directly proportional to the frequency integral of imaginary part of $w$, ie. to the system dielectric function times the Coulombian -Im$\epsilon^{-1}_{\rm tot}v$ \cite{Giustino2017,PhysRevX.13.031026}, where -Im$\epsilon^{-1}_{\rm tot}$ contains the electrodynamic excitations of the system (phonons, plasmons and electron-hole pairs in our case). We then use the form of $\tau^{-1,\rm{el}}$ to deduce the collisional integral for the electronic BTE and, in light of the detailed balance principle, for the BTE of the excitations. The only ingredient missing at this point is the inclusion of anharmonicity for the modes that appear as excitations of -Im$\epsilon^{-1}_{\rm tot}$, i.e. polar phonons in our model. Recent developments of first-principles treatments of anharmonicity \cite{PhysRevB.86.115203,PhysRevB.88.045430,PhysRevB.87.214303,PhysRevB.106.024312,PhysRevX.12.041011,Monacelli_2021,PhysRevB.87.104111} allow for a precise account of phonon-phonon interactions to be inserted in the excitations BTE. That is, if phonon excitations are well define quasiparticles.

Here, we encounter a more fundamental problem: polar phonons excitations are not always well defined Lorentzians in the momentum-frequency region that is relevant for transport regime, i.e. when electron-hole pair creation is possible. This is due to strong interference between electronic and phononic polarizabilities, that sum additively in the determination of $\epsilon_{\rm tot}$, but not in $\epsilon^{-1}_{\rm tot}$ (notice that interference in interactions is also naturally embedded in the electronic self-energy \cite{PhysRevB.90.115134,PhysRevX.13.031026}). This effect is strong in our system due to reduced dimensionality: in three dimensions one single plasmon frequency can be used to model the electronic dielectric function and catch the screening effects on the phonons \cite{PhysRevLett.133.116402}, whereas in two dimensions the plasmon is `acoustic' \cite{RevModPhys.54.437} (i.e. energy vanishes at zero momentum), so there is not a well defined reference frequency to expand the electronic response (as in the so-called plasmon pole approximation \cite{PhysRevB.88.125205}). In two dimensions, the continuum of electron-hole excitations strongly modifies both the phonon dispersion and its spectral shape in a wide range of doping levels. 
The effect on the spectral shape is mitigated when the carrier density is very high, so that the metallic-like static screening limit is approached \cite{Macheda2022,Macheda2023} and a Lorentzian is recovered. Conversely, one can of course neglect the screening mechanism altogether when the carrier density is very low.
Nevertheless, many realistic cases correspond to intermediate cases. In all the previous treatments, the phonon has been always considered as a well-defined excitation \cite{PhysRevB.86.165422,Sanborn1995,Kim1978,Hauber2017a}, and anharmonicity was modeled based on the dispersion and spectral shape of unscreened phonons.

Our approach is to individuate instead a frequency region where anharmonic dissipation is reasonably expected to take place. This is done in the following way: we consider electrons and phonons systems as separate entities, and consider how they screen each other when merged together. The regions where electrons screen phonons are associated to phonon excitations, and are quantified by the introduction of a `phonon content' function that differs from previous definitions~\cite{Ong2012,Hauber2017a}. 
in the excitations BTE. The details of this procedure are given in the main text. 
This procedure is intended to give a reasonable estimate of the dissipation effects  in the system, in the same spirit of employing density-dependent dissipative scattering rate in fluidodynamic approaches \cite{PhysRevB.92.165433}. 
With our approach, we implicitly assume that quasi-momentum dissipation is contributed also by the electron-hole pairs that dress the polar phonon. While this is customarily taken into account in the standard approach of anharmonic coupling in absence of carriers, where the energy required to create the longitudinal optical (LO) phonon is larger than the transverse optical (TO) frequency due to the medium's polarization, this is not as clear for the case where free-carrier screening is taken into account dynamically. We assume here that the energy to create a phonon should include the energy to create the electron-hole pairs dressing it, and leave the investigation of this assumption for future works.

We finally mention a couple of interesting points. The nonequilibrium dynamics of the coupled system under the influence of an electric field could be in principle treated beyond the BTE approach, perturbatively computing the electronic current-current response in Matsubara formalism or on the Keldysh contour \cite{Stefanucci_vanLeeuwen_2013} with conserving approximations, further introducing realistic anharmonic effects that break momentum conservation. The presence of phonon-related scattering events and of a phonon out-of-equilibrium population in presence of an electric field here implies, at steady state, lattice-mediated heat flow (phonon Peltier effect) \cite{Protik2022}. This effect may be relevant for the determination of thermoelectric coefficients, depending on the experimental condition under which they are determined and the magnitude of anharmonic scattering\cite{PhysRevB.94.085204,Grasselli2021}. An alternative route would be to study the time-evolution of the statistical distributions of quasiparticles on the Keldysh contour \cite{MAHAN1987251}. In these ways, one shall be able to theoretically take into account the exact spectral representation of Green's functions in a coherent manner, as done for example in Ref. \onlinecite{PhysRevLett.134.186401} for the electron-phonon case, for all kind of interactions. On a more practical note, an interesting avenue is to study the impact of the electronic component of $w$ on transport beyond the RPA approximation, following e.g. the recent developments of Refs. \onlinecite{PhysRevB.111.075118,PhysRevB.111.075137,y1dn-m6pc}. We leave all these interesting aspect for future studies.

\section{The scattering time}
\label{sec:scat}
Let's consider a quasi-2d system with in-plane periodicity, made of $N_{\rm el}$ electrons and $N_{\rm N}$ nuclei. The non-interacting electronic Hamiltonian describes Bloch states with band and quasi-momentum indexes $(n,\mathbf{k})$, energies $\varepsilon_{n\mathbf{k}}-E_{\rm F}$, $E_{\rm F}$ being the Fermi level, and wavefunctions $\psi_{n\mathbf{k}}(\mathbf{r},z)=e^{i\mathbf{k}\cdot \mathbf{r}}u_{n\mathbf{k}}(\mathbf{r},z)$, where $u_{n\mathbf{k}}(\mathbf{r},z)$ is left invariant by a lattice translation. In this section, we are interested in evaluating the scattering rate $\tau^{-1,\rm el}_{n\mathbf{k}}$ (or inverse lifetime), intended as the decay rate for the retarded electronic Green's function (see App. \ref{app:Mahan} for the comparison with definitions based on the Fermi golden rule). We assume in the following that the real part of the self-energy is negligible \footnote{In principle, for the model that we will consider, its influence can be thought as reabsorbed in the electronic effective mass}, therefore the quasi-particle renormalization factors are taken as unity. We can then write  \cite{Giustino2017}
\begin{align}
\tau^{-1,\rm el}_{n\mathbf{k}}=-\frac{2}{\hbar}\rm{Im} \Sigma^{\rm el, R}_{n\mathbf{k}},
\label{eq:taueldef}
\end{align}
where $\Sigma^{\rm el, R}$ is the diagonal component of the retarded electronic self-energy in the Bloch representation.
To obtain an expression for $\Sigma^{\rm el, R}$, we start from the expression for the time-ordered self-energy $\Sigma^{\rm el, T}$ in real space, i.e. the one expressed via the Hedin's equations that involve the time-ordered interacting electronic Green's function $G^{\rm T}$. With the same notation as Ref. \onlinecite{Giustino2017}, 
\begin{align}
\Sigma^{\rm el, T}(12)=i\hbar\int d(34) G^{\rm T}(13) \Gamma(324) w(41^+),
\label{eq:selfenint}
\end{align}
where we introduced the notation $(\mathbf{r_{1}}t_1)=1,(\mathbf{r_{1}}t_1+\eta)=1^+$, for small $\eta$. $\Gamma$ is the vertex, and $w$ is the screened Coulomb interaction
\begin{align}
w(12)=\int d3 \epsilon^{-1}_{\rm tot}(13)v(32).
\end{align}
In the above equation $\epsilon^{-1}_{\rm tot}$ is the response function that describes the screening of the Coulomb kernel due to the interactions of the system. When split between its clamped nuclei component and the reminder, it represents the screened electron-electron (e-e) and electron-phonon (e-ph) interactions. $\epsilon^{-1}_{\rm tot}$ is a key central item of this work, and its form will be specified later. 

We will perform the following simplifications to Eq. \ref{eq:selfenint}:
\begin{enumerate}
\item $G$ is taken as the non-interacting Green's function $G^0$, i.e. the Green's function involving independent particle energies and wavefunctions, whose explicit expression is
\begin{align}
G(\mathbf{r},\mathbf{r'},z,z',\omega)=2\sum_{n\mathbf{k}}\frac{\psi_{n\mathbf{k}}(\mathbf{r},z)\psi^*_{n\mathbf{k}}(\mathbf{r'},z')}{\omega-\varepsilon_{n\mathbf{k}}+E_{\rm F}-i\delta_{n\mathbf{k}}},
\end{align}
where $\delta_{n\mathbf{k}}=0^+$ if $\varepsilon_{n\mathbf{k}}<E_{\rm F}$ and $\delta_{n\mathbf{k}}=0^-$ if $\varepsilon_{n\mathbf{k}}>E_{\rm F}$ and we have separated in-plane ($\mathbf{r}$) and out-of-plane ($z$) variables; 
\item the vertex term is neglected, i.e. $\Gamma(324)=\delta(34)\delta(24)$; i.e., we disregard both electronic and vibrational (Migdal) vertex corrections in both the e-e and e-ph interaction vertexes in the self-energy estimation: \cite{Hedin1970}
\begin{align}
\Sigma^{\rm el, T}(12)=i\hbar G^{0,\rm T}(12) w(1^+2);
\label{eq:selfen0int}
\end{align}
\item $\epsilon^{-1}_{\rm tot}$ is evaluated neglecting vertex corrections in the linear response, i.e. in the RPA approximation;
\item the self-energy is taken as diagonal in the Bloch's basis set;
\item our systems display time-reversal symmetry.
\end{enumerate}
Within the above approximations, we evaluate the self-energy expression in Matsubara frequency, with the aim of obtaining the retarded self-energy with the substitution $i\omega_n \rightarrow \omega+i\delta$ \cite{Mahan1990}. Since we are mostly interested in the theoretical treatment of long-range macroscopic effects, we first disregard in-plane local field effects, and reintroduce them later in Sec. \ref{sec:localfields}. Specifying the Bloch functions basis set \cite{Reining_16}, the self-energy in Matsubara representation reads
\begin{align}
\Sigma_{n\mathbf{k}}(i\omega_{s})=-\frac{1}{\beta}\sum_{i\nu_p}\frac{1}{N_{\mathbf{q}}}\sum_{\mathbf{q}m}G^0_{m\mathbf{k+q}}(i\omega_s+i\nu_p)\times \nonumber \\
 w_{n\mathbf{k}m\mathbf{k+q}}(i\nu_p),
\label{eq:sigmaMatsu}
\end{align}
where the expressions for $G^0$ and $w$ are given below. We start from the spectral representation of the non interacting Green's function in the Bloch's basis set
\begin{align}
G^0_{m\mathbf{k+q}}(i\omega_s+i\nu_p)=\int_{-\infty}^{\infty}\frac{d\omega'}{2\pi}\frac{A_m(\mathbf{k+q},\omega)}{i\omega_s+i\nu_p-\omega'},\\
A_m(\mathbf{k+q},\omega)=2\pi\delta(\omega-\varepsilon_{m\mathbf{k+q}} + E_{\rm F}).
\label{eq:Am}
\end{align}
To express the self-energy, we now need
only the diagonal part of the RPA screened Coulomb interaction (which would usually have four quasi-momentum indexes)
\begin{align}
w_{n\mathbf{k}m\mathbf{k+q}}(i\nu_p)= \int_{0}^{\infty}\frac{d\omega'}{2\pi}\frac{2\omega'B^{\mathbf{k}}_{nm}(\mathbf{q},\omega')}{(i\nu_p)^2-(\omega')^2},\label{eq:wnolf} \\
B^{\mathbf{k}}_{nm}(\mathbf{q},\omega') =\int d\mathbf{r} d\mathbf{r'}dz dz'd\bar{z} F^{nm}_{\mathbf{k}\mathbf{k+q}}(\mathbf{r},z,\mathbf{r'},z') \nonumber \\
\times \frac{2}{A}\textrm{Im}\left[-\epsilon_{\rm tot}^{-1}(\mathbf{q},z,\bar z,\omega'+i0^+)v(\mathbf{q},\bar z,z')\right],\label{eq:Bnm}
\end{align} 
where we have defined
\begin{align}
F^{nm}_{\mathbf{k}\mathbf{k+q}}(\mathbf{r},z,\mathbf{r'},z')=u^*_{n\mathbf{k}}(\mathbf{r},z)u_{n\mathbf{k}}(\mathbf{r'},z')\times \nonumber \\
u_{m\mathbf{k+q}}(\mathbf{r},z)u^*_{m\mathbf{k+q}}(\mathbf{r'},z').
\end{align}
Notice that the imaginary part above is the typical shorthand notation for the spectral representation of the matrix case, which would be written in terms of both retarded and advanced functions; since this is not crucial to the scopes of this work, we prefer to use a shorthand notation. We now use standard Matsubara summation techniques (chapter 3.5 of Ref. \onlinecite{Mahan1990}) and obtain
\begin{align}
&\Sigma_{n\mathbf{k}}(i\omega_{s})=\frac{1}{N_{\mathbf{q}}}\sum_{\mathbf{q}m}\int_{-\infty}^{\infty} \frac{d\omega'}{2\pi}  B^{\mathbf{k}}_{nm}(\mathbf{q},\omega') \nonumber\\
&\times \int_{-\infty}^{\infty} \frac{d\omega''}{2\pi}A_{m}(\mathbf{k+q},\omega'')I(i\omega_s,\omega',\omega''),
\end{align}
where
\begin{align}
I(i\omega_s,\omega',\omega'')=\frac{n_{\omega'}+f_{\omega''}}{i\omega_s+\omega'-\omega''},
\end{align}
$n_{\omega}$ and $f_{\omega}$ being the Bose-Einstein and the Fermi-Dirac distributions. By substituting Eq. \eqref{eq:Am} and $i\omega_s\rightarrow \varepsilon_{n\mathbf{k}}-E_{\rm F}+i0^+$ we obtain
\begin{align}
\Sigma^{\rm R}_{n\mathbf{k}}(\varepsilon_{n\mathbf{k}})=\frac{1}{N_{\mathbf{q}}}\sum_{m\mathbf{q}}\int_{-\infty}^{\infty}  \frac{d\omega'}{2\pi} B^{\mathbf{k}}_{nm}(\mathbf{q},\omega') I(\varepsilon_{n\mathbf{k}},\omega'+i0^+,\varepsilon_{m\mathbf{k+q}})
\end{align}
We now pass to the scattering time using Eq. \eqref{eq:taueldef}; we notice that we can apply $-2\rm Im$ of Eq. \eqref{eq:taueldef} directly to $I$. Application of $-2\textrm{Im}$ to $I$  gives $2\pi\delta(\omega'+\varepsilon_{n\mathbf{k}}-\varepsilon_{m\mathbf{k+q}})$, so that 
\begin{align}
\tau^{-1,\rm el}_{n\mathbf{k}}=\frac{2\pi}{\hbar}\frac{1}{N_{\mathbf{q}}}\sum_{m\mathbf{q}}\int_{-\infty}^{\infty} \frac{d\omega'}{2\pi}  B^{\mathbf{k}}_{nm}(\mathbf{q},\omega')\nonumber \\
\times (n_{\omega'}+f_{m\mathbf{k+q}})\delta(\omega'+\varepsilon_{n\mathbf{k}}-\varepsilon_{m\mathbf{k+q}}).
\label{eq:prePz}
\end{align}

 Further, to bring the formula to the same form as Ref. \onlinecite{PhysRevB.110.115407}, we now perform a change of variable $\omega' \rightarrow -\omega$, use that $n_{-\omega}=-1-n_{\omega}$ and that $B^{\mathbf{k}}_{nm}(\mathbf{q},\omega')=-[B^{\mathbf{k}}_{nm}(\mathbf{q},-\omega')]^*$ (see App. \ref{app:real}) to finally obtain:
\begin{widetext}
\begin{align}
\tau^{-1,\rm el}_{n\mathbf{k}}=\frac{2\pi}{\hbar}\frac{1}{ A N_{\mathbf{q}}}\sum_{m\mathbf{q}}\int_{-\infty}^{\infty} d\omega \int d \mathbf{r}d \mathbf{r'}dzdz' d\bar z  \textrm{sign}(\omega)\mathcal{A}(\omega)  
 F^{nm}_{\mathbf{k}\mathbf{k+q}}(\mathbf{r},z,\mathbf{r'},z') \times \nonumber \\
\textrm{Im}\left[-\epsilon_{\rm tot}^{-1}(\mathbf{q},z,\bar z,\omega+i0^+)\frac{v(\mathbf{q},\bar z,z')}{\pi} \right]\delta(\omega+\varepsilon_{m\mathbf{k+q}}-\varepsilon_{n\mathbf{k}}), \quad 
\mathcal{A}(\omega)= \left[n_{|\omega|}+\frac{1}{2}+\textrm{sign}(\omega)\frac{1}{2}-\textrm{sign}(\omega)f_{m\mathbf{k+q}}\right],
\label{eq:Pz}
\end{align}
\end{widetext}
The above results are in accordance with the results of Ref. \onlinecite{ECHENIQUE20001}, and has been used e.g. in the study of angle-resolved electron reflection spectroscopy \cite{PhysRevB.111.125113} and hot carrier properties \cite{PhysRevB.106.195434}. Notice that, in Eq. \eqref{eq:Pz}, $\epsilon^{-1}$ is the retarded inverse dielectric function; in the following, retardation is always implicit and we will drop $i0^+$ specifications. The integral over frequencies running on negative frequencies gives positive contributions to the scattering rate, following from the relation shown in App. \ref{app:real} \footnote{Notice that in Ref. \cite{PhysRevB.110.115407} there was a typo in the expression of $A(\omega)$, with $n_{|\omega|}$ wrongly replaced by $n_\omega$}. Positive and negative frequencies correspond to emission and absorption, respectively, and we have used that to pass from Eq. \eqref{eq:prePz} to Eq. \eqref{eq:Pz} by coherently changing the conserving delta function. In this work we present two different versions of the above formula, in dependence of the investigated system, presented here below.

\textit{Parabolic case---} We suppose that there is an energy band $n^*$ that is the first conduction band in the neutral system, well separated from the valence ones by a large gap $\Delta_{\rm g}$ and also from the other conduction bands, that gets occupied as soon as external carriers are inserted into the system (the valence case can be treated analogously). With this assumption, the $n^*$ band totally determines the scattering rate via intraband processes. We further simplify assuming that:
\begin{enumerate}
\item the band $n^*$ has a parabolic band dispersion characterized by a certain effective mass $m_{\rm eff}$ via
$\varepsilon_{n^*\mathbf{k}}=\frac{\hbar^2 k^2}{2 m_{\rm eff}}$;
\item the Bloch functions are constant both in-plane and out-of-plane, or equivalently we approximate $\langle u_{n^*\mathbf{k}}| u_{n^*\mathbf{k+q}}\rangle \sim 1$;
\item the screening and the Coulombian do not present out-of-plane dependence, i.e. $\epsilon^{-1}_{\rm tot}(\mathbf{q},z,z')=\epsilon^{-1}_{\rm tot}(\mathbf{q})$ and $v(\mathbf{q},z,z')=v({\mathbf{q}})$, denoted for simplicity from now on as $v_{\mathbf{q}}=\frac{2\pi e^2}{q}$. 
\end{enumerate}
Eq. \ref{eq:Pz} then simplifies to
\begin{widetext}
\begin{align}
\tau^{-1, \rm el}_{n^*\mathbf{k}}=\frac{2\pi}{\hbar}\frac{1}{A N_{\mathbf{q}}}\sum_{\mathbf{q}}\int_{-\infty}^{\infty} d\omega  \textrm{sign}(\omega)\mathcal{A}(\omega)  
\textrm{Im}\left[-\epsilon_{\rm tot}^{-1}(\mathbf{q},\omega)\frac{v_{\mathbf{q}}}{\pi} \right]\delta(\omega+\varepsilon_{n^*\mathbf{k+q}}-\varepsilon_{n^*\mathbf{k}}).
\label{eq:P}
\end{align}
\end{widetext}
For the parabolic case, from now on we will drop the band index. We remind that in Eq. \eqref{eq:P} $\epsilon^{-1}_{\rm{tot}}(\mathbf{q},\omega)$ contains scattering from electrons and long-range infrared active phonons, since we took out the local fields (whose treatment is devoted to Sec. \ref{sec:localfields}). To highlight the role of interactions, Eq. \ref{eq:P} can  be conveniently rewritten in the following form
\begin{widetext}
\begin{align}
\tau^{-1, \rm el}_{\mathbf{k}}=\frac{2\pi}{\hbar}\frac{1}{ N_{\mathbf{q}}}\sum_{\mathbf{q}}\int_{-\infty}^{\infty} d\omega  \textrm{sign}(\omega)\mathcal{A}(\omega)  
g^2(\mathbf{q},\omega)\delta(\omega+\varepsilon_{\mathbf{k+q}}-\varepsilon_{\mathbf{k}}),\label{eq:Pg2}\\
\epsilon_{\rm tot}^{-1}(\mathbf{q},\omega)=1+v_{\mathbf{q}}\chi_{\rm tot}(\mathbf{q},\omega)\, , g^2(\mathbf{q},\omega)=\textrm{Im}\left[-v_{\mathbf{q}}\chi_{\rm tot}(\mathbf{q},\omega)\frac{v_{\mathbf{q}}}{A \pi}\right].
\end{align}
\end{widetext}
In the above equation, we expect the imaginary part to display peaks around the excitations of the total density-density-response function of the system. Then, the peak intensity represents the strength of the interaction between the electron and the excitations. Finally, notice that $g^2(\mathbf{q},\omega)$ is a coupling in units of energy: the physical coupling is obtained by integrating over the frequency spectrum.

\textit{VED case---} For the VED case we follow the developments of Ref. \onlinecite{PhysRevB.110.115407}. Here we are interested in vdW heterostructures. The VED model builds the electrodynamic (density-density) response of the heterostructure $\chi_{kl}$ from the response of single layers computed in DFT. The electron-mode coupling $g^2_{kl}$ is then expressed from this quantity. 
Response functions are computed within a dual-basis set approach, that takes into account both the in-plane and out-of-plane response of each single layer ($k=1,..,N$) using a set of perturbing potentials $\phi^i_k \,(i=0,1)$ and a set of response densities $f^i_k \,(i=0,1)$. The natural object in this framework is the density-response of the heterostructure, so that the inverse lifetimes is expressed as a generalization of Eq. \ref{eq:Pg2}. Indeed, under the approximation that single-layer Bloch functions can be taken as even with respect to the out-of-plane direction, the inverse lifetime for a quasi-twodimensional Bloch state reads
\begin{widetext}
\begin{align}
\tau^{-1,\textrm{el}}_{n\mathbf{k}}=\frac{2\pi}{\hbar} \frac{1}{N_{\mathbf{q}}}\sum_{\mathbf{q}}\int^{\infty}_{-\infty} d\omega \textrm{sign}(\omega)\mathcal{A}(\omega)
\sum_{mkl}\langle u_{n\mathbf{k}}|\phi^0_k|u_{m\mathbf{k+q}}\rangle \langle u_{m\mathbf{k+q}}|\phi^0_l|u_{n\mathbf{k}}\rangle 
 g^2_{kl}(\mathbf{q},\omega) \delta(\hbar\omega+\varepsilon_{m\mathbf{k+q}}-\varepsilon_{n\mathbf{k}}),\\
 g^2_{kl}(\mathbf{q},\omega)=\sum_{ijk'l'}\textrm{Im}\left[-   v^{0i}_{kk'}(\mathbf{q}) \chi^{ij}_{k'l'}(\mathbf{q}, \omega)  \frac{v^{j0}_{l'l}(\mathbf{q})}{A\pi} \right]
\label{eq:lifetimeVED}
\end{align}
\end{widetext}
where $\phi^0_k$ simply evaluates to unity in a properly chosen box centered around the layer $k$, and the $v^{ij}_{kl}$ and $\chi^{ij}_{kl}$ are the Coulombian and the density-density response expressed in their appropriate basis representation \cite{PhysRevB.110.115407}. For the case of BN-encapsulated graphene, the lifetime of a graphene state is obtained by considering $k,l$ as indexes corresponding to the graphene layer, since the wavefunctions are localized on such layer.

\section{The Boltzmann transport equations}
\label{sec:BTE}
$\tau^{-1,\rm el}$ given by Eqs. \ref{eq:Pz} and \ref{eq:lifetimeVED} is notoriously different from the exact transport scattering time $\tau^{-1,\rm tr}$ stemming from the BTE. An historical approximation to compute $\tau^{-1,\rm tr}_{\rm el}$ is to insert a term $(1-\frac{v_{\mathbf{k}}\cdot v_{\mathbf{k+q}}}{|v_{\mathbf{k}}|^2} )$ under the integral of Eq. \ref{eq:Pz}, which weights the scattering processes in favor of large angles and that for spherical bands reduces to $(1-\cos \theta)$ \cite{PhysRevB.3.305}. The most recents ab-initio developments\cite{Macheda2018,PhysRevMaterials.2.114010,Macheda2020,PhysRevB.94.085204,Sohier2021,Sohier2023} allowed the exact computation of $\tau^{-1,\rm tr}$ from the solution of the BTE, in the case where couplings are screened statically and the effect of electron-electron scattering is neglected. 

Only few attempts have been done to determine $\tau^{-1,\rm tr}$ in presence of dynamical interactions, and including electron-electron scattering in a consistent way, the most convincing being the work of Hauber et al. \onlinecite{Hauber2017a}. We approach the problem with their same spirit, but extend the treatment beyond simple parabolic bands and well-defined Einstein phonons, by using our VED formalism.

We consider two coupled Boltzmann equations for both the electrons and the excitations of the system, represented by the resonances of $-\rm{Im}\epsilon^{-1}_{\rm tot}(\mathbf{q},\omega)$. We start by expanding the out-of-equilibrium statistical occupations of the system excitations and of the electrons as
\begin{align}
N({\mathbf{q},\omega})=n_{\omega}-\frac{\partial n}{\partial \omega}\mathcal{G}(\mathbf{q},\omega),\\
F_{n\mathbf{k}}=f_{n\mathbf{k}}-\frac{\partial f}{\partial \varepsilon_{n\mathbf{k}}}\Phi_{n\mathbf{k}},
\end{align}
where we have assumed unitary electric fields, that in the following we will consider along the $\hat x$ direction without losing generality since our system is assumed to be isotropic in the in-plane direction. The electronic current, in the semiclassical BTE picture, is written as 
\begin{align}
J_x=\frac{2e}{AN_{\mathbf{k}}}\sum_{\mathbf{k}}[v_{n\mathbf{k}}]_x \frac{\partial f}{\partial \varepsilon_{n\mathbf{k}}}\Phi_{n\mathbf{k}}
\label{eq:current}
\end{align}
from which the conductivity can be determined if $\Phi$ is known. In order to determine $\Phi$ we consider the following generalization of the collisional integral of the coupled Boltzmann equations to the dynamical case \cite{Hauber2017a}
\begin{widetext}
\begin{align}
\frac{\partial{F}_{n\mathbf{k}}}{\partial t}\Big{|}^{\textrm{coll}}=-\frac{2\pi}{\hbar k_B T N_\mathbf{q}}\sum_{m\mathbf{q}}\int_{-\infty}^{+\infty}d\omega f_{m\mathbf{k+q}}[1-f_{n\mathbf{k}}][n_\omega+1]\Big[ \Phi_{n\mathbf{k}}-\Phi_{m\mathbf{k+q}}
+\mathcal{G}(\mathbf{q},\omega)\Big]\mathcal{T}^{\mathbf{k}}_{nm}(\mathbf{q},\omega)\delta(\varepsilon_{m\mathbf{k+q}}-\varepsilon_{n\mathbf{k}}-\omega), \label{eq:BTEel}\\
\frac{\partial{N}}{\partial t}\Big{|}^{\textrm{coll}}=\frac{2\times2\pi}{\hbar k_B T N_\mathbf{k}}\sum_{nm\mathbf{k}} \int^{+\infty}_{-\infty} d\omega \Big[\Phi_{n\mathbf{k}}-\Phi_{m\mathbf{k+q}}+\mathcal{G}(\mathbf{q},\omega)\Big]n_\omega[n_\omega+1][f_{m\mathbf{k+q}}-f_{n\mathbf{k}}]\mathcal{T}^{\mathbf{k}}_{nm}(\mathbf{q},\omega)\delta(\varepsilon_{m\mathbf{k+q}}-\varepsilon_{n\mathbf{k}}-\omega). \label{eq:BTEphomegaint}
\end{align}
\end{widetext}
\textit{Parabolic case---} In the parabolic case, in the BTE equations we set $m=n$, with
\begin{align}
\mathcal{T}^{\mathbf{k}}_{nm}(\mathbf{q},\omega)=\mathrm{Im}\left[-\epsilon_{\mathrm{tot}}^{-1}(\mathbf{q},\omega)\frac{v_{\mathbf{q}}}{A\pi}\right]\delta_{mn}.
\end{align}
\textit{VED case---} For the VED case, we recover
\begin{align}
\mathcal{T}^{\mathbf{k}}_{nm}(\mathbf{q},\omega)=\sum_{kl}\langle u_{n\mathbf{k}}|\phi^0_k|u_{m\mathbf{k+q}}\rangle \langle u_{m\mathbf{k+q}}|\phi^0_l|u_{n\mathbf{k}}\rangle g^2_{kl}(\mathbf{q},\omega)
\end{align}
Notice that the scattering time of Eq. \eqref{eq:BTEel} in the SERTA approximation \cite{Ponce2021}, i.e. disregarding repopulation terms proportional to $\Phi_{m\mathbf{k+q}}$, has occupations factors that are in the same form of Eqs. \eqref{eq:prePz} or \eqref{eq:lifetimeVED} by using that 1) $f_{m\mathbf{k+q}}(1-f_{n\mathbf{k}})(n_{\omega}+1)=f_{n\mathbf{k}}(1-f_{m\mathbf{k+q}})n_{\omega}$ and 2) $f_{n\mathbf{k}}(1-f_{m\mathbf{k+q}})n_{\omega}=(n_{\omega}+f_{m\mathbf{k+q}})f_{n\mathbf{k}}(1-f_{n\mathbf{k}})$ for energies that satisfy the conservation energy $\delta(\varepsilon_{n\mathbf{k}}-\varepsilon_{m\mathbf{k+q}}+\hbar \omega)$.
\subsubsection{Peierls' argument and $G$ expression}
As pointed out by Peierls for the single parabolic band case \cite{peierls1955quantum}, in stationary conditions functions of the form $\tilde{\Phi}_{n\mathbf{k}}=\lambda_n \mathbf{k}$ and $\tilde{G}(\mathbf{q},\omega)=\lambda(\omega) \mathbf{q}$, for proper $\lambda_n$ and $\lambda(\omega)$, can nullify the collisional integral of the electronic BTE of Eqs. \eqref{eq:BTEph} and \eqref{eq:BTEel}, i.e. can be non trivial stationary solutions of the coupled BTEs in absence of electric field. In other words, these functions allow for non-zero out-of-equilibrium populations that contribute to current via Eq. \eqref{eq:current}, even in absence of an external electric field. 
Physically, $\tilde{\Phi}$ and $\tilde{G}$ are admittable solutions if the quasi-momentum of the full system is conserved, i.e. in the case where Umklapp scattering is not present in our system; indeed, in presence of Umklapp scattering we would have that $\Phi_{\mathbf{k+q}}\propto \lambda[\mathbf{k+q+G}]$ were G is a generical reciprocal lattice vector (see Eq. (9.14.4) of \onlinecite{ziman2001electrons}). The success of Peierls' argument is inherently connected to the inability of a close system alone to degrade quasi-momentum in absence of Umklapp scatterings (see App. \ref{app:solconv} for the numerical proof of Peierl's argument in absence of Umklapp scattering). 

To close the Boltzmann Equations we need to equate the collisional integrals Eqs. \eqref{eq:BTEel} and \eqref{eq:BTEph} to the rate of population variation stemming from diffusion processes and the action of external forces. At steady state and for the electronic problem of Eq. \eqref{eq:BTEel}, in absence of thermal gradient and presence of a unitary  electric field along $\hat x$ we can equate \cite{ziman2001electrons}
\begin{align}
\frac{\partial{f_{n\mathbf{k}}}}{\partial t}\Big{|}^{\textrm{coll}}=-e\frac{\partial f}{\partial \varepsilon_{n\mathbf{k}}}[v_{n\mathbf{k}}]_x.
\end{align}
For Eq. \eqref{eq:BTEph}, the choice is less straightforward. As for the electronic case, we impose the absence of thermal gradients of any sort. Then, no other forces act directly on the bosonic excitations degrees of freedom, since even in presence of polar phonons the sum of the Born effective charges is zero and therefore the net force exerted by the electric field is zero---this holds exactly for semiconductors, and approximately for lightly doped semiconductors, otherwise a different sum rule for metals exists \cite{PhysRevLett.128.095901,Marchese2023}. We are left with the explicit time variation of the out-of-equilibrium distribution function, that we attribute to the decay induced by the anharmonic interaction between phonons. Importantly, the anharmonic interaction enables quasi-momentum degradation of the system via the Umklapp interaction with the crystal. This can in principle be treated from a first-principles perspective \cite{PhysRevB.96.014111,PhysRevB.87.104111,PhysRevB.87.214303}, but this goes beyond the scopes of our work. We instead model the anharmonic decay as
\begin{align}
\frac{\partial{N}}{\partial t}\Big{|}^{\textrm{coll}}=\frac{\partial{N}}{\partial t}\Big{|}_{\mathrm{ANH}}=-\int_{-\infty}^{\infty} d\omega \frac{\partial n}{\partial \omega}\mathcal{G}(\mathbf{q},\omega)\tau^{-1}_{\rm{ANH}}(\mathbf{q},\omega), \\
 \tau^{-1}_{\rm ANH}(\mathbf{q},\omega)= \tau^{-1}_{\rm anh}\mathcal{F}(\mathbf{q},\omega)
\label{eq:taum1anh},
\end{align}
where $\mathcal{F}(\mathbf{q},\omega)$ is a function which determines the `phonon content' of an excitation \cite{PhysRevB.86.165422}, and whose form will be discussed in Sec. \ref{sec:phcontent} along side with the values of $\tau^{-1}_{\rm anh}$---for which we dropped the eventual $\omega$ dependence, since in this work we take it as frequency independent, as discussed later. Quite generally, $\mathcal{F}(\mathbf{q},\omega)$ has to be zero at every frequency if no phonon excitations are present, and it is sharply peaked at the phonon frequency if a well-defined phonon quasi-particle excitation exists. $\tau^{-1}_{\rm ANH}(\mathbf{q},\omega)$ instead is the inverse of a decay time per unit energy. Eq. \eqref{eq:BTEphomegaint} can then be imposed frequency by frequency, obtaining

\begin{widetext}
\begin{align}
-\frac{\partial n}{\partial \omega}\mathcal{G}(\mathbf{q},\omega)\tau^{-1}_{\rm{ANH}}(\mathbf{q},\omega)=\frac{2\times2\pi}{\hbar k_B T N_\mathbf{k}}\sum_{nm\mathbf{k}} \Big[\Phi_{n\mathbf{k}}-\Phi_{m\mathbf{k+q}}+\mathcal{G}(\mathbf{q},\omega)\Big]n_\omega[n_\omega+1][f_{m\mathbf{k+q}}-f_{n\mathbf{k}}]\mathcal{T}^{\mathbf{k}}_{nm}(\mathbf{q},\omega)\delta(\varepsilon_{m\mathbf{k+q}}-\varepsilon_{n\mathbf{k}}-\omega). \label{eq:BTEph}
\end{align}
\end{widetext}

Notice that assuming a decay time for the population due to an external process (anharmonic scattering) is equivalent to consider anharmonic interactions explicitly as interactions within the system in the relaxation time approximation at the steady state, i.e. considering them into the collisional integral and approximating the scattering matrix to its diagonal components. This equivalence holds true only if the thermal gradient is always set to zero. 
This does exclude the rise of phonon Peltier effects, which are present  if realistic finite lifetimes are present (even though usually thought of secondary importance for the specific purpose of a routine electrical conductivity measurement). These could be evaluated computing anharmonic relaxation times via first-principles calculations, or better one could include the full anharmonic scattering matrix inside the collisional term \cite{Protik2022}.  A detailed investigation of this delicate point goes beyond the scopes of this work, and we leave it to further studies.

With Eq. \eqref{eq:taum1anh} we can rewrite Eq. \eqref{eq:BTEph} for our two cases of study.
\begin{widetext}
\textit{Parabolic case---}
\begin{align}
\mathcal{G}(\mathbf{q},\omega)=\frac{\frac{2\times2\pi}{\hbar A} \mathrm{Im}\left[-\epsilon_{\mathrm{tot}}^{-1}(\mathbf{q},\omega)\frac{v_{\mathbf{q}}}{\pi}\right]\frac{1}{N_{\mathbf{k}}}\sum_{\mathbf{k}}\left[\Phi_{\mathbf{k}}-\Phi_{\mathbf{k+q}}\right][f_{\mathbf{k+q}}-f_{\mathbf{k}}]\delta(\varepsilon_{\mathbf{k+q}}-\varepsilon_{\mathbf{k}}-\omega)}{\tau^{-1}_{\mathrm{ANH}}(\mathbf{q},\omega)+\frac{2}{\hbar} \mathrm{Im}\left[-\epsilon_{\mathrm{tot}}^{-1}(\mathbf{q},\omega)\frac{v_{\mathbf{q}}}{\pi}\right]\mathrm{Im}[\epsilon_{\mathrm{el}}(\mathbf{q},\omega)]/v_\mathbf{q}}
\label{eq:Gdet}
\end{align}
where we used that
\begin{align}
\mathrm{Im}[\epsilon_{\mathrm{el}}(\mathbf{q},\omega)]/v_\mathbf{q}=\frac{2\pi}{AN_{\mathbf{k}}}\sum_{\mathbf{k}}\left[f_{\mathbf{k}}-f_{\mathbf{k+q}}\right]\delta(\varepsilon_{\mathbf{k+q}}-\varepsilon_{\mathbf{k}}-\omega)
\end{align}
\textit{VED case---}
\begin{align}
\mathcal{G}(\mathbf{q},\omega)=\frac{\frac{2\times2\pi}{\hbar} \frac{1}{N_{\mathbf{k}}} \sum_{kl}\sum_{nm\mathbf{k}} g^2_{kl}(\mathbf{q},\omega)\left[\Phi_{n\mathbf{k}}-\Phi_{m\mathbf{k+q}}\right][f_{m\mathbf{k+q}}-f_{n\mathbf{k}}]\langle u_{n\mathbf{k}}|\phi^0_k|u_{m\mathbf{k+q}}\rangle \langle u_{m\mathbf{k+q}}|\phi^0_l|u_{n\mathbf{k}}\rangle \delta(\varepsilon_{m\mathbf{k+q}}-\varepsilon_{n\mathbf{k}}-\omega)}{\tau^{-1}_{\mathrm{ANH}}(\mathbf{q},\omega)+\frac{2A}{\hbar} \sum_{kl}  g^2_{kl}(\mathbf{q},\omega)\mathrm{Im}[\epsilon_{\mathrm{el}}(\mathbf{q},\omega)]_{kl}/v_\mathbf{q}}
\label{eq:Gdet}
\end{align}
\end{widetext}
where we used that 
\begin{align}
\mathrm{Im}[\epsilon_{\mathrm{el}}(\mathbf{q},\omega)]_{kl}/v_\mathbf{q}=\frac{2\pi}{AN_{\mathbf{k}}}\sum_{nm\mathbf{k}}\left[f_{n\mathbf{k}}-f_{m\mathbf{k+q}}\right]\times \nonumber \\
\sum_{kl}\langle u_{n\mathbf{k}}|\phi^0_k|u_{m\mathbf{k+q}}\rangle \langle u_{m\mathbf{k+q}}|\phi^0_l|u_{n\mathbf{k}}\rangle\delta(\varepsilon_{m\mathbf{k+q}}-\varepsilon_{n\mathbf{k}}\omega)
\end{align}
\subsubsection{Implications of Peierl's argument}
\textit{Parabolic case---} The form of Eq. \eqref{eq:taum1anh} is chosen because we have the following limiting behaviours. First, we consider the case where $\tau^{-1}_{\rm ANH}(\mathbf{q},\omega)=0$ everywhere. This is obtained if phonons are absent (zero phonon content everywhere) even in presence of doping, or if phonons are present but we set $\tau^{-1}_{\rm anh}=0$ $\forall \omega$ (absence of anharmonic interactions). In this case we have
\begin{align}
&\mathcal{G}(\mathbf{q},\omega)=\nonumber \\
&\frac{2\pi}{A}\frac{1}{N_{\mathbf{k}}}\frac{\sum_{\mathbf{k}}\left[\Phi_\mathbf{k}-\Phi_\mathbf{k+q}\right][f_{\mathbf{k+q}}-f_{\mathbf{k}}]\delta(\varepsilon_{\mathbf{k+q}}-\varepsilon_{\mathbf{k}}-\omega)}{\mathrm{Im}[\epsilon_{\mathrm{el}}(\mathbf{q},\omega)]/v_\mathbf{q}}.
\label{eq:Gelalone}
\end{align} 
In absence of phonons, by inserting this expression for $G$ inside Eq. \eqref{eq:BTEel} and considering that $\epsilon^{-1}_{\rm tot}=\epsilon^{-1}_{\rm el}$, the collisional integral of the electronic BTE is transformed in the form of an electron-electron scattering, once the integration over the frequency is performed \cite{Sanborn1995}, proportional to 
\begin{align}
\sum_{\mathbf{q}\mathbf{k'}} |\frac{v_{\mathbf{q}}}{\epsilon_{\rm el}({\mathbf{q},\varepsilon_{\mathbf{k+q}}-\varepsilon_{\mathbf{k}}})}|^2f^0_{\mathbf{k+q}}[1-f^0_{\mathbf{k}}]f^0_{\mathbf{k'}}(1-f^0_{\mathbf{k'+q}}) \times \nonumber \\
\Big[\Phi_\mathbf{k}-\Phi_\mathbf{k'}-\Phi_\mathbf{k+q}+\Phi_\mathbf{k'+q}\Big]\times \nonumber \\
\delta(\varepsilon_{\mathbf{k'+q}}-\varepsilon_{\mathbf{k'}}-\varepsilon_{\mathbf{k+q}}+\varepsilon_{\mathbf{k}}),\label{eq:collelonly}
\end{align}
which represents the collision integral typical for electron-electron interactions. Notice that the Perierl's argument here holds with $\tilde{\Phi}_\mathbf{k}=\lambda \mathbf{k}$ (and consequently $\tilde{G}(\mathbf{q},\omega)=\lambda \mathbf{q}$) so that conductivity is infinite---in the parabolic case, noting that $\Phi_{\mathbf{k}}\propto \tau_{\mathbf{k}}\mathbf{v}_{\mathbf{k}}$ \cite{PhysRevB.94.085204}, these solutions correspond to the case of a constant relaxation time, since the velocity is already proportional to the momentum. The presence of such solutions is inherently connected to the inability of electron-electron interactions alone to scatter momentum in absence of Umklapp scattering, which is correctly capture by the form of $\tau^{-1}_{\rm{ANH}}$ if $\tau^{-1}_{\rm anh}$ is zero, as confirmed by numerical results (see App. \ref{app:solconv}).

Secondly we consider the case where doping is disregarded, i.e. in absence of electron-hole pairs, and $\tau^{-1}_{\rm anh}$ is large. By large, we mean that the rate at which phonons relax due to anharmonic interactions is quick. As a consequence, the impact of the out of equilibrium phonon population on electronic transport properties is negligible. Notice that, in this sense, a large scattering rate for phonons is pretty typical at room temperature, so that historically calculations have been mostly performed in the Bloch's ansatz, i.e. by considering the phonons in perfect thermal equilibrium. Notice also that, despite having a large scattering rate, in this regime phonons can usually still be well described as quasi-particles, with a sharply peaked function around $\omega=\pm\omega_{\rm ph}$, since it still usually holds that $\tau^{-1}_{\rm ANH}(\omega_{\rm ph})\ll\omega_{\rm ph}$. Now, in this regime our choice of Eq. \eqref{eq:taum1anh} implies that $\mathcal{G}(\mathbf{q},\pm\omega_{\rm ph})=0$, as shown in Sec. \ref{sec:resp}. Moreover, since electron-hole excitation pairs are not present, $\mathcal{G}(\mathbf{q},\omega)=0$ $\forall \omega$. Therefore, Eq. \eqref{eq:BTEel} falls back on the frequency independent BTE obtained via the the Bloch's ansatz.

In the realistic case where both phonon and electron-hole pairs excitations are present, for large $\tau^{-1}_{\rm anh}$ and for frequencies near the phonon we have $\mathcal{G}(\mathbf{q},\omega \sim \omega_{\rm ph})=0$, while for all the other frequencies $G$ is given by Eq. \eqref{eq:Gelalone}. Electron-electron interactions are expected to mostly cancel, but not completely, giving rise to finite contributions to the conductivity. Indeed, now we have that $\tilde{\Phi}(\mathbf{k})=\lambda \mathbf{k}$ and $\tilde{G}(\mathbf{q},\omega)=\lambda \mathbf{q}$ are no more physically (and numerically) admittable solutions.

\textit{VED case---} The VED case is very similar to the parabolic case in terms of physical conclusions, with the adequate generalization to the multiband case. Indeed, Eqs. \eqref{eq:Gelalone} and \eqref{eq:collelonly} are now complemented with a sum over appropriate band indexes, and wavefunctions overlaps (we omit the full formula for brevity). The only difference concerns the application of Perierl's argument. Taking as an example the case of BN-encapsulated graphene, and concentrating on the graphene layer, it is clear that the Fermi velocity is no more proportional to the momentum, but rather it is constant in modulus and radial in direction, pointing inward/outward for the lower/upper cones.

Nevertheless, we still find a set of solutions $\tilde{\Phi}_{n\mathbf{k}}=\lambda_n \mathbf{k}$ and $\tilde{G}(\mathbf{q},\omega)=\lambda(\omega) \mathbf{q}$ that nullify the collisional integral. In the Dirac cones approximation for the band,  $\mathbf{v}_{n\mathbf{k}}= \pm v_{\rm F} \mathbf{k}/|\mathbf{k}|$ and $\tilde{\Phi}_{n\mathbf{k}} \propto  \pm\tau_{\mathbf{k}}v_{\rm F} \mathbf{k}/|\mathbf{k}|$. The lifetimes that comply with Peierl's argument are thus of the form $\tau_{\mathbf{k}} \propto |\mathbf{k}|$, and the electronic collisional integral evaluates to zero separately in both the intraband and interband regions.  This is all confirmed by numerical tests where we set $\tau^{-1}_{\rm ANH}$ to zero, as shown in App. \ref{app:solconv}. \footnote{Stating the argument more precisely: the modulus of the Fermi velocity $v_{\rm F}$ is fixed, and therefore $\mathbf{v}_{n\mathbf{k}}=(-1)^{n-1} v_{\rm F} \mathbf{k}/|\mathbf{k}|$. Then, $\tilde{\Phi}_{n\mathbf{k}} \propto \lambda_{n}(-1)^{n-1}\tau_{n\mathbf{k}}v_{\rm F} \mathbf{k}/|\mathbf{k}|$. The lifetimes form that can allow for the Peierl's argument to hold is $\tau_{n\mathbf{k}} \propto |\mathbf{k}|$. Then if we choose $\lambda_{n}=(-1)^{n-1}\lambda$ the electronic collisional integral evaluates to zero separately in both the intraband and interband regions. These arguments hold exactly for the zero temperature or the isotropic case, where a single Fermi velocity can be defined. The finite temperature or realistic bands cases has to be done numerically, but same physical conclusions shall apply.} Considerations regarding the other regimes for $\tau^{-1}_{\rm ANH}$ are analogous to the parabolic case. Notice that more complicated band dispersions will entail more complicated forms for the lifetime that satisfies Peierls' argument, and we leave that for analysis in further works.

\section{The response functions}
\label{sec:resp}
We here present the form of the response functions used in this work. They encode all the relevant interactions of the electron-phonon system.

\textit{Parabolic case---}As discussed in Sec. \ref{sec:scat}, we express $\epsilon^{-1}_{\rm tot}$ in RPA starting from the expression of $\epsilon_{\rm tot}$ \cite{PhysRev.137.A1896}:
\begin{align}
\epsilon^{-1}_{\rm tot}(\mathbf{q},\omega)=\frac{1}{\epsilon_{\rm tot}(\mathbf{q},\omega)}, \\
\epsilon_{\rm tot}(\mathbf{q},\omega)=\epsilon_{\rm el}(\mathbf{q},\omega)+\epsilon_{\rm ph}(\mathbf{q},\omega)-1,
\end{align}
where $\epsilon_{\rm el}(\mathbf{q},\omega)$ is the screening response at clamped nuclei, while $\epsilon_{\rm ph}(\mathbf{q},\omega)$ is the remainder \footnote{Notice that in this work we use a division where $\epsilon_{\rm ph}(\mathbf{q},\omega)=1+\ldots$, and $\epsilon_{\rm tot}(\mathbf{q},\omega)=\ldots-1$. One can perform a different division where $\epsilon_{\rm ph}(\mathbf{q},\omega)$ does not contain unity but it directly corresponds to $v_\mathbf{q}\chi_{\rm ph}(\mathbf{q},\omega)$, as done in Ref. \onlinecite{PhysRevB.110.115407}}. We use the following forms for the responses \cite{PhysRevB.94.085415,Sohier2017a,Macheda2022,Macheda2023}:
\begin{align}
\epsilon^{\infty}_{\rm el}(\mathbf{q})=1+r_{\rm eff}q \\
\epsilon_{\rm el}(\mathbf{q},\omega)=\epsilon^{\infty}_{\rm el}(\mathbf{q})-v_{\mathbf{q}}\chi^0_{\rm L}(\mathbf{q},\omega+i\eta_{\rm pl}) \label{eq:epsilonel}\\
\epsilon_{\rm ph}(\mathbf{q},\omega)=1-\epsilon^{\infty}_{\rm el}(\mathbf{q})\left( \frac{\omega^2_{\rm LO \mathbf{q}}-\omega^2_{\rm TO}}{(\omega+i\eta_{\rm ph})^2-\omega^2_{\rm TO}}\right) \label{eq:epsilonph}
\end{align}
where $\epsilon^{\infty}_{\rm el}(\mathbf{q})$ is the screening function of the neutral system, the expression for $\epsilon_{\rm ph}$ is an appropriate simplification of the first-principles one found in Ref. \onlinecite{PhysRevB.110.115407} (see App. \ref{app:Mahan} for a comparison with other models), and $\chi^0_{\rm L}$ is the non interacting polarizability for the band $n^*$ :
\begin{align}
\chi^0_{\rm L}(\mathbf{q},\omega)=\frac{2}{(2\pi)^2}\int_{0}^{k_{\rm max}} d^2\mathbf{k} \frac{f_{\mathbf{k}}-f_{\mathbf{k+q}}}{\varepsilon_{\mathbf{k}}-\varepsilon_{\mathbf{k+q}}+\hbar\omega+i\eta_{\rm pl}}
\label{eq:chi0L}
\end{align}
where the subscript L has been used to avoid confusion with other similar symbols, the factor 2 accounts for spin degeneracy, and $k_{\rm max}$ is a suitably cutoff for the modulus of the wavevector, as discussed in the computational details. $n_{\rm pl}$ and $\eta_{\rm ph}$ are needed to have retarded quantities, and can also be phenomenologically intepreted as the widths of the the plasmon and phonon excitations. $\epsilon_{\rm ph}$ is derived from the frequencies of LO and transverse-optical (TO) phonons. We model the relation between $\omega_{\rm LO \mathbf{q}}$ and $\omega_{\rm TO}$ as \cite{PhysRevB.94.085415,Sohier2017a}
\begin{align}
\omega_{\rm LO \mathbf{q}}=\omega_{\rm TO}+\frac{Sq}{1+r_{\rm eff}q},
\end{align}
where it is evident that the LO-TO splitting vanishes at zone center, as expected for two dimensional materials. Finally, in the following we will denote $v^{\infty}_{\mathbf{q}}=\frac{v_{\mathbf{q}}}{\epsilon^{\infty}_{\rm el}(\mathbf{q})}$. 

We now evaluate $\epsilon^{-1}_{\rm tot}(\mathbf{q},\omega)$ at zero doping (neutral case), for small enough $\eta_{\rm ph}$. We have 
\begin{align}
\epsilon^{-1}_{\rm tot}(\mathbf{q},\omega)=\frac{1}{\epsilon^{\infty}_{\rm el}(\mathbf{q})}\left[1+\frac{ \omega^2_{\rm{LO}\mathbf{q}}-\omega^2_{\rm TO}}{( \omega+i\eta_{\rm ph})^2-\omega^2_{\rm LO \mathbf{q}}}\right]
\end{align}
where we have approximated $\epsilon_{\rm el}(\mathbf{q},\omega)\sim \epsilon^{\infty}_{\rm el}(\mathbf{q})$, which holds for $\omega\ll \Delta_{\rm g}$ (the gap), i.e. in the physical region of interest.
The coupling that enters in Eq. \eqref{eq:P} is therefore
\begin{align}
\textrm{Im}[-\epsilon^{-1}_{\rm tot}(\mathbf{q,\omega})\frac{v_{\mathbf{q}}}{\pi}]\sim
v^{\infty}_{\mathbf{q}}\frac{\omega^2_{\rm LO \mathbf{q}}-\omega^2_{\rm TO}}{2\omega_{\rm LO 
 \mathbf{q}}}[\delta(\omega-\omega_{\rm LO \mathbf{q}})+\nonumber \\
 -\delta(\omega+\omega_{\rm LO \mathbf{q}})]=Ag^{2}_{\mathbf{q}}[\delta(\omega-\omega_{\rm LO \mathbf{q}}) -\delta(\omega+\omega_{\rm LO \mathbf{q}})].
 \label{eq:staticg}
\end{align}
$g^{2}_{\mathbf{q}}$, i.e. the residue at the pole, is interpreted here as the interaction vertex between electrons and phonons; see App. \ref{app:Mahan} for a comparison with previous models. In the following we will refer to this coupling as 'unscreened', meaning that it is unscreened with respect to screening from free-carriers, but screened by the semiconducting dielectric constant. Finally, notice that we had to take a small $\eta_{\rm ph}$ to deduce the unscreened coupling. If we took the limit after multiplying $v\epsilon^{-1}$ by the Bose distribution in Eq. \ref{eq:P}, we would have encountered the problems discussed in App. \ref{app:boselorentz}.

\textit{VED case---} The general treatment of the response functions for the VED case is described in Sec. II of Ref. \onlinecite{PhysRevB.110.115407}. 
The response of individual layers is determined \textit{ab initio}. The response of the heterostructure is then found then via a mean-field approach over the layers composing it. The response functions thus acquire layer indices that identifies which subsystem is responding. In this framework, electronic and atomic responses are naturally separated because they come from different layers. 
Notably, in BN-encapsulated graphene the macroscopic phonon response is entirely due to BN layers, while the free-carrier electronic one is due to graphene (although BN also contains an electronic contribution in the form of standard dielectric screening).

With respect to Ref. \onlinecite{PhysRevB.110.115407}, already containing LO phonons, we introduce in this work and in the accompanying paper \cite{accpaper} the response of BN's ZO phonons. This amounts to modifying BN's out-of-plane response, Eq. (24) of Ref. \onlinecite{PhysRevB.110.115407}, in the following way: 
\begin{align}
Q^1(q,\omega)=Q^1_{\rm el}(q,\omega=0)+Q^1_{\rm ph}(q,\omega).
\end{align}
To deduce the expression for $Q^1_{\rm ph}, (q,\omega)$ for a bidimensional layer, we follow Ref. \onlinecite{PhysRevX.11.041027}, given that $Q$ connects induced densities ($\rho$) and external potentials ($V_{\rm ext}$) on a single layer as $\rho=QV_{\rm ext}$. The induced density can be split in two components (notation of Ref. \onlinecite{PhysRevX.11.041027} for the left hand side and of Ref. \onlinecite{PhysRevB.110.115407} on the right hand side)
\begin{align}
\rho^{\parallel}(q)=\int dz \rho^q(z) \cosh(qz)\approx \int dz \rho^q(z) = \rho^0(q) \\
\rho^{\perp}(q)=\int dz \rho^q(z) \sinh(qz) \approx q \int dz \rho^q(z) z = q \rho^1(q)
\label{eq:tosteng}
\end{align}
Then we notice that, analogously to Eq. (A22) of Ref. \onlinecite{PhysRevB.110.115407},  $Q^1_{\rm ph}$ is related to the long-range contribution to the dynamical matrix as
\begin{align}
v^{11}Q^1_{\rm ph}(q,\omega)=[\epsilon^{-1}_{\rm el}]^1(q) \frac{\mathbf{e}_{\rm ZO q}D^{\rm L}(q) \mathbf{e}^*_{\rm ZO q}}{(\omega+i\eta_{\rm ph})^2-\omega^2_{\rm ZO}}
\end{align}
We further assume that ZO phonons are dispersionless for the BN monolayer, and define the reduced mass as $\mu=\frac{M_{B}M_{N}}{M_{B}+M_{N}}$ (where $B,N$ stands for boron and nitrogen).
Using Eq. (45) of Ref. \onlinecite{PhysRevX.11.041027} with the identification Eq. \eqref{eq:tosteng} we finally obtain
\begin{align}
v^{11}Q^1_{\rm ph}(q,\omega)= \frac{v^{11}}{A \mu \epsilon^2_{\perp}(q)}\frac{Z^2_{\perp}}{(\omega+i\eta_{\rm ph})^2-\omega^2_{\rm ZO}},\\
\epsilon_{\perp}(q)=1+\frac{4\pi}{q} Q^1_{\rm el}(q)q^2.
\end{align}
To link with the language and notation of Ref. \onlinecite{PhysRevX.11.041027}, in the above expression $Z^2_{\perp}$ is computed in zone-boundary electrostatic conditions, and we have that $2Q^1_{\rm el}(q=0)=-\alpha_{\perp}$.

\subsection{Dynamical response with doping}

\textit{Parabolic case---} We now study the effect of adding carriers in the parabolic band $n^*$. Two sets of parameters, listed in Tab. \ref{tab:materials}, correspond to two cases: h-BN and 2H-MoS$_2$. We stress that this is a very simplified model, meant to be illustrative rather than quantitative. Notably, the multivalley nature of the bands and spin-orbit coupling are disregarded, which is relevant for MoS$_2$ \cite{PhysRevB.94.085415}. 

\begin{table*}
\begin{center}
\begin{tabular}{| c | c | c | c | c | c | c |}
\hline
Material & $r_{\textrm{eff}}$ ($\AA$)& $\omega_{\rm TO}$ (cm$^{-1}$) & S (eV$^2\AA$)& $\eta_{\rm ph}$ (cm$^{-1}$) & $m_{\rm eff}$ (e) & A ($\AA^{2}$)\\[1ex]
\hline
BN & 7.64 & 1387.2 & 8.4 $\times 10^{-2}$& 2.8  & 0.63 & 5.52\\[1ex]
MoS$_2$ & 46.5 & 373.7 & 1.13 $\times 10^{-3}$& 1.0  & 0.465 & 8.67\\[1ex]
\hline
\end{tabular}
\end{center}
\caption{Parameters used to model the electronic and vibrational responses, and the parabolic bands structure, of BN and MoS$_2$. We use a phenomenological representative for $\eta_{\rm ph}$ of the same order of magnitude as the width of polar phonon peak seen in experiments.}
\label{tab:materials}
\end{table*}

\begin{figure*}[t!]
    \centering
    \includegraphics[width=0.45\linewidth]{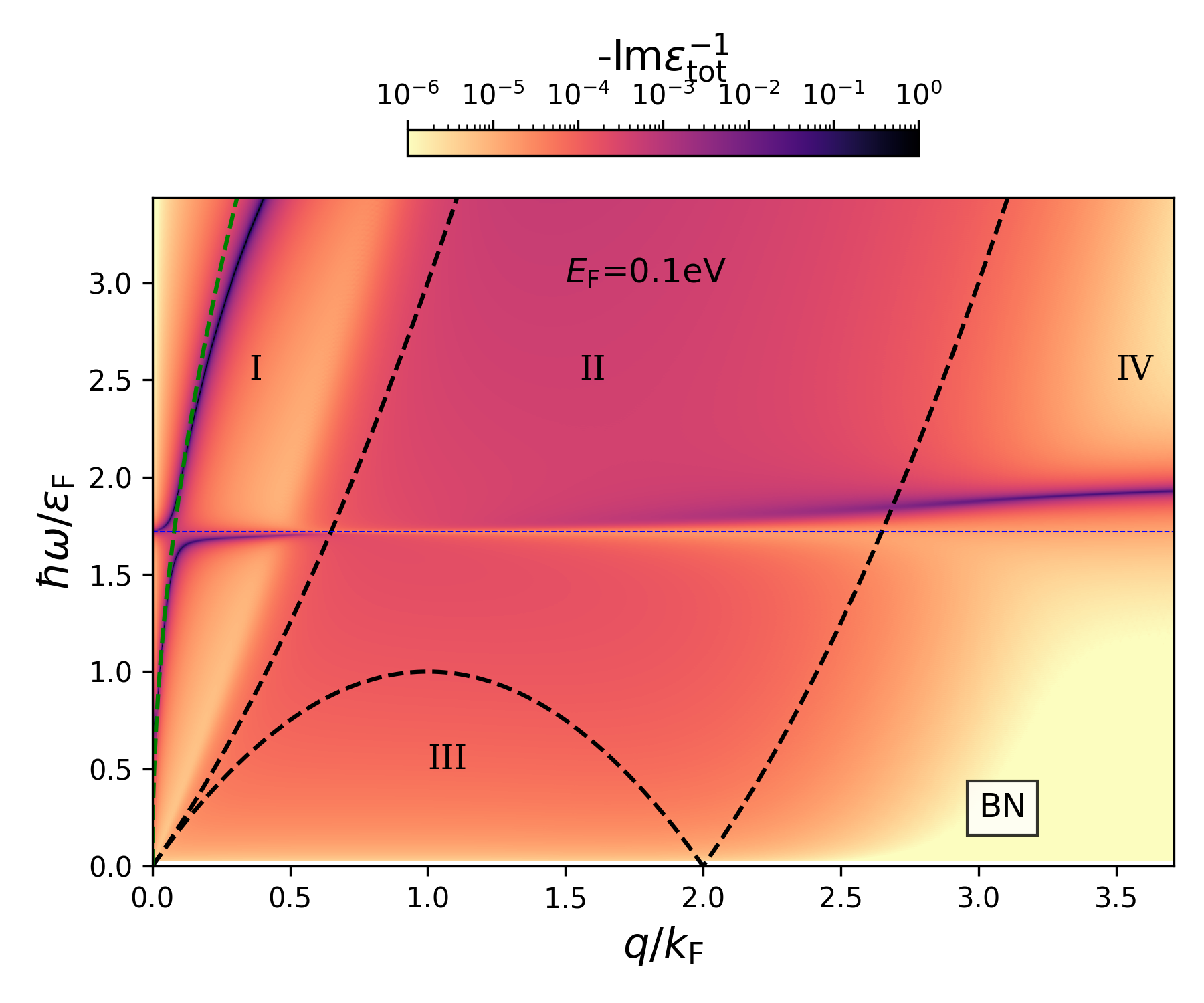}
    \includegraphics[width=0.45\linewidth]{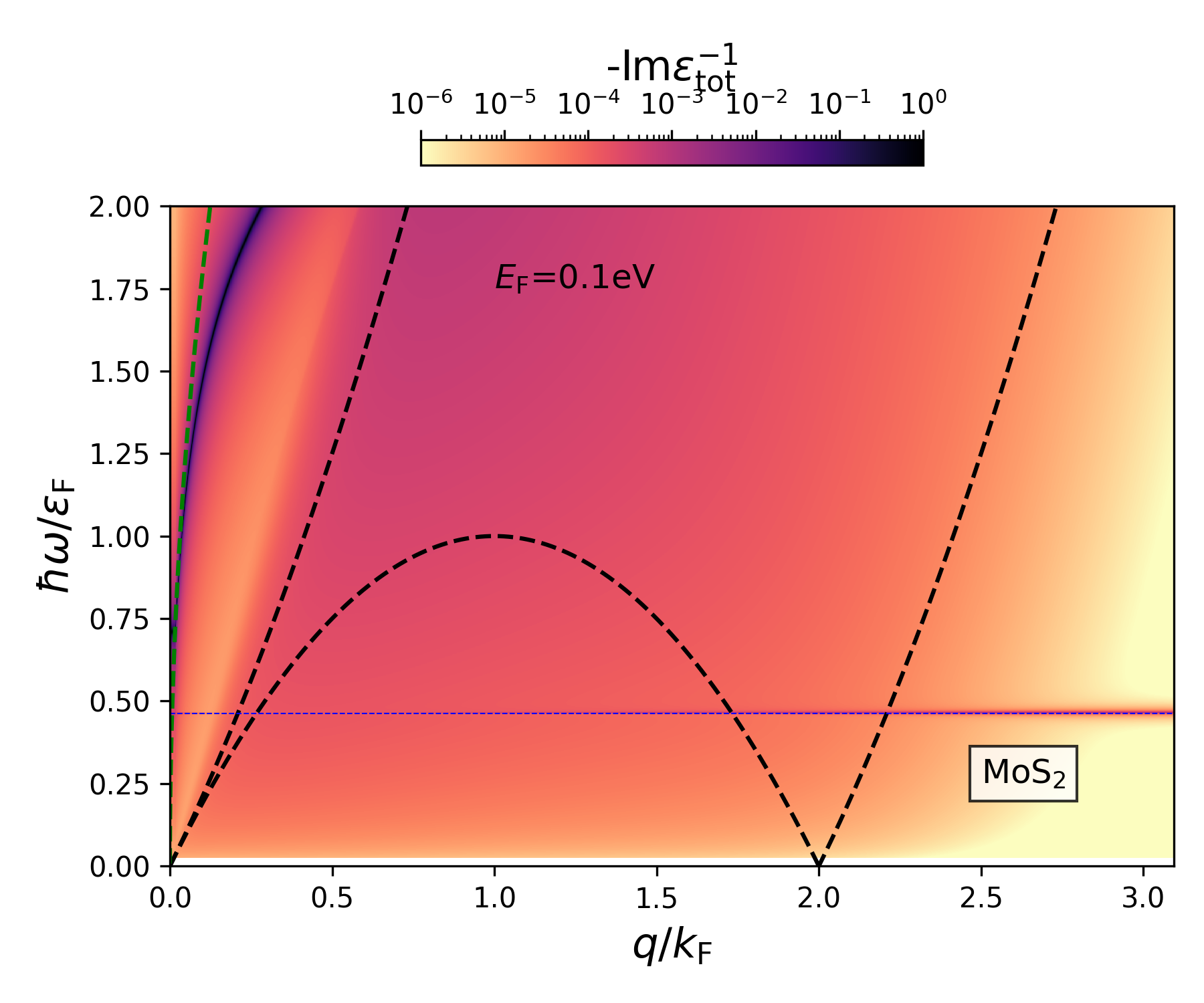}   \vspace{0.5cm} 
    \includegraphics[width=0.45\linewidth]{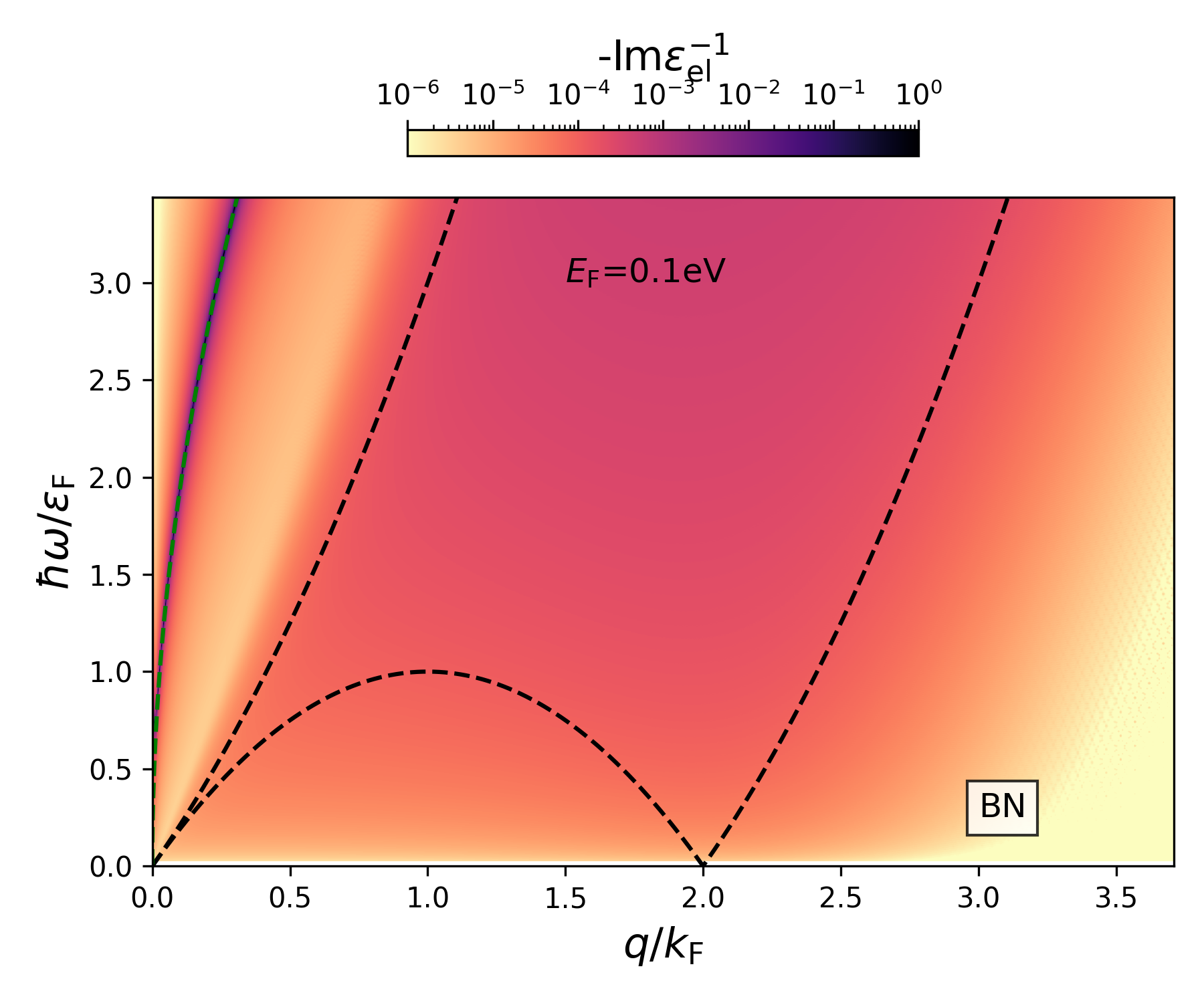}
    \includegraphics[width=0.45\linewidth]{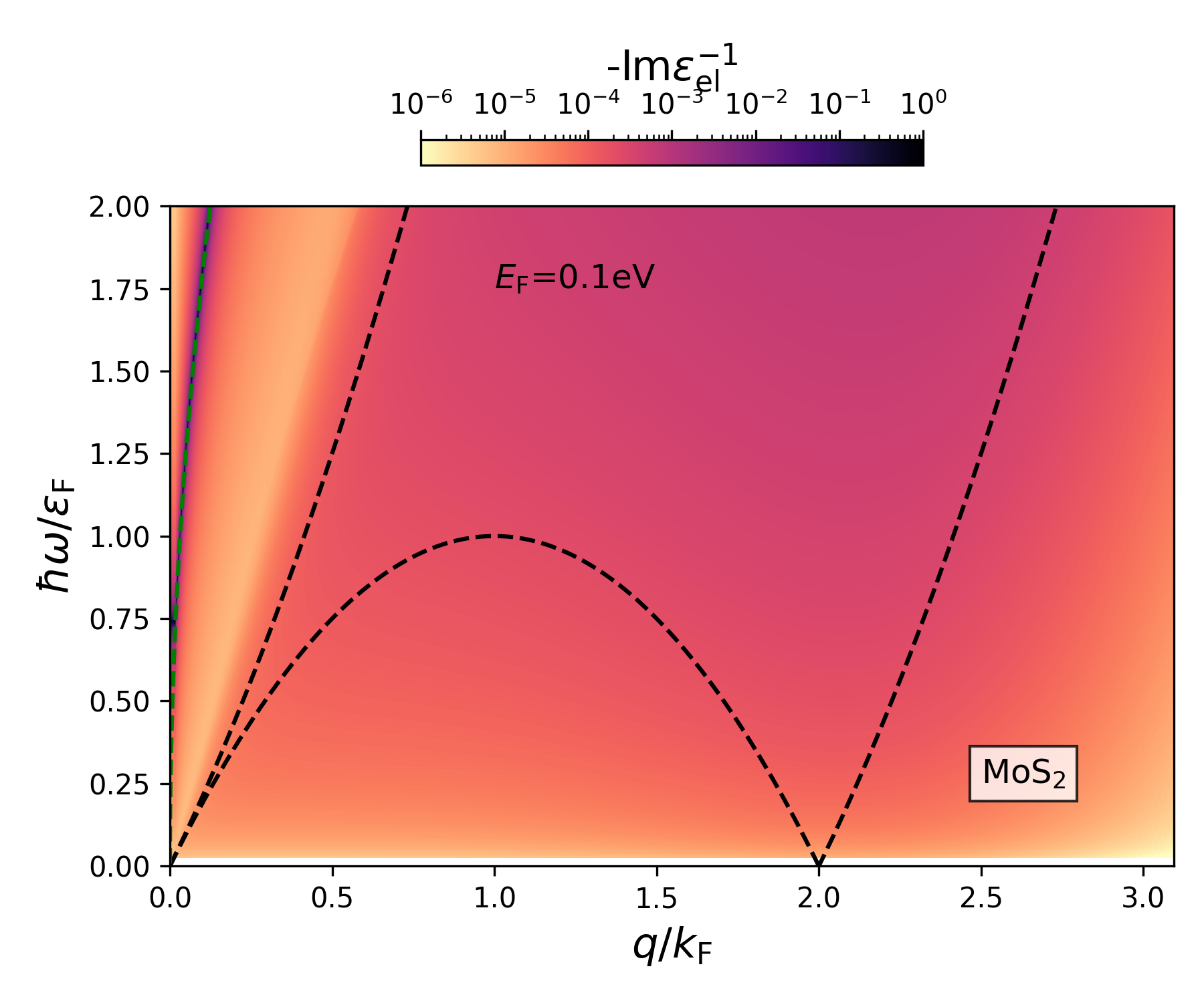}   
    \caption{(Upper panels) -Im$\epsilon^{-1}_{\rm tot}$ for a doping level of 0.1 eV, for BN (left) and MoS$_2$ (right). (Lower panels) same but for -Im$\epsilon^{-1}_{\rm el}$. Materials are modeled with an isotropic parabolic band dispersion with parameters given in Tab. \ref{tab:materials}. Upwards parabolae drawn as black dashed lines represent the limits of the electron-hole continuum at zero temperature (regions II and III, while in regions I and IV no single electron excitations are possible). The blue dashed horizontal lines are the uncoupled TO modes, while green dashed lines represent the uncoupled plasmons.}
    \label{fig:minvepstot}
\end{figure*}

Fig. \ref{fig:minvepstot} shows a color-plot of -Im$\epsilon^{-1}_{\rm tot}(\mathbf{q},\omega)$ in the upper panels and -Im$\epsilon^{-1}_{\rm el}(\mathbf{q},\omega)$ in the lower panels, for BN and MoS$_2$. The temperature is 300K, $\eta_{\rm pl}$ is set to $k_{B}T/50$ as discussed in App. \ref{app:etaph}, the doping levels are $E_{\rm F}=0.1$ eV for both systems, and the vertical and horizontal axis are scaled, respectively, with respect to the Fermi energy and to the Fermi wavevector $k_{\rm F}$. The horizontal blue dashed line is drawn for $\omega=\omega_{\rm TO}$, while the green dashed line represents the analytical dispersion of the two-dimensional acoustic plasmon given by \cite{RevModPhys.54.437}
\begin{align}
\hbar \omega_{\rm pl}=\sqrt{4E_{\rm F} q+\frac{3}{4}q^2k_{\rm F}}.
\end{align}
The black dashed lines instead represent the usual partition of the frequency-momentum plane, analytically obtained at zero temperature. In the regions I and IV, at zero temperature no electron-hole excitation is possible, contrary to zones II and III. The excitations of $\epsilon^{-1}_{\rm tot}$, i.e of the coupled electronic and vibrational system, are multiple. We find a continuum of electron-hole excitations in regions II and III, 
as well as phonon- and plasmon-like peaks. The electron-hole continuum mediates the e-e interaction. 

The plasmon peak is entirely contained in region I for the range of frequencies studied. For both BN and MoS$_2$, it dispersion is different from the one seen in $\epsilon^{-1}_{\rm el}$ for mainly two reasons: the presence of an anti-crossing with the phonon peak at around $\omega \sim \omega_{\rm TO}$ and the bending of the plasmon dispersion at high frequencies, due to the screening induced by the phonon system. As it regards the phonon excitation instead, in both compounds it gradually shifts from the TO frequency value near the anticrossing with the plasmon to the LO frequency of the neutral case at large momenta. In BN this behaviour is more pronounced than in MoS$_2$ due to the larger LO-TO splitting found at finite momentum with the parameters of Tab. \ref{tab:materials}. This phenomenon, which is akin to the one described in BN-encapsulated graphene in Ref. \onlinecite{PhysRevB.110.115407}, is due to different regimes of interaction with electronic excitations, from hybridization with the plasmon in region I, to free-carrier screening being effective inside the electron-hole continuum of region II, then ineffective outside in region IV.

\begin{figure*}[t!]
    \centering
    \includegraphics[width=0.45\linewidth]{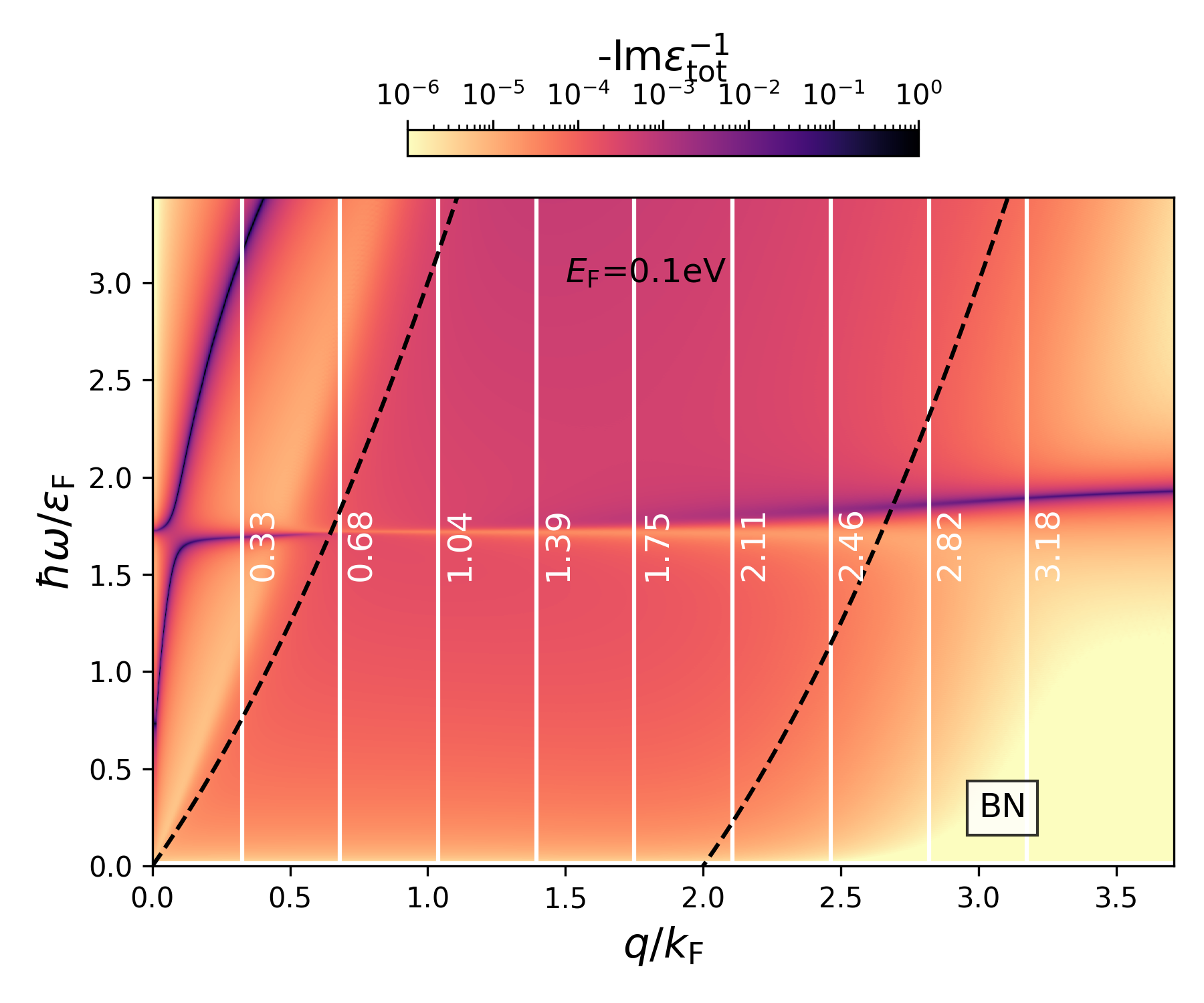}
    \includegraphics[width=0.45\linewidth]{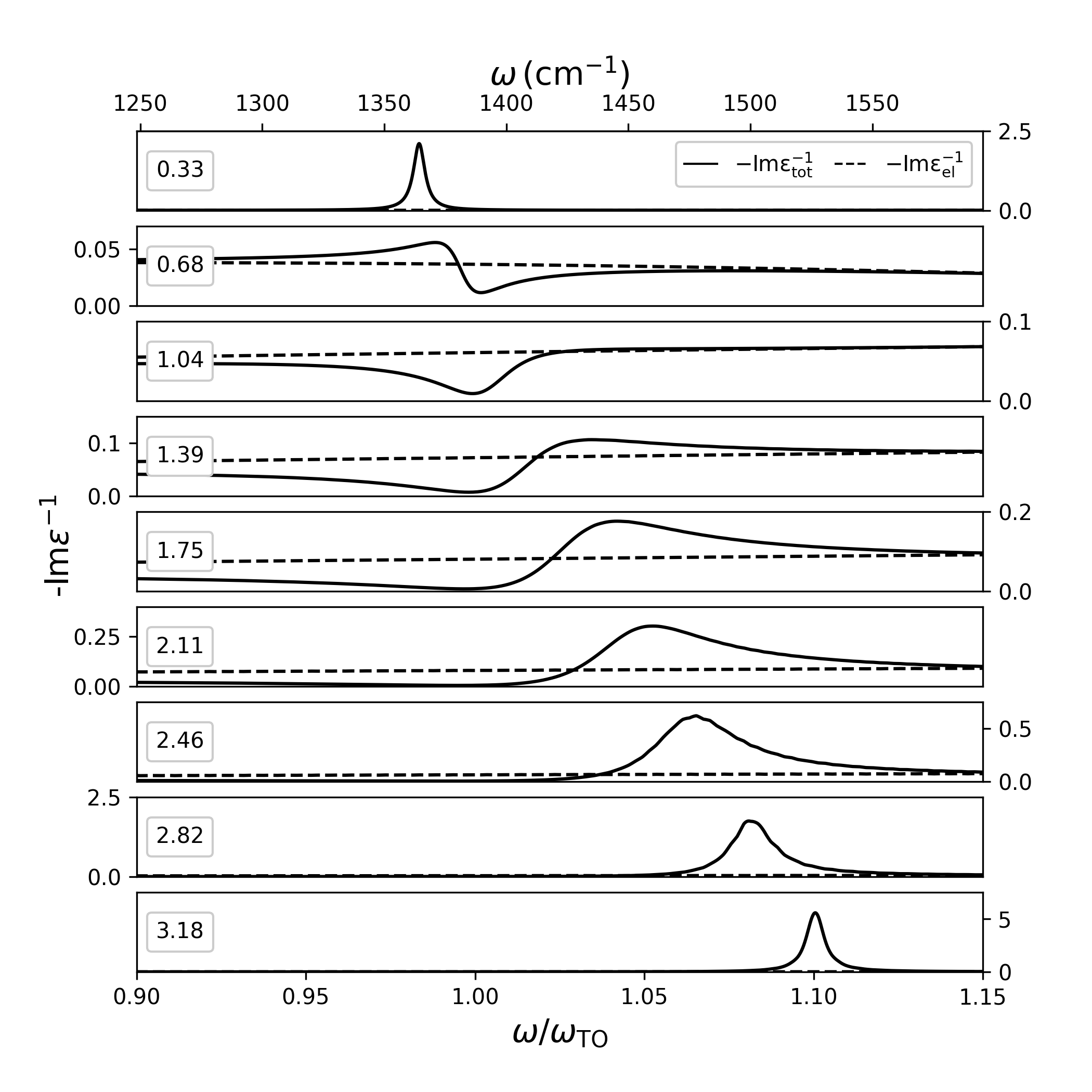}
    \caption{-Im$\epsilon^{-1}_{\rm tot}$ as a function of $\omega$ for given cuts at fixed $q/k_{\rm F}$, for BN. The cuts are presented in the left panel as white vetical lines, with the corresponding $q/k_{\rm F}$ value reported, and the value of -Im$\epsilon^{-1}_{\rm tot}$ is reported in the right panel as black continuous line. Dashed lines represent instead the value of -Im$\epsilon^{-1}_{\rm el}$. The emergent excitations are not of Lorentzian nature.}
    \label{fig:cuts}
\end{figure*}

While from a general perspective the situation is quite clear and intuitive, closer analysis of the excitations reveals a more complex picture. In Fig. \ref{fig:cuts} we evaluate -Im$\epsilon^{-1}_{\rm tot}$ and -Im$\epsilon^{-1}_{\rm el}$ along cuts performed at fixed values of $q/k_{\rm F}$. We concentrate on a range of frequencies around the TO phonon frequency, ranging $0.9<\omega/\omega_{\rm TO}<1.15$. Evidently, the peak of -Im$\epsilon^{-1}_{\rm tot}$ along the cut is of Lorentzian shape only when -Im$\epsilon^{-1}_{\rm el}$ is almost zero, i.e. where electron-hole excitations are not relevant (first and last cuts). In all other cases, the peak has large interference shapes (spanning tenths of cm$^{-1}$) due to the interplay between 
electrons and phonons. For $q/k_{\rm F}=1.04$ we even witness -Im$\epsilon^{-1}_{\rm tot}<-\rm{Im}\epsilon^{-1}_{\rm el}$ in almost all the studied frequency range. We stress that with our choice for $\eta_{\rm pl}$ and $\eta_{\rm ph}$ this is not an effect due to the finite smearing value used in the polarizabilities, but it is a true interference effect (see App. \ref{app:etaph}). The witnessed effect is not problematic for the evaluation of the scattering time from Eq. \eqref{eq:P}, but it is problematic for transport calculations, as discussed in the next section. 

\begin{figure}
    \centering
    \includegraphics[width=\linewidth]{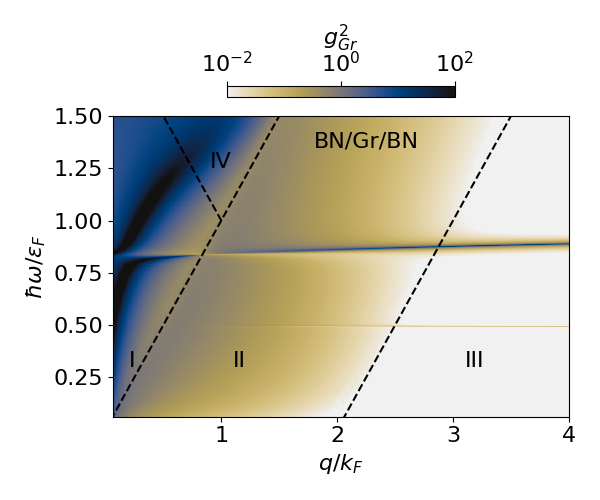}
    \includegraphics[width=\linewidth]{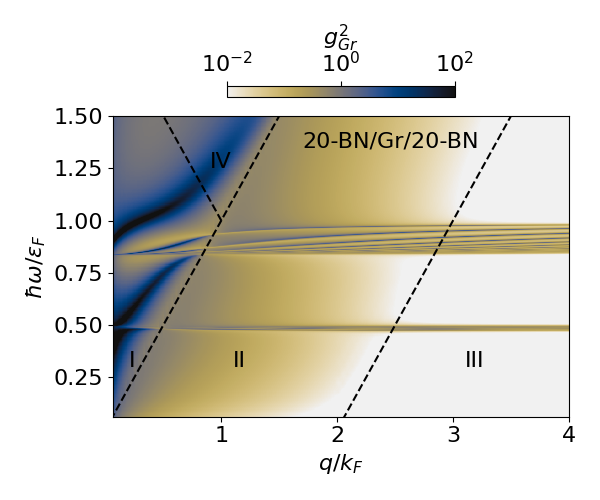}
    \caption{
    Electron-mode coupling in BN-encapsulated graphene. The coupling is computed for an electron in graphene $g^2_{\rm Gr} = g^2_{\rm kk}$ with $k$ the layer index of graphene. The number of BN layers is 1 and 20 on each side for the top and bottom panel, respectively. The electrodynamic active modes are graphene's plasmon and BN's remote LO and ZO phonons. 
    Zones II and IV are the intra- and interband electron-hole continua. Zone I is where the plasmon excitation is undamped. There are no electron-hole excitations in zone III. The limits between those zones (dashed lines) are smeared by temperature ($300$ K here). }
    \label{fig:vchivved}
\end{figure}

\textit{VED case---} For the VED case, we consider BN-encapsulated graphene. This time, we do aim for realistic results. We present the interaction strength between electrons and the system excitations, containing the dynamical response function of the system, in Fig. \ref{fig:vchivved}. We consider the cases of graphene sandwiched within 1 and 20 layers of BN, per side.
The frequency-momentum plane is in this case divided in different regions with respect to the parabolic case, due to the linearity of graphene conic bands, that is the metallic layer where electron-hole excitations are possible. In particular, zones II and IV are the intra- and interband electron-hole continua while zone I is where the plasmon excitation is undamped. There are no electron-hole excitations in zone III. If we concentrate on the interplay between BN's LO phonons and graphene's electrons, Figs. 3, 4, 5, 7 and 8 of Ref. \onlinecite{PhysRevB.110.115407} already explain the qualitative and quantitative features of the response functions. The interplay between graphene's plasmon and BN's phonons follows the same qualitative behavior as the parabolic case, with some complicated anti-crossing patterns appearing when the number of BN layers is increased. 

The same analysis can be performed for the ZO modes, that shows similar qualitative trends, but with the important difference that $\omega_{\rm ZO}<\omega_{\rm LO}$.
Importantly, the deviation of the phonon excitations from the Lorentzian shape -- as observed in Fig. \ref{fig:cuts} for the parabolic case -- remains. As observed in Fig. \ref{fig:vchivved}, the number of LO and ZO modes increases as the number of BN layers, and also their relative coupling strength with graphene electrons changes with the number of layers involved. Notice also that each branch corresponds to different relative phases of the movement of the BN atoms in the various layers. 
As evident, many remote phonon modes couple to the graphene electrons, rather than a unique interfacial mode.

\subsection{The nature of the phonon excitations and the phonon content}

\label{sec:phcontent}
\begin{figure*}[t!]
    \centering
    \includegraphics[width=0.45\linewidth]{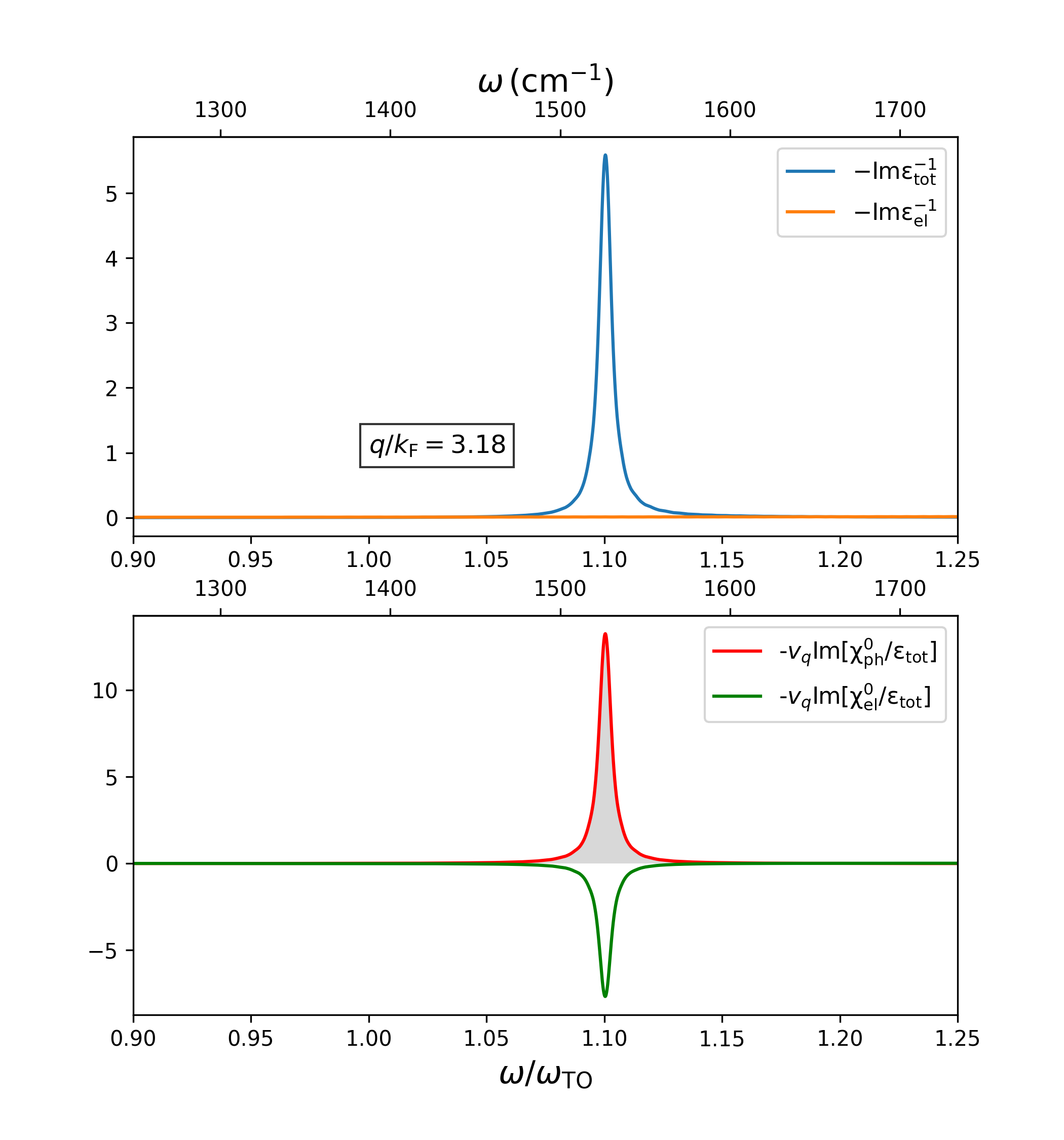}
    \includegraphics[width=0.45\linewidth]{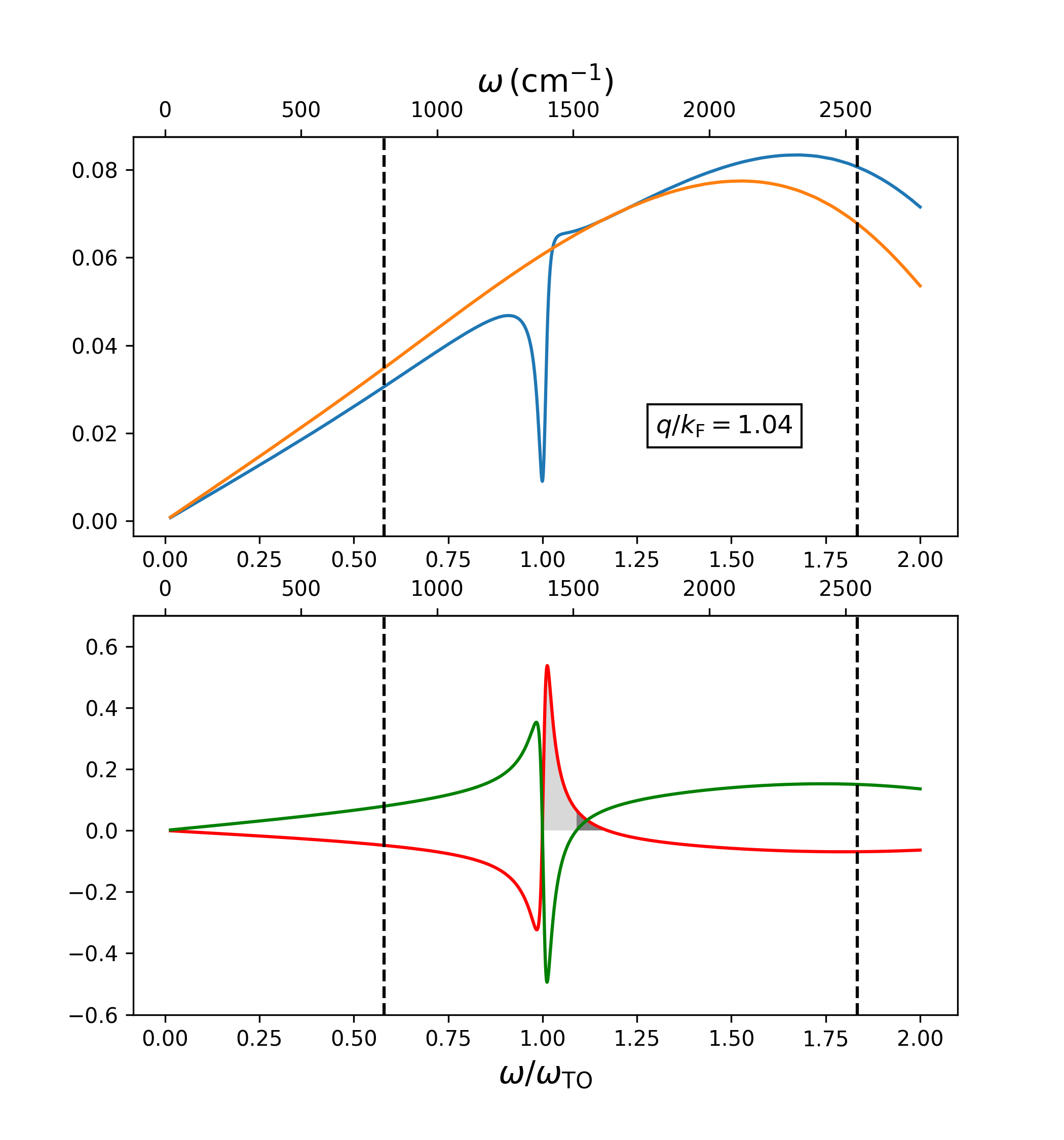} 
    \caption{(Upper panels) $-\rm{Im}\epsilon^{-1}_{\rm tot}$ and $-\rm{Im}\epsilon^{-1}_{\rm el}$ and (lower panels) $\mathcal{F}_{\rm el/ph}=-\rm{Im} v\chi^0_{\rm el/ph}/\epsilon_{\rm tot}$ as function of $\omega$, for BN, along those cuts of Fig. \ref{fig:cuts} situated at $q/k_{\rm F}=3.18$ (left panels) and 1.04 (right panels). Gray dashed areas are regions where $\mathcal{F}_{\rm ph}>0$: light gray if $\mathcal{F}_{\rm el}<0$ and dark gray if $\mathcal{F}_{\rm el}>0$.}
    \label{fig:division}
\end{figure*}

In a closed system made up of electrons and phonons, and assuming negligible Umklapp scattering, neither e-e or e-ph interactions are able to degrade momentum/current, hence lead to finite conductivity \cite{peierls1955quantum}. Only anhamornic decay, where Umklapp scattering is almost never negligible, can represent the main intrinsic physical channel through which momentum degradation happens \cite{peierls1955quantum}, and indirectly determine electronic transport coefficients. 
The anharmonic interactions of a given phonon mode with other phonons can be taken into account in the BTE for phonons, and its effect on electronic transport coefficients can be deduced by coupling the BTE for phonon and electrons (see e.g. Ref. \onlinecite{PhysRevLett.134.226301} for a recent computation of such effects). 
However, this entire procedure relies on having well-defined phonon quasi-particles, usually represented by sharp Lorentzian spectral functions.
This is not what we observe for polar phonons interacting with the electron-hole continuum. At first glance, the spectral shapes of Fig. \ref{fig:cuts} do not offer a clear boundary between purely electronic and dressed phonon excitations.
At the very least, one needs to identify a region of phase space where anharmonic decay can reasonably take place. Therefore, a `phonon content' needs to be associated to the excitations.

\textit{Parabolic case---} To propose a definition of phonon content, we first note that in RPA it holds
\begin{align}
\chi_{\rm tot}(\mathbf{q},\omega)=\frac{\chi^0_{\rm el}(\mathbf{q},\omega)}{\epsilon_{\rm tot}(\mathbf{q},\omega)}+\frac{\chi^0_{\rm ph}(\mathbf{q},\omega)}{\epsilon_{\rm tot}(\mathbf{q},\omega)}.
\label{eq:division}
\end{align}
where 
\begin{align}
\chi_{\rm tot}(\mathbf{q},\omega)=\frac{1}{v_{\mathbf{q}}}\left[-1+\epsilon^{-1}_{\rm tot}(\mathbf{q},\omega)\right],\\
\chi^0_{\rm el/ph}(\mathbf{q},\omega)=\frac{1}{v_{\mathbf{q}}}\left[1-\epsilon_{\rm el/ph}(\mathbf{q},\omega)\right].
\end{align}
The physical meaning of the separation performed in Eq. \eqref{eq:division} is the division of the total system into electronic and phononic subsystems coupled only via mean-field interactions. Indeed, when such subsystems are perturbed by an external scalar potential $V_{\rm{el/ph},\rm ext}$, the induced densities read
\begin{align}
\rho_{\rm el}(\mathbf{q},\omega)=\chi_{\rm el}(\mathbf{q},\omega)\left[V_{\rm el,\mathrm{ext}}+v_{\mathbf{q}}\rho_{\rm ph}(\mathbf{q},\omega)\right],\\
\rho_{\rm ph}(\mathbf{q},\omega)=\chi_{\rm ph}(\mathbf{q},\omega)\left[V_{\rm ph,\mathrm{ext}}+v_{\mathbf{q}}\rho_{\rm el}(\mathbf{q},\omega)\right],
\end{align}
which represents a $2\times2$ coupled system where the unknowns are the densities (omitting momentun and frequency dependence):
\begin{align}
\begin{pmatrix}
\rho_{\mathrm{el}} \\
\rho_{\mathrm{ph}}
\end{pmatrix}=\frac{1}{1-v\chi_{\mathrm{el}}v\chi_{\mathrm{ph}}}
\begin{pmatrix}
\chi_{\mathrm{el}} & v\chi_{\mathrm{el}}\chi_{\mathrm{ph}}\\
v\chi_{\mathrm{el}}\chi_{\mathrm{ph}} & \chi_{\mathrm{ph}}\\
\end{pmatrix}
\begin{pmatrix}
V^{\mathrm{el}}_{\mathrm{ext}} \\
V^{\mathrm{ph}}_{\mathrm{ext}}
\end{pmatrix}.
\label{eq:2x2}
\end{align}
If in the above system we take $V_{\rm el,ext}=V_{\rm ph,ext}=V_{\rm ext}$, then we obtain
\begin{align}
\rho_{\mathrm{tot}}(\mathbf{q},\omega)=\rho_{\mathrm{el}}(\mathbf{q},\omega)+\rho_{\mathrm{ph}}(\mathbf{q},\omega),\\
\rho_{\mathrm{el/ph}}(\mathbf{q},\omega)=\frac{\chi^0_{\rm el/ph}(\mathbf{q},\omega)}{\epsilon_{\rm tot}(\mathbf{q},\omega)}V_{\mathrm{ext}},
\end{align}
which is equivalent to Eq. \eqref{eq:division}. It is clear that $\chi_{\rm el/ph}$ takes into account the contribution of the single subsystem to the global excitations, once they are renormalized by the screening of both subsystems $\epsilon_{\rm tot}(\mathbf{q},\omega)$, and not just $\epsilon_{\rm el/ph}(\mathbf{q},\omega)$. Notice that the above reasoning is generalizable to multicomponent (or multilayer) systems, as we performed in Ref. \onlinecite{PhysRevB.110.115407} and discussed later. 

The separation of the response of Eq. \eqref{eq:division} provides an insightful framework to study the excitations. It immediately follows that
\begin{align}
-\rm{Im}\epsilon^{-1}_{\rm tot}(\mathbf{q},\omega)=-\rm{Im}\left[ v_{\mathbf{q}}\frac{\chi^0_{\rm el}(\mathbf{q},\omega)}{\epsilon_{\rm tot}(\mathbf{q},\omega)}+v_{\mathbf{q}}\frac{\chi^0_{\rm ph}(\mathbf{q},\omega)}{\epsilon_{\rm tot}(\mathbf{q},\omega)}\right], 
\end{align}
and the two terms $-\rm{Im} v\chi^0_{\rm el/ph}/\epsilon_{\rm tot}$ are plotted in Fig. \ref{fig:division} for BN along the cuts of Fig. \ref{fig:cuts} situated at $q/k_{\rm F}=3.18$ (lower left panel) and 1.04 (lower right panel).  We also report, in the upper panels, the value of -Im$\epsilon^{-1}_{\rm tot/el}(\mathbf{q},\omega)$ along the same cuts.  
For $q/k_{\rm F}=3.18$, as anticipated in the previous sections, -Im$\epsilon^{-1}_{\rm tot}(\mathbf{q},\omega)$ shows a well defined peak because the electron-hole pair excitations are essentially absent. The peak is situated at a higher frequencies with respect to TO due to LO-TO splitting. The phonon contribution to the total peak is positive, while the electronic contribution is negative, meaning that the electrons screen the peak intensity. We are here in the case of Eq. \eqref{eq:staticg}, where the electronic system screens the phonon system and changes the values of the interaction only via the semiconducting static screening. 
In contrast, for $q/k_{\rm F}=1.04$, -Im$\epsilon^{-1}_{\rm tot}(\mathbf{q},\omega)$ displays an interference shape somewhat localized around the TO frequency, with a reversed peak of FWHM around $50$cm$^{-1}$, and a small shoulder at frequencies above TO. Both phonon and electronic contributions rapidly change sign across the TO frequency, and non-trivially add up to create the full excitation. In particular, the phonon contribution is negative for $\omega<\omega_{\rm TO}$, positive for $\omega_{\rm TO}<\omega\lesssim 1.15 \omega_{\rm TO}$ and goes back to negative for $\omega \gtrsim 1.15 \omega_{\rm TO}$. The electronic contribution instead is positive for $\omega<\omega_{\rm TO}$, negative for $\omega_{\rm TO}<\omega\lesssim 1.10 \omega_{\rm TO}$ and goes back to positive for $\omega \gtrsim 1.10 \omega_{\rm TO}$. Also, notice that the absolute value of the intensity of each single contribution is much larger than the final peak, and the tails decay slower than a Lorentzian. 
We interpret the excitation in different frequencies regimes in the following way:
\begin{enumerate}
\item for $\omega<\omega_{\rm TO}$ it consists in electronic excitations screened by the phonon
\item for $\omega_{\rm TO}<\omega\lesssim 1.10 \omega_{\rm TO}$ it is a phonon excitation screened by electron-hole pairs, consistent with the fact that the renormalized phonon frequency is always expected to be larger than the TO frequency
\item for $ 1.10 \omega_{\rm TO}\lesssim \omega \lesssim 1.15 \omega_{\rm TO}$, there is a small crossover regime where the electron-hole pairs anti-screen the phonon excitation (both contributions are positive)
\item for $\omega \gtrsim 1.15 \omega_{\rm TO}$ we again have electron-hole pairs screened by the phonon
\end{enumerate}

In light of the previous discussion, we define the phonon content $\mathcal{F}(\mathbf{q},\omega)$ as
\begin{align}
\mathcal{F}_{\rm el}(\mathbf{q},\omega)=-\rm{Im}\left[ v_{\mathbf{q}}\frac{\chi^0_{\rm el}(\mathbf{q},\omega)}{\epsilon_{\rm tot}(\mathbf{q},\omega)}\frac{v_{\mathbf{q}}}{A\pi}\right], \nonumber \\
\mathcal{F}_{\rm ph}(\mathbf{q},\omega)=-\rm{Im}\left[ v_{\mathbf{q}}\frac{\chi^0_{\rm ph}(\mathbf{q},\omega)}{\epsilon_{\rm tot}(\mathbf{q},\omega)}\frac{v_{\mathbf{q}}}{A\pi}\right], \nonumber \\
\mathcal{F}=\alpha_q \times 
\begin{cases}
\mathcal{F}_{\rm ph}+\mathcal{F}_{\rm el} \quad \rm{if}  \, \mathcal{F}_{\rm ph}>0 \,\rm{and}\, \mathcal{F}_{\rm el}<0\\
\mathcal{F}_{\rm ph} \quad \rm{if} \, \mathcal{F}_{\rm ph}>0 \,\rm{and}\, \mathcal{F}_{\rm el}>0 \\
0 \quad \rm{otherwise}
\end{cases},\label{eq:phcontent}
\end{align}
where we have omitted obvious momentum and frequency dependencies for brevity, and we added a Coulombian factor on the right, to conveniently write the phonon content in terms of quantities that represent the strength of the interaction between the electron and the excitations. Further, $\alpha_q > 0$ is a generic function whose form is determined by the normalization condition 
\begin{align}
\int d\omega \mathcal{F}(\mathbf{q},\omega)=1,
\end{align}
since we have only one IR active phonon in our model. 

\begin{figure*}[t!]
    \centering
    \includegraphics[width=0.45\linewidth]{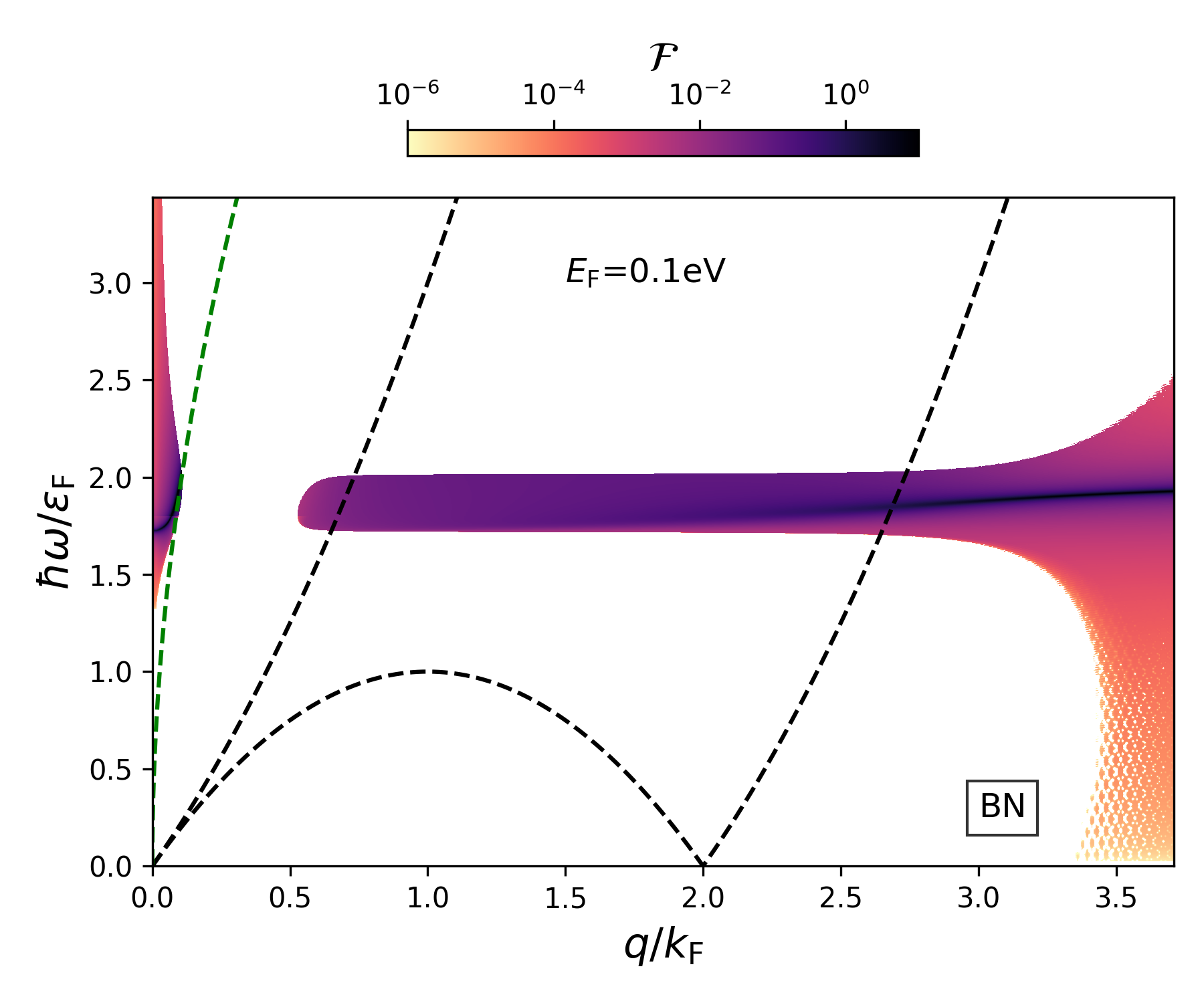}
    \includegraphics[width=0.45\linewidth]{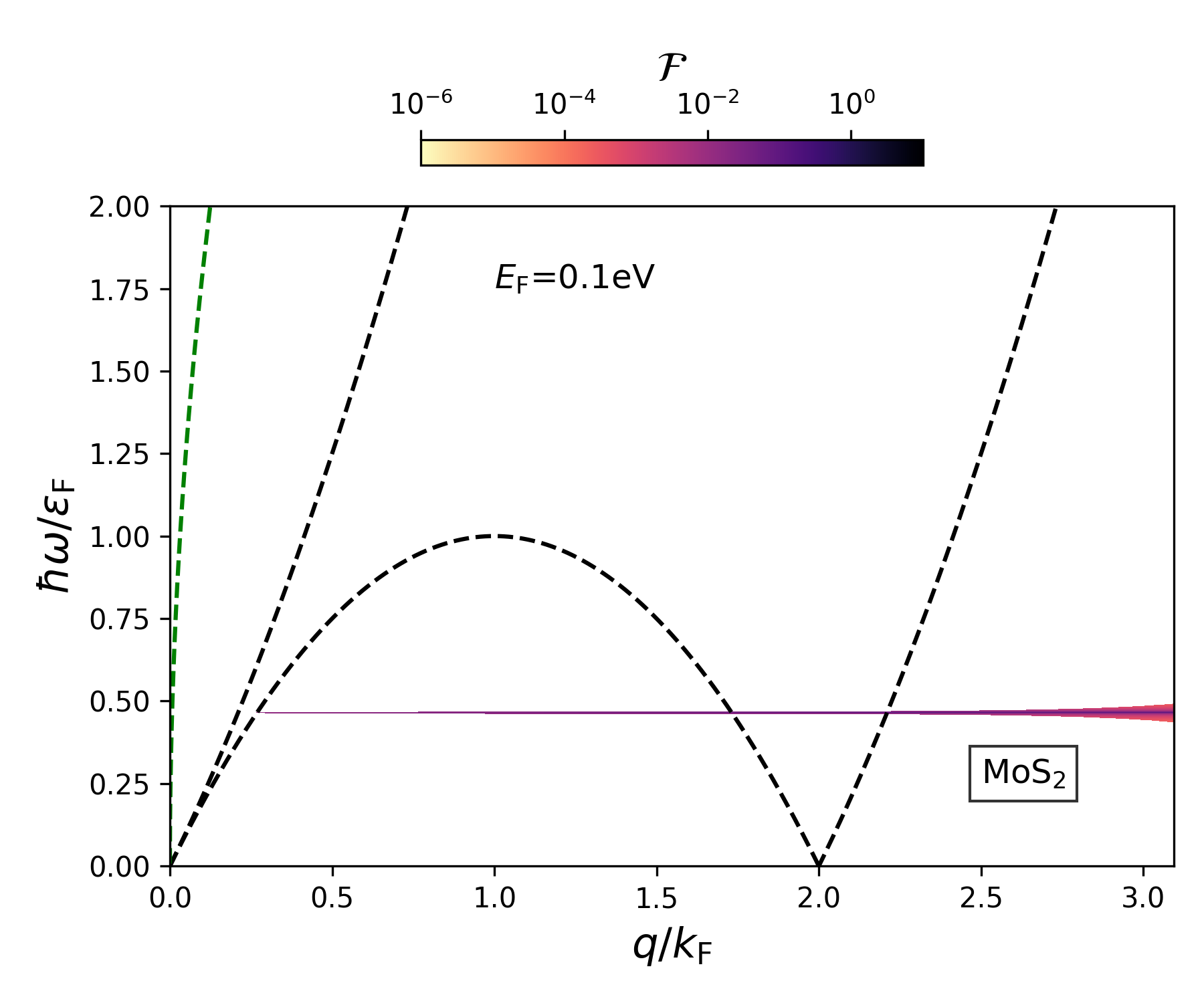} 
    \caption{Phonon content $\mathcal{F}$ of Eq. \eqref{eq:phcontent} for BN (left panel) and MoS$_2$ (right panel), which shows localization around the phonon mode.  }
    \label{fig:phononcontent}
\end{figure*}

The phonon content contains the portion of the total inverse dielectric function that we attribute to phonon excitations. 
Notice that we avoid to account for anti-screening effects due to $\mathcal{F}_{\rm el}$, which are expected to be limited to small portions of the momentum-frequency plane overall, since it is not intuitively clear that such effects should enter in the determination of transport properties. 
We plot $\mathcal{F}$ in Fig. \ref{fig:phononcontent} for BN (left panel) and MoS$_2$ (right panel), which shows localization around the phonon excitation as expected. The spread of $\mathcal{F}$ around the excitation is larger for BN than MoS$_2$, due to the stronger hybridization of electronic and vibrational responses. 
Note that in most of the previous treatments the phonon content was defined starting from the assumption that the phonon frequency is a narrow Lorentzian peak, as e.g. in Refs. \onlinecite{Kim1978,10.1063/1.1405826}. Only Ref. \onlinecite{Hauber2017a} gives a comparable definition of the region where anharmonic scattering is possible. In the notation of this work, their phonon content corresponds to $\mathcal{F}= \rm{Im}\epsilon_{\rm ph}/\rm{Im}\epsilon_{\rm tot}$. However, we find this definition problematic since such phonon content peaks at the frequency of the bare phonon without long-range polar interactions ($\omega_{\rm TO}$, the pole of $\epsilon_{\rm ph}$), while anharmonic interactions are expected to occur at the frequency of the (screened) LO phonon (the pole of $\epsilon^{-1}_{\rm tot}$), which is higher due to polar interactions via the well-known LO/TO splitting mechanism. We will properly see in the following section how $\mathcal{F}$ enters the determination of transport properties.

\textit{VED case---}
We apply the same strategy used to extract an electron-phonon coupling from the electron-electrodynamic mode coupling in Eq. (51) of Ref. \cite{PhysRevB.110.115407}. The procedure is summarized here for the case of BN-encapsulated graphene. 
The general idea is that electronic and phonon excitation originate from different types of layers (graphene and BN, respectively). Since within VED, all quantities are layer-resolved, one can easily separate electronic and phonon excitations by separating contributions of different types of layer. 
We are interested in the transport of electrons situated in the graphene layer, and therefore in the couplings
\begin{align}
\label{eq:el-EDMcoupling}
    g^2_{\textrm{Gr}}(q, \omega) = \sum_{ij k' l'} \frac{-1}{A \pi} \textrm{Im}{  \left[ v^{0i}_{\textrm{Gr}k'}(q) \chi^{ij}_{k'l'}(q, \omega)  v^{j0}_{l'\textrm{Gr}}(q) \right]}.
\end{align}
We divide the contribution to the coupling as 
\begin{align}
g^2_{\textrm{Gr}}&=g^{2,\textrm{BN}}_{\textrm{Gr}}+g^{2,\textrm{Gr}}_{\textrm{Gr}}, \\
    g^{2,\textrm{BN}}_{\textrm{Gr}}(q, \omega) &= \frac{-1}{A\pi} \sum_{k = \textrm{BN}} \sum_{ijl} v^{0,i}_{\textrm{Gr}, k} \textrm{Im}\left[\chi^{ij}_{kl}\right] v^{j0}_{l, \textrm{Gr}},\nonumber\\
    g^{2,\textrm{Gr}}_{\textrm{Gr}}(q, \omega) &= \frac{-1}{A\pi} \sum_{k = \textrm{Gr}} \sum_{ijl} v^{0i}_{\textrm{Gr}, k} \textrm{Im}\left[\chi^{ij}_{kl}\right] v^{j0}_{l, \textrm{Gr}}.
    \label{eq:gsepar}
\end{align}
We finally define the phonon content as 
\begin{align}
\mathcal{F}=\alpha_{q} \times 
\begin{cases}
g^2_{\textrm{Gr}} & \text{ if } g^{2,\textrm{BN}}_{\textrm{Gr}}>0 \text{ and} \ g^{2,\textrm{Gr}}_{\textrm{Gr}}<0\\
g^{2,\textrm{BN}}_{\textrm{Gr}} & \text{ if } g^{2,\textrm{BN}}_{\textrm{Gr}}>0 \text{ and} \ g^{2,\textrm{Gr}}_{\textrm{Gr}}>0  \\
0 & \text{  if } g^{2,\textrm{BN}}_{\textrm{Gr}}<0  \\
\end{cases}.
\label{eq:gphcases}
\end{align}
The definition of Eq. \eqref{eq:gphcases} is equivalent to Eq. \eqref{eq:phcontent} if considering the graphene/BN system as a 2x2 system that interacts as in Eq \eqref{eq:2x2} with the substitution $(\chi_{\rm el},\chi_{\rm ph}) \rightarrow (\chi_{\rm Gr},\chi_{\rm BN})$.

\subsubsection{Choice of $\tau^{-1}_{\rm anh}$}
We have seen in the previous sections that the existence of a finite $\tau^{-1}_{\rm anh}$ is of fundamental importance to allow for a finite conductivity, when Umklapp scattering is neglected. Here we detail the choice of $\tau^{-1}_{\rm anh}$ for the systems under study.

\textit{Parabolic case---} For the parabolic case studied in this work, we choose a very large value of $\tau^{-1}_{\rm anh}$, corresponding to a very large anharmonic scattering, so that we  effectively use
\begin{align}
 \tau^{-1}_{\rm ANH}(\mathbf{q},\omega)=
 \begin{cases}
 \infty \quad \textrm{if} \, \mathcal{F}(\mathbf{q},\omega) \neq0\\
 0 \quad \textrm{if} \, \mathcal{F}(\mathbf{q},\omega) =0
\end{cases}.
\end{align}
This is a simplifying choice (for an already oversimplified model) leading to an overestimation of resistivity values. Nevertheless, it allows for a simple quantitative assessment of the effect of electron-electron interaction on the carrier mobility. 

\textit{VED case---} In the more realistic VED model, we first fix $\alpha_q$ such that
\begin{align}
\int d\omega \mathcal{F}(\mathbf{q},\omega)=N_{\rm ph},
\end{align}
where $N_{\rm ph}$ is the number of IR active phonons, i.e. we impose that the phonon content integrates to the number of actual active phonon modes present in the system. We can then study the very large anharmonicity regime like for the parabolic case. We leave the description of realistic value to further works.

\section{Local fields effects}
\label{sec:localfields}
In realistic descriptions of a material such as the VED case, macroscopic long-range fields are just one component of the total electrostatic field inside the material. Short-range fields, i.e. local fields effects stemming from considering the dependence of quantities on the reciprocal lattice vectors $\mathbf{G}$ and $\mathbf{G'}$, are also very important mechanisms from a quantitative description of transport properties of materials. In this section we present the scattering time and the BTEs formulation including local fields effects, and show how a decoupling of long and short range interactions is performed. We start from Eq. \eqref{eq:wnolf}. With local fields, we have (Eq. (27) of Ref. \onlinecite{ECHENIQUE20001})
\begin{align}
w_{n\mathbf{k}m\mathbf{k+q}}(i\nu_p)= \frac{1}{N_{\mathbf{q}}}\sum_{\mathbf{qGG'}}\int_{-\infty}^{\infty}\frac{d\omega'}{2\pi}\frac{[B^{\mathbf{k}}_{nm}]_{\mathbf{GG'}}(\mathbf{q},\omega')}{i\nu_p-\omega'}\nonumber \\
[B_{nm}]^{\mathbf{k}}_{\mathbf{GG'}}(\mathbf{q},\omega') = \int d\mathbf{r} d\mathbf{r'}dzdz'd\bar{z} [F^{nm}_{\mathbf{k}\mathbf{k+q}}]_{\mathbf{GG'}}(\mathbf{r},z,\mathbf{r'},z') \nonumber \\
\times \frac{2}{A}\textrm{Im}\left[- [\epsilon_{\rm tot}^{-1}]_{\mathbf{G'G}}(\mathbf{q},z,\bar z,\omega'+i0^+)v(\mathbf{q+G},\bar z,z')\right] \label{eq:BG},
\end{align}
with
\begin{align}
[F^{nm}_{\mathbf{k}\mathbf{k+q}}]_{\mathbf{GG'}}(\mathbf{r},z,\mathbf{r'},z')=e^{-i\mathbf{G}\cdot\mathbf{r}}e^{i\mathbf{G'}\cdot\mathbf{r'}}u^*_{n\mathbf{k}}(\mathbf{r},z)u_{n\mathbf{k}}(\mathbf{r'},z')\times \nonumber \\
u_{m\mathbf{k+q}}(\mathbf{r},z)u^*_{m\mathbf{k+q}}(\mathbf{r'},z').
\label{eq:Fkk+qGGpdef}
\end{align}
Notice that the imaginary part above is the typical shorthand notation for the spectral representation in the matrix case; in reality, it represent the expression of the spectral function in terms the retarded and advances response functions. As for the macroscopic case, we can again pass the imaginary part in front of the self-energy directly to $I$ to obtain:
\begin{align}
\tau^{-1,\rm el}_{n\mathbf{k}}=\frac{2\pi}{\hbar}\frac{1}{N_{\mathbf{q}}}\sum_{\mathbf{qGG'}m}\int_{-\infty}^{\infty} \frac{d\omega'}{2\pi} [B_{nm}]^{\mathbf{k}}_{\mathbf{GG'}}(\mathbf{q},z,z',\omega')\nonumber \\
\times (n_{\omega'}+f_{m\mathbf{k+q}})\delta(\omega'+\varepsilon_{n\mathbf{k}}-\varepsilon_{m\mathbf{k+q}}).
\label{eq:taulocf}
\end{align}

In the approximation where the quantities entering Eq. \eqref{eq:taulocf} are weakly dependent on the out-of-plane coordinate, we can separate the contributions coming from $\mathbf{G/G'}=0$, i.e. long-range interactions, from the ones coming from $\mathbf{G,G'}\neq0$, i.e. short range interactions, in an additive way. The $\mathbf{G,G'}\neq0$ terms come from the `body' of the $W$ matrix \cite{Macheda2023}, i.e. from the part of the interaction where the macroscopic Coulombian interaction is set to zero.

In a realistic description, materials have $3N_{\rm{N}}$ phonon modes, embedded in the poles of the phonon propagator inside $\epsilon_{\rm{ph}}$. In a simple biatomic polar material, one can assume that the electron-phonon couplings related to the polar LO and ZO phonons are mostly of long-range nature, while all couplings with other phonon modes are due to local fields (and therefore short-ranged). One can then sum the contributions from polar and other modes to the scattering rate separately, as usually done in a Fermi Golden Rule scheme. This separation can be assumed to hold true, within a reasonable precision, even in the RPA+xc approximation typical of DFT calculations\cite{PhysRev.136.B864,PhysRev.140.A1133}. 

The above derivation for $\tau^{-1,\rm el}$ is not directly applicable to $\tau^{-1,\rm tr}$. The tempting assumption is to consider local field interactions as unaffected by free-carrier screening effects, especially dynamical ones, so that they can be computed in the adiabatic approximation; in other words that $[\epsilon^{-1}_{\rm el}]_{\mathbf{GG'}\neq0}(\omega)\approx[\epsilon^{-1}_{\rm el}]_{\mathbf{GG'}\neq0}(\omega=0)$, which is real and has no imaginary part. It turns out that this assumption is not correct if performed directly in the BTE containing both short- and long-range interactions, since it violates quasi-momentum conservation. Indeed, the dynamical electronic response at the lattice vectors $\mathbf{G,G'}\neq 0$ is crucial to ensure quasi-momentum circulation between short- and long-interactions. If this circulation is not taken into account, we find that the computed resistivity is very large and non-physical.

Therefore we proceed by considering short- and long-range interactions as separate scattering processes, adopting the Matthiessen's rule
\begin{align}
\tau^{-1,\rm tr}(\mathbf{q},\omega)=\tau^{-1,\rm tr}_{\rm LR}(\mathbf{q},\omega)+\tau^{-1,\rm tr}_{\rm SR}(\mathbf{q},\omega),
\end{align}
where LR means long-range and SR means short-range. Quasi-momentum exchange is treated independently for the two cases since $\tau^{-1,\rm tr}_{\rm SR/LR}(\mathbf{q},\omega)$ is obtained from the solution of the electronic BTE with only short- or long-range interactions separately. At this point, for the short-range case we can safely disregard the dynamical electronic excitations at $\mathbf{G,G'}\neq 0$, i.e. we can now set $[\epsilon^{-1}_{\rm el}]_{\mathbf{GG'}\neq0}(\omega)\approx[\epsilon^{-1}_{\rm el}]_{\mathbf{GG'}\neq0}(\omega=0)$, and employ the Bloch's ansatz  $\mathcal{G}(\mathbf{q},\omega)=0$. For the long-range case, we solve Eqs. \eqref{eq:BTEel} and \eqref{eq:BTEph}. Note that local fields effects are taken into account only for the VED case, as presented in an accompanying paper \cite{accpaper}.

\section{Computational methodology}
\label{sec:compmeth}
In this section we describe the implementation of the solution of the coupled BTEs of Eqs. \eqref{eq:BTEel} and \eqref{eq:BTEph}. The solution to the problem mainly consists in four operations:
\begin{enumerate}
\item Evaluation of $-\rm{Im}\epsilon^{-1}_{\rm tot}(\mathbf{q},\omega)$;
\item Solution of the electronic BTE Eq. \eqref{eq:BTEel} to find $\Phi(\mathbf{k})$;
\item Determination of $\mathcal{G}(\mathbf{q},\omega)$ from Eq. \eqref{eq:Gdet};
\item Iterations between Eq. \eqref{eq:BTEel} and Eq. \eqref{eq:Gdet} until convergence is achieved.
\end{enumerate}
All of the above points are computationally expensive, and require a careful parallelization. As a first approximation and accordingly to what is presented in the theory section, we assume isotropy for the response functions, meaning that e.g. $\epsilon^{-1}_{\rm tot}(\mathbf{q},\omega)=\epsilon^{-1}_{\rm tot}(q,\omega)$. Moreover, since the quantity $\mathbf{k+q}$ often appear in our formulae, we practically assume throughout all the calculations $\mathbf{q}=q\hat{q_x}$.

\textit{Parabolic case---}The $(q,\omega)$ responses are first computed on a uniform grid for $q$ in the interval $[10^{-3}\rm{Bohr}^{-1},1.4 \left(k_{\rm F}+\sqrt{k_{\rm F}^2+2m_{\rm eff}\omega_{\rm TO}}\right)]$ and a uniform grid for $\omega$ in the interval $[1\rm{cm}^{-1},\rm{max}(2E_{\rm F},2\omega_{\rm TO})]$. They are then interpolated on a fine frequency grid when required. 

The bottleneck for the computations of $-\rm{Im}\epsilon^{-1}_{\rm tot}(q,\omega)$ is the evaluation of the non-interacting polarizability of the excess electrons given by Eq. \eqref{eq:chi0L}. For this task, we adopt telescopic $\mathbf{k}$ meshes that are denser around the Fermi energy, that start from a uniform grid of dimension $400\times400$ in the coarse region and are denser by a factor $p=10$ along each axis in the region critical for transport. Also, for numerical stability, we exploit time-reversal to rewrite Eq. \eqref{eq:chi0L} in the following more numerically accurate form
\begin{align}
\chi^0_{\rm L}(\mathbf{q},\omega)=\frac{2}{(2\pi)^2}\int_{0}^{k_{\rm max}} d^2\mathbf{k} \frac{(f_{\mathbf{k}}-f_{\mathbf{k+q}})(\varepsilon_{\mathbf{k}}-\varepsilon_{\mathbf{k+q}})}{(\varepsilon_{\mathbf{k}}-\varepsilon_{\mathbf{k+q}})^2-(\hbar\omega+i\eta_{\rm pl})^2}.
\end{align}
The cutoff value is set to $k_{\rm max}=\sqrt{2m_{\rm eff}\rm{max}(4E_{\rm F},0.6\rm{eV})}$; a large value of $k_{\rm max}$ is important to describe the real part of the electronic response in our considered frequency range. Integration, here and below, is performed via the replacement
\begin{align}
\frac{A}{(2\pi)^2}\int d^2\mathbf{k} \leftrightarrow \frac{1}{N_{\mathbf{k}}} \sum_{\mathbf{k}} 
\end{align}

We now introduce, as done for the VED case, the screened total coupling
\begin{align}
g^2_{\rm tot}(\mathbf{q},\omega)=-\frac{1}{A}\rm{Im}[\epsilon^{-1}_{\rm tot}(\mathbf{q},\omega)\frac{v_{\mathbf{q}}}{\pi}].
\label{eq:scrtotcoup}
\end{align}
We also use isotropy to say that our solution has to be in the form \cite{Sohier2014}
\begin{align}
\Phi(\mathbf{k})=\tau^{\rm {tr}}(\varepsilon_{\mathbf{k}})v_{\mathbf{k}x},
\end{align}
where for parabolic bands $
v_{\mathbf{k}x}=\frac{k_x}{m_{\rm eff}}$.
Since the lifetime depends only on the energy, we first change the summation variable of Eq. \eqref{eq:BTEel} as $\mathbf{q}\rightarrow \mathbf{k'}=\mathbf{k+q}$, and then to the energy variable $\varepsilon'=k'^2/2m_{\rm eff}$. By calling for simplicity $\varepsilon_{\mathbf{k}}=\varepsilon$, Eq. \eqref{eq:BTEel} is recast in the following form
\begin{widetext}
\begin{align}
\frac{\frac{\partial{f}}{\partial t}\Big{|}_{\mathbf{E}}}{\tau^{\rm tr}(\varepsilon)}=-\frac{A}{\hbar 2\pi k_B T}\int_{0}^{\varepsilon_{\rm max}} d\varepsilon'\int_{0}^{2\pi}d\theta\int_{-\infty}^{+\infty}d\omega f_{\varepsilon'}[1-f_{\varepsilon}][n_\omega+1]\Big[ k_x-k'\cos\theta\frac{\tau^{\rm tr}(\varepsilon')}{\tau^{\rm tr}(\varepsilon)}+\frac{\mathcal{G}(\mathbf{k'-k},\varepsilon'-\varepsilon)}{\tau^{\rm tr}(\varepsilon)}\Big]\nonumber \\
\times g^2_{\rm tot}(\mathbf{k'-k},\varepsilon'-\varepsilon)\delta(\varepsilon'-\varepsilon-\hbar\omega),
\end{align}
\end{widetext}
where $\varepsilon_{\rm{max}}$ is obtained from $k_{\rm{max}}$. We solve the above equation on uniform grids, with numerical precautions when $\tau(\epsilon)$ evaluates to zero. For energy integration, we use a 500 points uniform grid for $\epsilon \in [0,\varepsilon_{\rm {max}}]$, while the integration is performed on 10000 points for $\varepsilon'$ and 10000 points for $\theta \in [0,2\pi]$. The solution is obtain by an iterative method that starts from a trial constant-in-energy solution and ends when the relative values of the $i+1$-th and $i$-th solutions differ by less that $10^{-8}$. 

We apply the same approach for the determination of $G$ via Eq. \eqref{eq:Gdet}. Then, we iterate between the two coupled Boltzmann equations, starting from the solution of the electronic BTE in presence of only static screening (see Sec. \ref{sec:results}) and then continuing until convergence is reached (see App. \ref{app:solconv} for numerical convergence). 

Finally, we determine the conductivity and the mobility via the following formula
\begin{align}
\sigma=\frac{2e^2}{k_{\rm B} T(2\pi)^2}\int_{0}^{\varepsilon_{\rm max}}d\varepsilon \int_{0}^{2\pi}d\theta \left[\frac{k}{m_{\rm eff}}\cos\theta\right]^2 \nonumber \\
\times f_{\varepsilon} (1-f_{\varepsilon}) \tau^{\rm tr}(\varepsilon), \quad \mu=\frac{\sigma}{ne}.
\label{eq:sigmamucomp}
\end{align}
where $n$ is the carrier concentration.

\textit{VED case---} For the case of BN-encapsulated graphene we setup a uniform q grid spanning the interval  $[10^{-3}\textrm{Bohr}^{-1},1.4\left(2k_{\rm F}+\omega_{\rm TO}/v_{\rm F}\right)]$, while the frequency grid is not uniform, but refined around the $\omega_{\rm LO}$ and $\omega_{\rm ZO}$ frequencies---the distance between two different point is as small as $\eta_{\rm LO}/10$. The graphene $\chi^0$ is treated as in the parabolic case but with the important addition of the overlaps between Bloch wavefunctions, and it is computed on uniform grids with a size that changes with doping and temperature for optimal computational results. At each $q$, the maximal wavevector considered is $k_{\rm max}=4k_{\rm F}+10q$. The electronic Boltzmann equation for the long-range interactions is in the same form of the parabolic case, with the important difference that we have here 2 electronic bands and the sum over layer indices. Further, we use $\varepsilon_{\rm max}=\textrm{max}\left(4E_{\rm F},0.4 \textrm{eV}\right)$. The Boltzmann equation for short-range interactions is instead solved as discussed in Ref. \onlinecite{Sohier2014}. Finally, the conductivity is computed with Eq. \eqref{eq:sigmamucomp}. The main difference with respect to the parabolic case is the way in which the solution is determined. 
Indeed, to speed up the convergence we implement a scanning algorithm that works in the following way. 
We start from a trial solution $\tau_0$ obtained from the (easily computed) solution of the electronic BTE with unscreened electron-phonon couplings.
We run a few cycles ($n \sim 5 $) of the iterative coupled BTE, at which point the general behavior of $\tau_n$ as a function of $\varepsilon$ is similar to the final solution but with a different magnitude. Then we rescale $\tau_{n+1} = s_0 \tau_n$ by a factor $s=\pm 0.75$ with a $+$ ($-$) sign if $\tau_{n}>\tau_{n-1}$ ($\tau_{n}<\tau_{n-1}$). We repeat this process and decrease the magnitude of the scaling factor $s \to s/2$ every time its sign changes, until the value of $s$ is below a predetermined value. The whole of this procedure has been implemented in an open-source code\cite{linkcode}

\section{Results}
\label{sec:results}
\textit{Parabolic case---} We will compute the mobility, only due to long-range fields, for the following treatments of the interactions:
\begin{enumerate}
\item `Unscreened coupling': we take $\epsilon_{\rm el}(\mathbf{q},\omega)=\epsilon^{\infty}_{\rm el}(\mathbf{q})$, corresponding to the neutral case, without screening from the electrons in the band. The response function contains only the e-ph interaction,  i.e. using Eqs. \eqref{eq:staticg} and \eqref{eq:scrtotcoup}:
\begin{align}
g^2_{\rm tot}(\mathbf{q},\omega)=g^{2}_{\mathbf{q}}[\delta(\omega-\omega_{\rm LO \mathbf{q}}) -\delta(\omega+\omega_{\rm LO \mathbf{q}})].
\end{align}
We here only solve the electronic BTE equation, with $\mathcal{G}(\mathbf{q},\omega)=0$ $\forall \omega$. In other words, there is no electronic electrodynamic mode, and phonon excitations are considered are thermal equilibrium because the anharmonic decay is very fast.
\item `Statically screened coupling': we take $\epsilon_{\rm el}(\mathbf{q},\omega)=\epsilon_{\rm el}(\mathbf{q},\omega=0)$, i.e.  we consider the effect of screening of electrons in the static approximation modifying Eq. \eqref{eq:epsilonel} as
\begin{align}
\epsilon_{\rm el}(\mathbf{q},\omega)=\epsilon^{\infty}_{\rm el}(\mathbf{q})-v_{\mathbf{q}}\chi^0_{\rm L}(\mathbf{q},\omega=0),
\end{align}
and determine $g^2_{\rm tot}(\mathbf{q},\omega)$ accordingly. Taking the electronic response at $\omega=0$ means that the perturbation (phonons) is considered much slower than the response time of electrons. This can thus be qualitatively interpreted as `perfect screening', since electrons have ample time to react to phonons. 
We here only solve the electronic BTE equation, with $\mathcal{G}(\mathbf{q},\omega)=0$ $\forall \omega$. Both electronic and phonon electrodynamic modes are at thermal equilibrium. 
\item `Dynamically screened coupling': we use Eq. \eqref{eq:scrtotcoup} with all the ingredients computed dynamically, electronic and phonon excitations being treated on the same footing. Their populations are driven out of thermal equilibrium and we solve the coupled BTEs.
\item `No phonons, e-e only': we take $\epsilon_{\rm ph}(\mathbf{q},\omega)=1$, i.e. we set to zero the vibrational contribution to the electrodynamic response, determine $g^2_{\rm tot}(\mathbf{q},\omega)$ accordingly, and then solve the coupled BTEs. In this way, only electron-electron interactions between carriers are present. We retain the full phonon content obtained without setting $\epsilon_{\rm ph}(\mathbf{q},\omega)=1$, such that electronic excitations around the phonon, presumably those that would dress the phonon excitation, contribute in momentum scattering.
\end{enumerate}

\begin{figure*}[t!]
    \centering
    \includegraphics[width=0.45\linewidth]{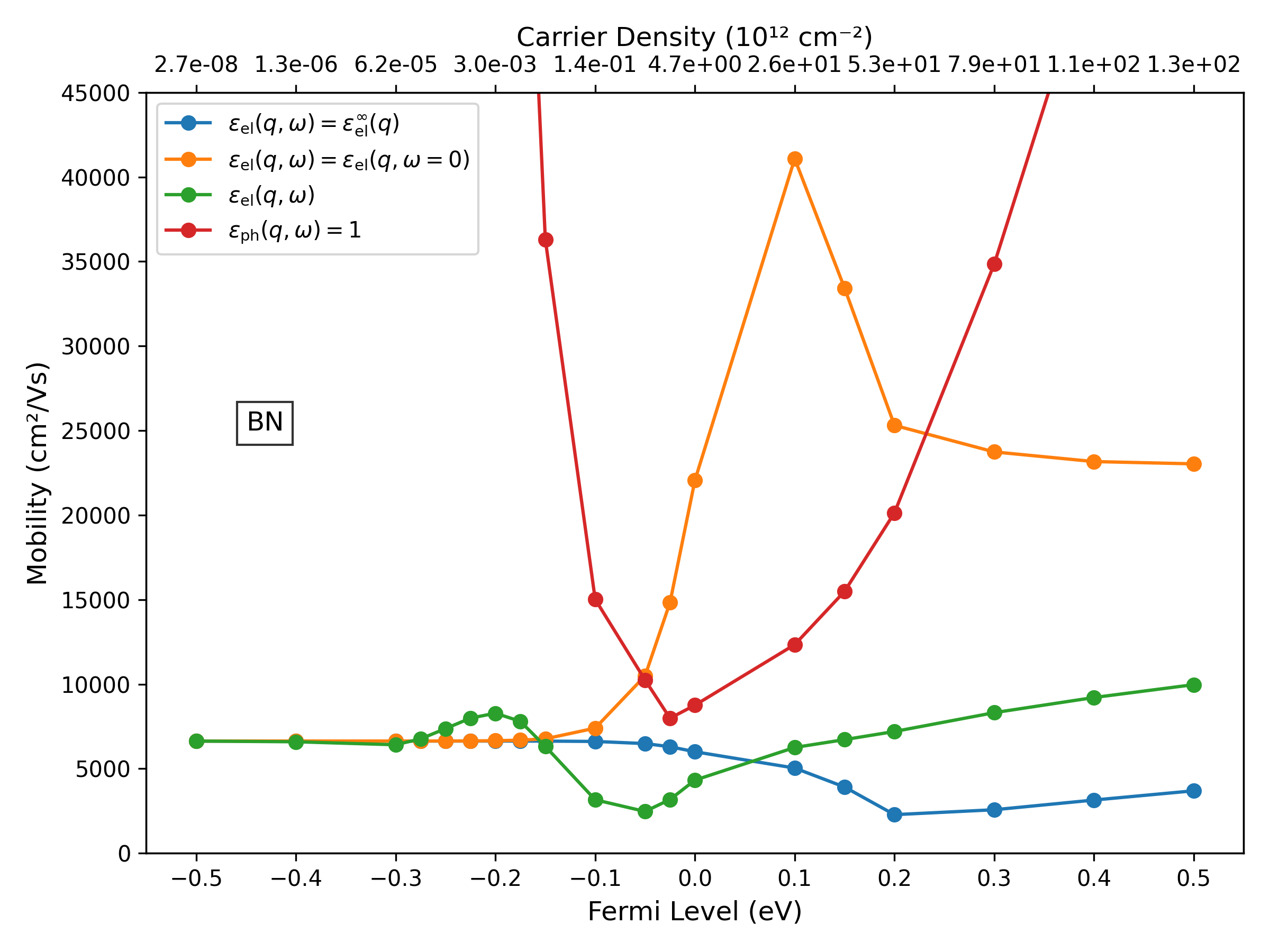}
    \includegraphics[width=0.45\linewidth]{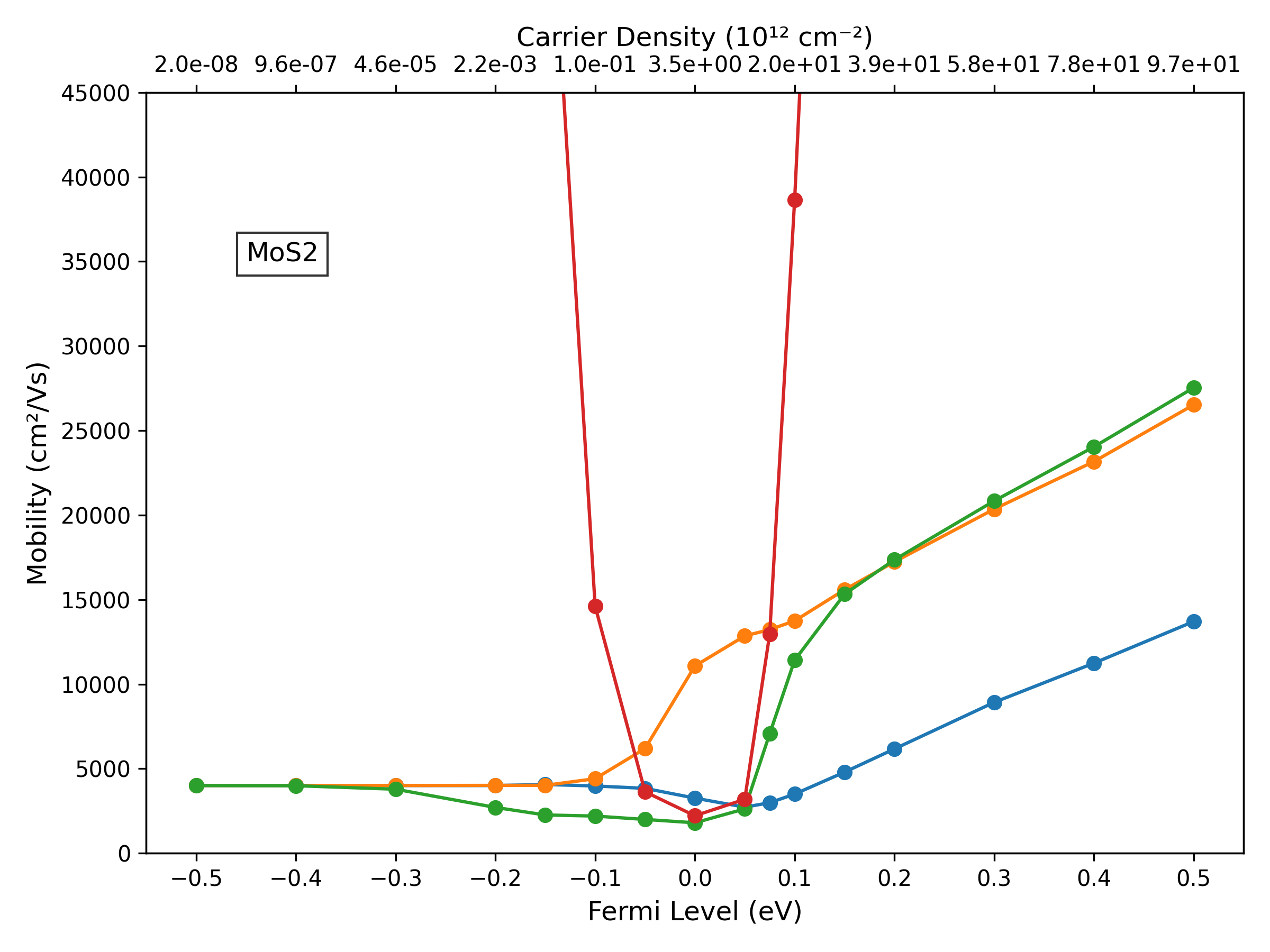} 
    \caption{Room temperature long-range limited mobilities for BN (left panel) and MoS$_2$ (right panel) using the dynamically screneed coupling and solving the coupled BTEs Eqs. \eqref{eq:BTEel} and \eqref{eq:BTEph} (green) with both e-ph and e-e interactions, or considering the system with no phonons and e-e interaction only (red). We compare to the solution of the electronic BTE Eqs. \eqref{eq:BTEel} with $G=0$ considering only the e-ph coupling in absence of free-carriers' screening (blue) or with static electronic screening (orange). }
    \label{fig:mobility}
\end{figure*}

We compute the mobilities for BN and MoS$_2$, as modeled with the parameters of Tab. \ref{tab:materials}. We show the results in Fig. \ref{fig:mobility}, as a function of Fermi level and corresponding carrier densities, at room temperature, for the several approximations discussed above. 

Note that most ab initio calculations of electronic transport properties in the literature\cite{Ponce2020} correspond to the `unscreened coupling' ($\epsilon_{\rm el}(\mathbf{q},\omega)=\epsilon^{\infty}_{\rm el}(\mathbf{q})$) approximation, while some\cite{Macheda2018,PhysRevMaterials.2.114010,Macheda2020,PhysRevB.94.085204,Sohier2021,Sohier2023} are done in the `statically screened coupling' ($\epsilon_{\rm el}(\mathbf{q},\omega)=\epsilon_{\rm el}(\mathbf{q},\omega=0)$) one. The full dynamically screened solution developed in this work differs from both.

As expected, for very low doping $E_{\rm F}<-0.3$ eV, all the solutions converge to the same asymptotic value, the so-called `intrinstic' mobility (substantially determined only by phonon absorption). Indeed, in this regime there are not enough electrons to have significant effects on phonon scattering, whether they are treated statically or dynamically. Note however that this asymptotic regime happens at very low carrier densities, corresponding in practice to an insulating regime, where transport is likely to be dominated by disorder anyway \cite{Castillo_2023}. 
For large doping $E_{\rm F}>0.2$eV, we approach the metallic-like behaviour, where phonon emission becomes relevant if $E_{\rm F}>\omega_{\rm TO}$. Here, the unscreened coupling leads to a mobility which is lower by a factor of $\sim~5$ than the statically screened one for both materials, as expected due to the large static screening of coupling. The dynamically screened coupling result behaves differently between BN and MoS$_2$. In the first case it falls in between the results with perfect (static) or absent (unscreened) screening. In the latter case it is much closer to the static result. This is due to the frequency of the MoS$_2$ phonon being much smaller, thus closer to a static perturbation of electrons.
Finally, in the intermediate regime $-0.3\rm{eV}<E_{\rm F}<0.2\rm{eV}$ (i.e. where the doping is near the bottom/top of the band) the relative value of dynamically screened with respect to unscreened/static one behaves in a non trivial way, and it is determined by the interplay between e-ph and e-e interactions. Crucially, such intermediate regime coincides with the region where most of the experiments are conducted \cite{Castillo_2023}.

Clearly, the behaviour of the dynamically screened solution in both the intermediate and high-doping regimes is strongly influenced by the presence of electronic electrodynamic excitations.
We individuate two mechanisms, a direct one with momentum scattering and an indirect one without, through which they contribute to transport. 

The first and direct mechanism is linked to the dynamical screening of the phonon, and is due to the e-e interactions in presence of a dissipative channel in the form of Eq. \eqref{eq:taum1anh}, which allows momentum dissipation in a selected frequency range. 
The electron-hole pairs that dress the polar phonons are considered part of the phonon excitation and as such they scatter charge carriers with the same momentum dissipation.
The corresponding e-e coupling term can be described by a Coulomb interaction that is dynamically screened by electrons themselves, as in Eq. \eqref{eq:collelonly}. 
This e-e scattering mechanism happens at all energies, but momentum is dissipated only if the interaction happens for $(q, \omega)$ where the phonon content is non-zero.
To isolate and quantify this effect, we eliminate direct phonon contributions to the response function, but we keep the same phonon content. 
The resulting solution of the coupled BTEs is represented by the red curve in Fig. \ref{fig:mobility}. At low dopings, the mobility stemming from this mechanism alone is very high (therefore its impact on full mobility is negligible), as expected from a vanishing number of carriers and therefore a strong reduction of the phase space for e-e scattering. The minimum mobility is found in the intermediate regime, where electronic screening is small but the number of carriers start to be relevant. At high dopings, this direct mechanism becomes weaker due to the strengthening of the screening towards metallic behavior, that overcompensate the growth of carriers. In particular, for MoS$_2$ mobility almost immediately becomes very high while for BN it gradually increases. 

\begin{figure*}
    \centering
    \includegraphics[width=\linewidth]{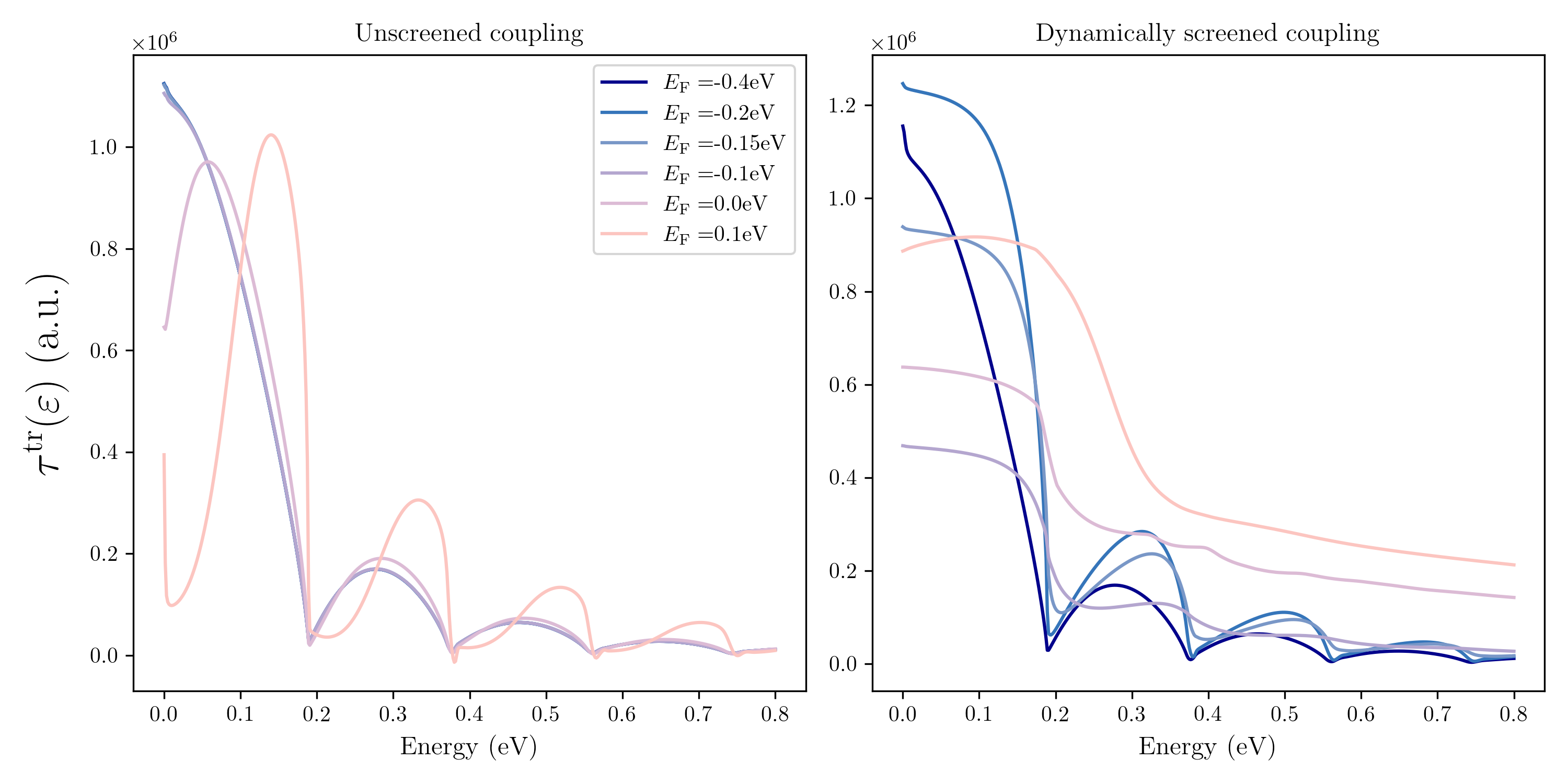}
    \caption{Evolution of the electronic lifetime $\tau^{\rm{tr}}(\varepsilon)$ for the uscreneed (left panel) and the dynamically screened coupling (right panel) as a function fo the Fermi level $E_{\rm F}$. }
    \label{fig:solutionsef}
\end{figure*}

\begin{figure*}
    \centering
    \includegraphics[width=\linewidth]{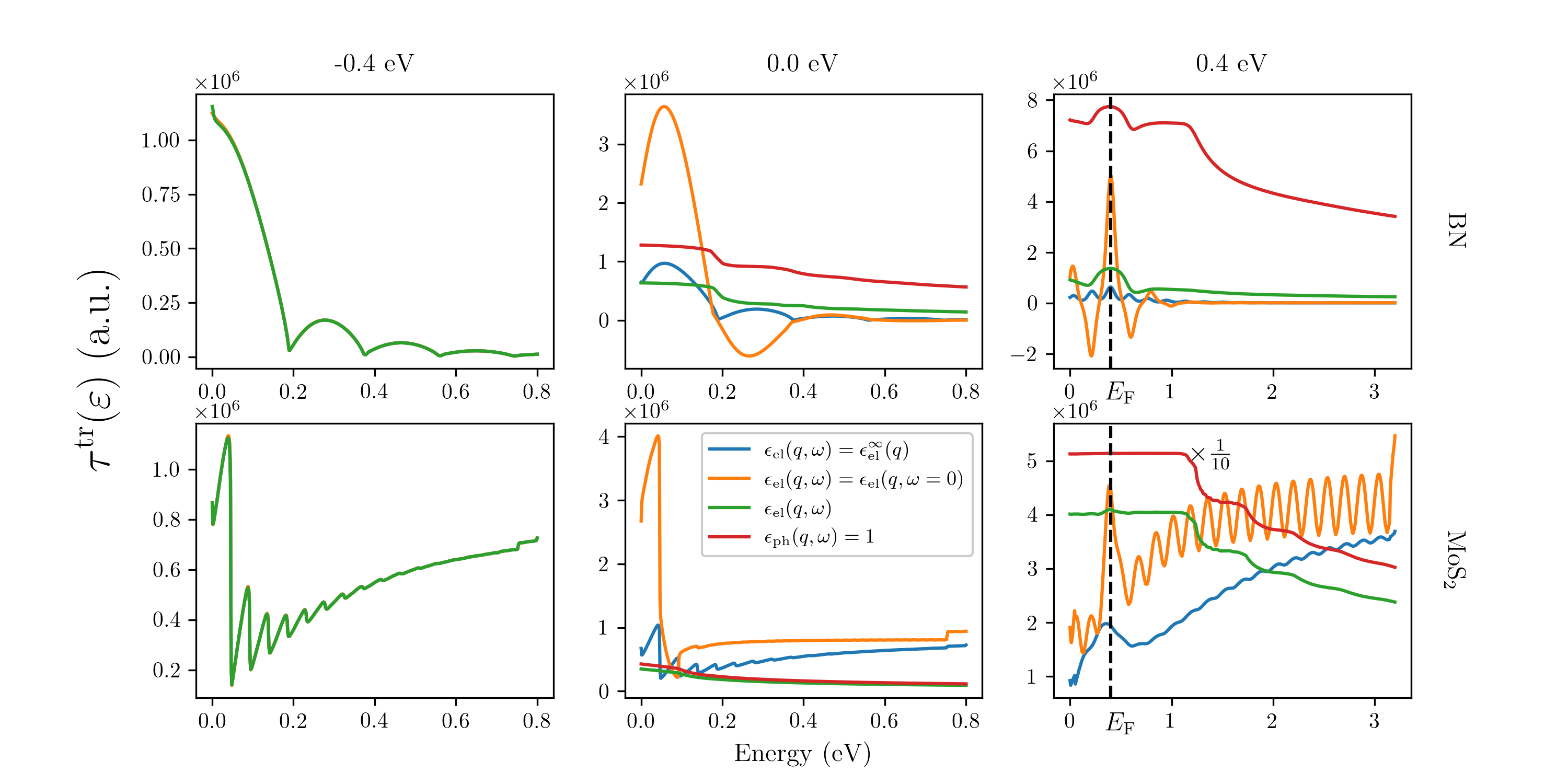}
    \caption{Electronic transport lifetime $\tau^{\rm tr}(\epsilon)$ for the different approximations for mobility calulations shown in Fig. \ref{fig:mobility}, for BN (upper panels) and MoS$_2$ at room temperature for different different Fermi levels ($-0.4$ eV in left column, $0.0$ eV central column, $0.4$ eV right column). For $E_{\rm F}=-0.4$eV we avoid plotting curve relative to no phonons and e-e interaction only, because it's severely out of scale, while at $E_{\rm F}=0.4$eV for MoS$_2$ we reduced it by a tenth to make it comparable to the others.}
    \label{fig:solutions}
\end{figure*}

The second and indirect mechanism is also due to e-e interactions, and follows from the presence of a continuum of electronic excitations that satisfy the energy conservation in the coupled BTEs. 
Those redistribute the energy among electrons, changing the occupation function and the scattering integrals in the process. 
To elucidate this mechanism, let's consider for example a single electron in an empty parabolic band. In presence of electron-phonon interaction alone, the electronic lifetime $\tau^{\rm el}(\varepsilon)$ (and similarly the transport lifetime $\tau^{\rm tr}(\varepsilon)$) is expected to display a minimum in correspondence of the phonon frequency, i.e. as soon as emission becomes allowed by energy conservation. Therefore, the population of the state at $\varepsilon=\omega_{\rm ph}$ is expected to be strongly depleted. As soon as e-e interactions are present, these tend to replenish the depopulated state. Similarly, they tend to depopulate more highly populated states. Therefore, the indirect effect is expected to induce a flattening of the shape of $\tau^{\rm tr}(\varepsilon)$. We complement this intuitive description with the plot of $\tau^{\rm tr}(\varepsilon)$, as found from the solution of the coupled BTEs, in Figs. \ref{fig:solutionsef} and \ref{fig:solutions}. We start from Fig. \ref{fig:solutionsef}, where on the left panel we study the evolution of $\tau^{\rm tr}(\varepsilon)$ with the unscreened coupling while increasing $E_{\rm F}$. The solution resemble the typical solution for Fr\"olich interaction, where as anticipated $\tau^{\rm tr}(\varepsilon)$ presents minima at integer multiples of the phonon frequency. At higher doping levels, Fermi-Dirac statistic of occupied states around the Fermi energy starts to become important and the minima of $\tau^{\rm tr}(\varepsilon)$ gradually shift, but the shape is overall changing only slightly. On the right panel instead we show the solution with dynamical screening, where the shape of $\tau^{\rm tr}(\varepsilon)$ strongly changes alongside the Fermi level. In particular, the minima of $\tau^{\rm tr}(\varepsilon)$ gradually disappear and the solutions flattens.

We now compare the shape of $\tau^{\rm tr}(\varepsilon)$ for different solutions in Fig. \ref{fig:solutions}. We here see that at small doping, where we do not represent the solution obtained with e-e interaction only because it is out of scale, the unscreneed/statically/dynamically screened coupling solutions are the same. At large dopings, the solutions are instead peaked around the Fermi energy. The dynamically screened solution has peak values that are comparable in magnitude with the statically screened one, but the shape is much more similar to solution with no phonons and e-e interaction only. Finally, intermediate regimes are again of more difficult interpretation, but the tendency of the dynamically screened solution is still to be quite flat in shape and more similar to the case where only e-e is present rather than the static or unscreened ones.

\textit{VED case---} The results for the realistic case of BN-encapsulated graphene, treated within the VED framework and including short-range interactions, are shown in an accompanying paper \cite{accpaper}.

\section{Conclusions}
\label{sec:concl}
We have implemented the solution of the coupled BTEs equations for the electrons and excitations in a simple parabolic band and single phonon model for two dimensional h-BN and 2H-MoS$_2$, taking into account both e-ph and e-e interactions, and dynamical screening. We have determined that anharmonic coupling is fundamental to dissipate electronic quasi-momentum, and consequently devised a way to treat anharmonic scattering for polar phonon excitations, that are not well defined quasi-particles when free carriers are introduced in the system, as previously assumed. We find that dynamical screening and e-e interactions, usually discarded in state-of-the-art calculations, have a strong non-trivial impact on the electronic mobility, especially for carrier densities where most of experimental investigations are conducted. The major qualitative effect of dynamical screening is to introduce a continuum of electron-hole  transitions that can satisfy energy conservation, thereby flattening the electronic out-of-equilibrium population, while direct e-e interactions contribute to momentum dissipation in the region where dissipation through anharmonic scattering is allowed. Our methodologies can all be generalized to realistic materials, as shown in the accompanying paper, where polar phonons are one of the many scattering channels that limit mobility. We hope that our findings will spur interest for both theory development and materials study.
\section{Ackwnoledments}
This problem was brought to the attention of F. Macheda, at the beginning of his PhD, by his supervisor N. Bonini, to whom we dedicate the findings of this work. We thank F. Mauri, A. Guandalini, G. Caldarelli and J. Fiore for fruitfull discussions.
\appendix
\section{Model vs first principles description}
\label{app:Mahan}
We here review two models that have been used to study the interplay between electronic and phononic responses, mainly for three dimensional materials, that are the Mahan model \cite{Mahan1990} and the Kim-Das-Senturia (KDS) model \cite{Kim1978}, and indicate in what they are similar or differ from our treatment.
\subsection{The Mahan model}
Mahan \cite{Mahan1990} introduces a phenomenological model consisting in a single isotropic electronic parabolic band dispersion $
\varepsilon_{\mathbf{k}}=\frac{\hbar^2 k^2}{2 m_{\rm eff}}$ and a single flat isotropic transverse/longitudinal optical mode $\omega_{\rm{TO/LO} \mathbf{q}}=\omega_{\rm TO/LO}$, where $\mathbf{k}$ and $\mathbf{q}$ are the electronic and phonon quasi-momenta, respectively. The electronic and phononic systems interact via an electron-phonon Fr\"olich coupling. All the effects of the other bands are included via the static screening $\epsilon_{\infty}$. The Hamiltonian reads \cite{Mahan1990,Sanborn1995} 
\begin{align}
H=\sum_{\mathbf{k}} \varepsilon_{\mathbf{k}}c^{\dagger}_{\mathbf{k}}c_{\mathbf{k}}+\sum_{\mathbf{q}}\omega_{\rm TO}a^{\dagger}_{\mathbf{q}}a_{\mathbf{q}}+\frac{1}{\sqrt{\Omega}}\sum_{\mathbf{q}\mathbf{k}}g_{\mathbf{q}}c^{\dagger}_{\mathbf{k+q}}c_{\mathbf{k}}A^{\dagger}_{\mathbf{q}}\nonumber\\
+\frac{1}{2}\sum_{\mathbf{q}}\frac{\left(\omega^2_{\rm LO \mathbf{q}}-\omega^2_{\rm TO}\right)}{2\omega_{\rm TO}} A^{\dagger}_{\mathbf{q}}A_{\mathbf{q}}+\frac{1}{2\Omega}\sum_{\mathbf{q,k,k'}}v^{\infty}_{\mathbf{q}}c^{\dagger}_{\mathbf{k+q}}c_{\mathbf{k}}c^{\dagger}_{\mathbf{k'-q}}c_{\mathbf{k'}}
\label{eq:startH}
\end{align}
where 
\begin{align}
A_{\mathbf{q}}=a_{\mathbf{q}}+a^{\dagger}_{\mathbf{-q}},
\end{align}
where $c^{\dagger},c/a^{\dagger},a$ are the electronic/phononic creation and destruction operators. The role of the fourth term in Eq. \eqref{eq:startH} is to shift the pole of the phonon propagator from the TO frequency to the value of the longitudinal optical (LO) one. In this model, for a diatomic system the electron-phonon couplig matrix element is given as
\begin{align}
g_{\mathbf{q}}=v^{\infty}_{\mathbf{q}}Zq \sqrt{\frac{\hbar}{2\Omega M_{\rm red}\omega_{\rm TO}}},
\end{align}
where $\Omega$ is the unit cell volume, $M_{\rm red}$ is the reduced atomic mass and $Z$ is the Born effective charge. Using the Lyddane–Sachs–Teller relation and the form of the long-range force constant
\begin{align}
\frac{\epsilon_0}{\epsilon_{\infty}}=\frac{\omega^2_{\rm LO}}{\omega^2_{\rm TO}}, \quad \omega^2_{\rm LO}-\omega^2_{\rm TO}=v^{\infty}_{\mathbf{q}}\frac{Z^2 q^2}{\Omega M_{\rm red}}
\end{align}
one finds 
\begin{align}
g^2_{\mathbf{q}}=v^{\infty}_{\mathbf{q}}\frac{\omega^2_{\rm LO}-\omega^2_{\rm TO}}{2\omega_{\rm TO}}
\end{align}
Notice the presence of $\omega_{\rm TO}$ at the denominator of the electron-phonon coupling, due to the inability of the model to catch the renormalization of the coupling following the migration of the pole from the TO to the LO frequency. In fact, notice that the fourth term of Eq. \ref{eq:startH} can be considered as a phonon-self energy that, if evaluated at $\omega=\omega_{\rm TO}$, shifts the pole of the phonon progator to $\omega_{\rm LO}$ \cite{Mahan1990}
\begin{align}
D(\mathbf{q},\omega)=\frac{2\omega_{\rm TO}}{\omega^2_{\rm LO \mathbf{q}}-(\omega+i\eta)^2}
\end{align}
but that gives a wrong spectral function for the phonon. Even if the full frequency dependence of the self-energy is taken into account \cite{PhysRevB.51.14247}, the phonon self-energy does not properly reconstruct the phonon propagator for the LO phonon. The model is therefore too simplicistic to allow for a quantitative assement of the problem.

\subsection{The KDS model}
The KDS Hamiltonian takes into acocunt all the bands of the crystal as
\begin{align}
H=\sum_{\mathbf{k}} \varepsilon_{n\mathbf{k}}c^{\dagger}_{n\mathbf{k}}c_{n\mathbf{k}}+\frac{1}{2}\sum_{\mathbf{q}}\left(P^{\dagger}_{\mathbf{q}}P_{\mathbf{q}}+\omega^2_{\rm TO}Q^{\dagger}_{\mathbf{q}}Q_{\mathbf{q}}\right) \nonumber \\
+\frac{1}{2}\sum_{\mathbf{q}}(\Omega^0_{\rm L})^2 Q^{\dagger}_{\mathbf{q}}Q_{\mathbf{q}}+\frac{1}{2}\sum_{\mathbf{q}}v_{\mathbf{q}}\rho_{\mathbf{q}}\rho^{\dagger}_{\mathbf{q}} \nonumber \\
-i\Omega^0_{\rm L}\sum_{\mathbf{q}}v^{1/2}_{\mathbf{q}}\left(\hat{\mathbf{q}}\cdot\mathbf{e_{q}}\right)\rho_{\mathbf{q}}Q^{\dagger}_{\mathbf{q}} \\
\Omega^0_{\rm L}=[\epsilon_{\infty}\left(\omega^2_{\rm LO}-\omega^2_{\rm TO}\right)]^{1/2}, \rho_{\mathbf{q}}=\sum_{n,n',\mathbf{k}}F^{nn'}_{\mathbf{k},\mathbf{k-q}}c^{\dagger}_{n'\mathbf{k-q}}c_{n\mathbf{k}} \nonumber \\
F^{nn'}_{\mathbf{k},\mathbf{k-q}}=\langle n'\mathbf{k-q}|e^{-i\mathbf{q}\cdot \mathbf{r}}|n\mathbf{k} \rangle
\end{align}
where
\begin{align}
P_{\mathbf{q}}=\sqrt{\frac{\omega_{{\rm LO}\mathbf{q}}}{2}}(a_{\mathbf{q}}+a^{\dagger}_{-\mathbf{q}})\quad Q_{\mathbf{q}}=\frac{a_{\mathbf{q}}+a^{\dagger}_{\mathbf{-q}}}{\sqrt{2\omega_{{\rm LO}\mathbf{q}}}}
\end{align}
A clearer approach is found in Ref. \onlinecite{Sellati2025}. KDS rewrote the FGR for intraband scattering within the RPA and using the fluctuation-dissipation theorem in a form that agrees with our Eq. \eqref{eq:Pz} 
restricted to intraband scattering if divided by $k_BT \frac{\partial f_{n\mathbf{k}}}{\partial \varepsilon_{n\mathbf{k}}}$, because of the factors difference between FGR results and lifetime; also for this reason there is a factor 2 with respect to our expression. Importantly, the phonon propagator has the correct pole at $\omega_{{\rm LO}}$ and the spectral function has the correct shape. Also, the expression of the electron-phonon coupling agrees with ours:
\begin{align}
g^2_{\mathbf{q}}=v^{\infty}_{\mathbf{q}}\frac{\omega^2_{\rm LO}-\omega^2_{\rm TO}}{2\omega_{\rm LO}}
\end{align}

\section{Independence of the interference peak from $\eta_{\rm ph}$ and $\eta_{\rm pl}$}
\label{app:etaph}
\begin{figure}[h!]
    \centering
    \includegraphics[width=\linewidth]{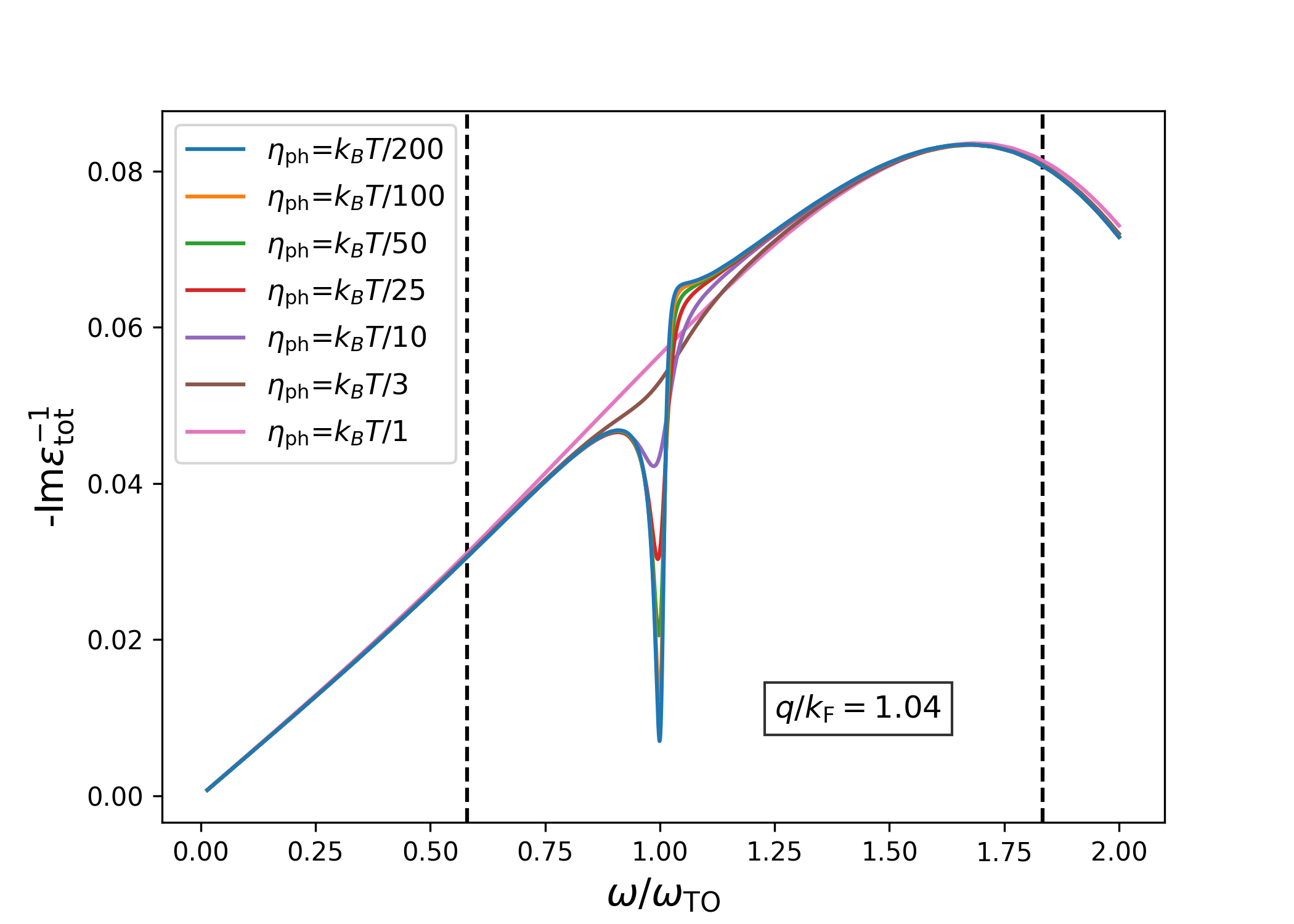}
    \caption{-Im$\epsilon^{-1}_{\rm tot}$ as a function of frequency  for the cut at $q/k_{\rm F}=1.04$, for different values of $\eta_{\rm ph}$. Vertical lines represent the limits of the electron-hole continuum at 0K.}
    \label{fig:etaph}
\end{figure}

We here show that the interference features of Fig. \ref{fig:cuts} are not artifacts due to the use of a finite smearing in the calculations. We consider BN, fixing $\eta_{\rm pl}=k_BT/50$, and vary $\eta_{\rm ph}$ from $k_{B}T$ to $k_{B}T/200$, where $T=300$K. We plot -Im$\epsilon^{-1}_{\rm tot}$ along the cut at $q/k_{\rm F}=1.04$ (the same of Fig. \ref{fig:division}) in Fig. \ref{fig:etaph}. We find that the interference pattern tends toward a well defined shape while the smearing decreases. Notice that our choice of $\eta_{\rm ph}$ from Tab. \ref{tab:materials} is around $\eta_{\rm ph}\sim k_BT/100$ and reasonably near to the limiting shape, and allows for practical convergence of calculations.

\begin{figure}[h!]
    \centering
    \includegraphics[width=\linewidth]{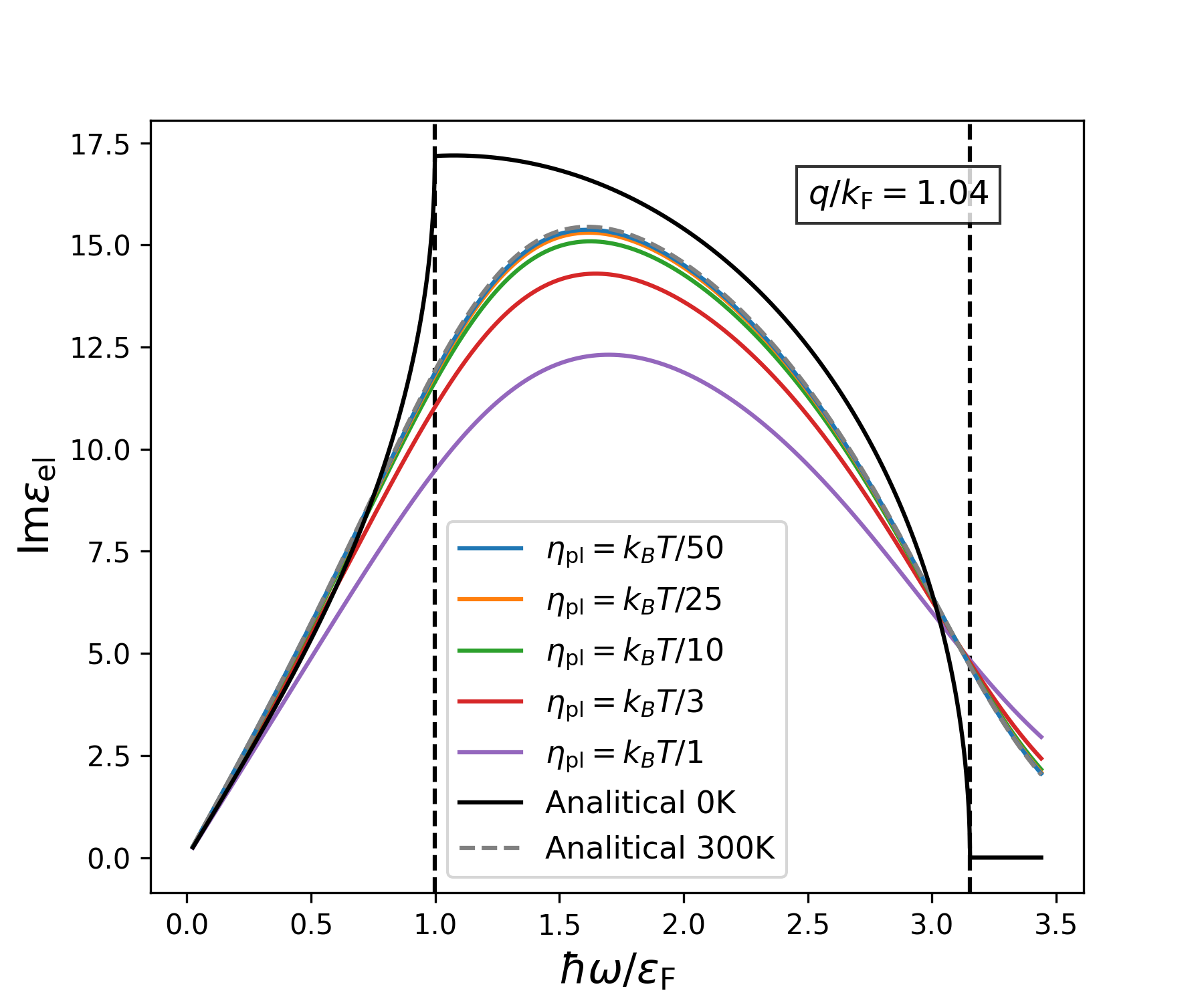}
    \caption{Im$\epsilon$ as a function of frequency for the cut at $q/k_{\rm F}=1.04$, for different values of $\eta_{\rm ph}$. Analytical results at 0K and 300K are shown as black continuous lines and gray dashed lines. Vertical lines represent the limits of the electron-hole continuum at 0K.}
    \label{fig:etapl}
\end{figure}

Next, we consider the convergence of the electronic response with respect to $\eta_{\rm pl}$. In Fig. \ref{fig:etapl} we plot the values of Im$\epsilon_{\rm el}$ for different values of $\eta_{\rm pl}$ ranging from $k_{B}T$ to $k_{B}T/50$, together with analytical expressions of the imaginary part of the Lindhard function at 0K and 300K. We see that our chosen value of $\eta_{\rm pl}=k_{B}T/50$ rightly reproduces the analytical results within excellent precision. We finally notice that, even though calculations are not reported in the manuscript our chosen value of $\eta_{\rm pl}$ ensures that solution of the coupled BTEs with and without the simplifying assumptions for the integration used in Sec. \ref{sec:compmeth} agree within a 1\% error.

\section{Lorentzian times Bose distribution}
\label{app:boselorentz}
In Eq. \eqref{eq:P} under integration we have products of quantities in the following form
\begin{align}
F(\omega,\eta)=-\frac{1}{\pi}\rm{Im}\left[\frac{2\omega_{\rm ph}}{(\omega+i\eta)^2-\omega^2_{\rm ph}}\right] (n_{\omega}+1),
\end{align}
where the first term comes from the Lorentzian description of a quasiparticle and the second is the Bose-Einstein occupation. In Eq. \eqref{eq:staticg} we take the limit of $n\rightarrow 0$ before multiplying the Lorentzian with the occupation factor, leading to
\begin{align}
\lim_{\eta \rightarrow 0}F(\omega,\eta)=\delta(\omega-\omega_0)n_{\omega_0}.
\end{align}
This is a correct procedure in the clean limit, as deduced from the $i0^+$ substitution present in Eq. \eqref{eq:Pz}. But if we consider the phonon to have some small width and therefore consider a finite $\eta$, then one has that
\begin{align}
\lim_{\omega\rightarrow 0}F(\omega,\eta)=\frac{4k_BT\eta}{\pi \omega^3_{\rm ph}}.
\end{align}
If $k_BT\eta \ll 1$ there is no problem. If on the other hand $k_BT\eta \sim 1$ or even greater, one can get non-negligible contributions to the conductivity from the low-energy spectrum of the total response, which is a  counter-intuitive result. In reality, this is expected to be cured by self-energy effects on the excitation spectral function that modify its shape, and tails. 

This low-frequency contribution to the conductivity is even more problematic if the phonon features have the shapes of Fig. \ref{fig:cuts}. Nonetheless, notice that our definition of phonon content completely avoid this problem. In fact, as shown in Fig. \ref{fig:division}, the only energy sectors selected by our definition of the phonon content is in the neighborhood of the phonon peak.
\section{Convergence of the solutions of the coupled BTE}
\label{app:solconv}
We here discuss the convergence of the solution of the coupled BTEs. In particular, the iteration $i$ consists in the determination of Eq. \eqref{eq:Gdet} using the solution of the electronic BTE of the previous step $i-1$, and the solution of the electronic BTE. This provides the mobility or the conductivity as a function of the iterations $\mu_{i}$ or $\sigma_{i}$.

\begin{figure}
    \centering
    \includegraphics[width=\linewidth]{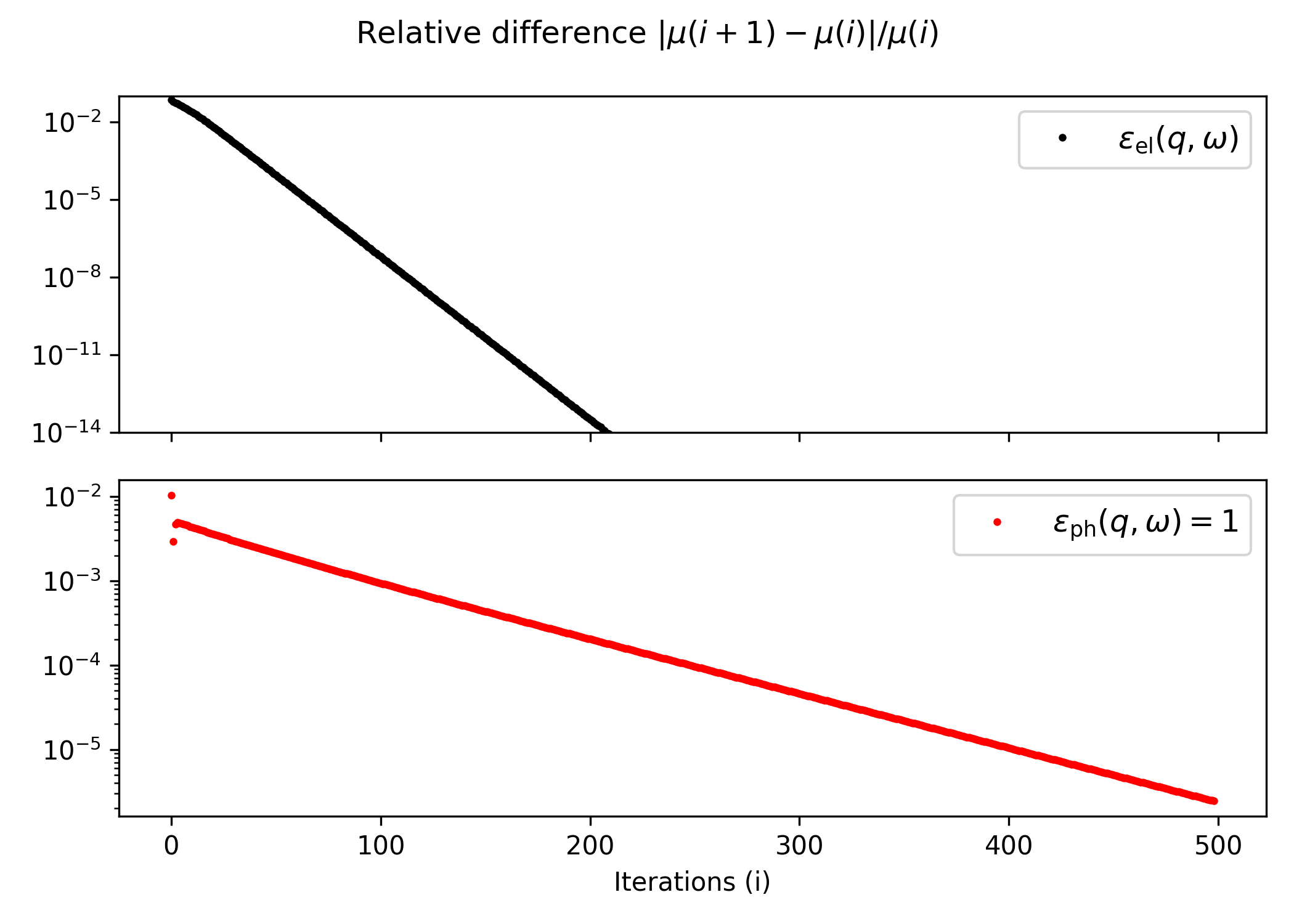}
    \caption{Convergence of the full solution (upper panel, black line) and the one with e-e interactions only (lower panel, red line) as a functions of iterations, for the parabolic case.}
    \label{fig:convcoupled}
\end{figure}

\begin{figure}
    \centering
    \includegraphics[width=\linewidth]{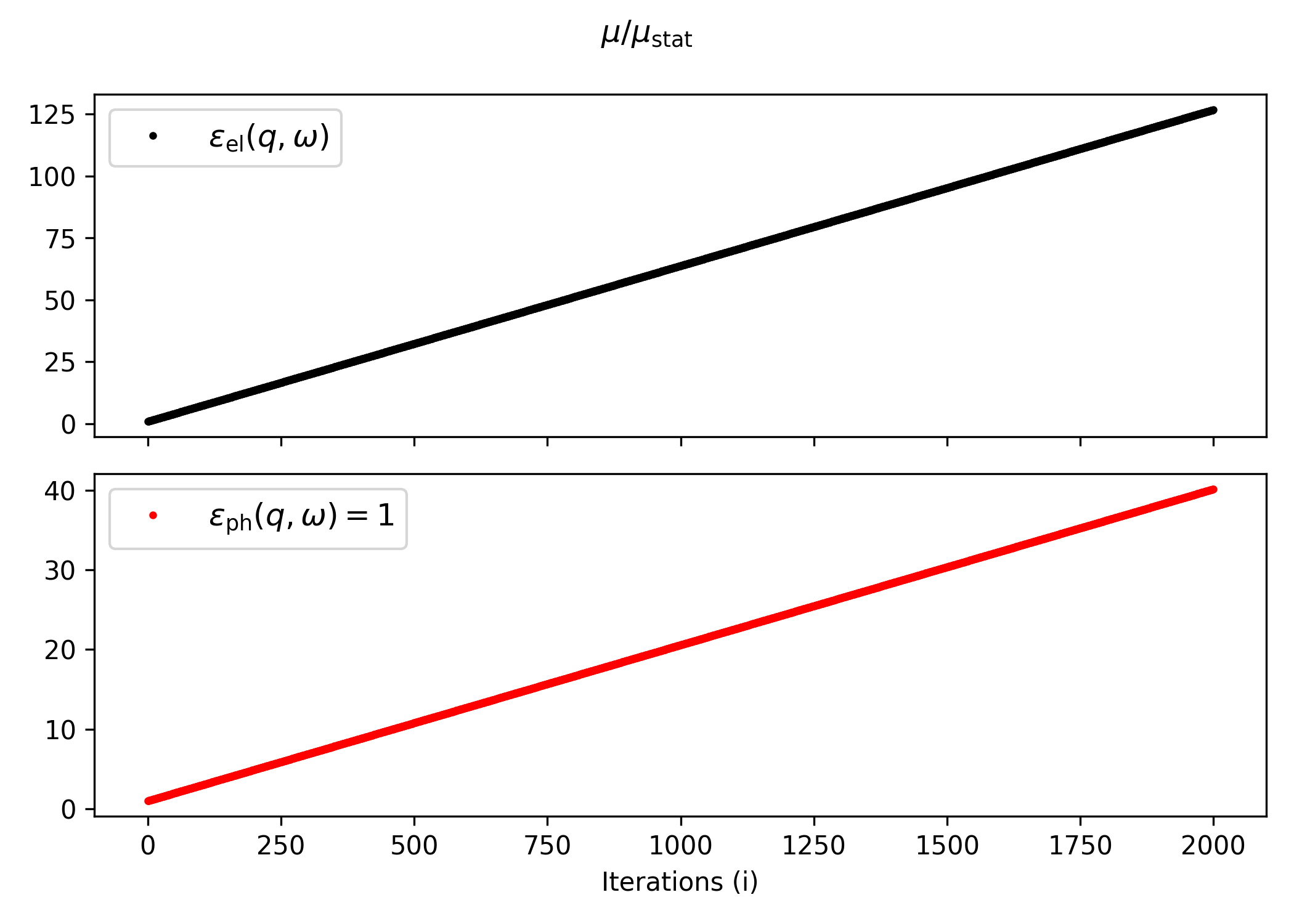}
    \caption{Un-convergence of the full solution (upper panel, black line) and the one with e-e interactions only (lower panel, red line), as a functions of iterations, if $\tau^{-1}_{\rm ANH}=0$ everywhere, for the parabolic case.}
    \label{fig:nonconvcoupled}
\end{figure}

\textit{Parabolic case---} We plot in Fig. \ref{fig:convcoupled} the ratio $|\mu_{i+1}-\mu_{i}|/\mu_{i}$, for the case where the interaction is taken as full or just the e-e contribution. As it is seen, the convergence is exponential because it is linear in a semi logarithmic scale.
We also plot in Fig. \ref{fig:nonconvcoupled} the case where $\tau^{-1}_{\rm ANH}$ of Eq. \ref{eq:taum1anh} is taken as 0 every where, i.e. we have a closed system where momentum conservation in enforced. In this case, we plot the ratio $\mu/\mu_{\rm{stat}}$, where $\mu_{\rm{stat}}$ is the mobility of the static screening case, and as it is evident the mobility diverges accordingly to the observations of Sec. \ref{sec:BTE}.

\begin{figure}
    \centering
    \includegraphics[width=\linewidth]{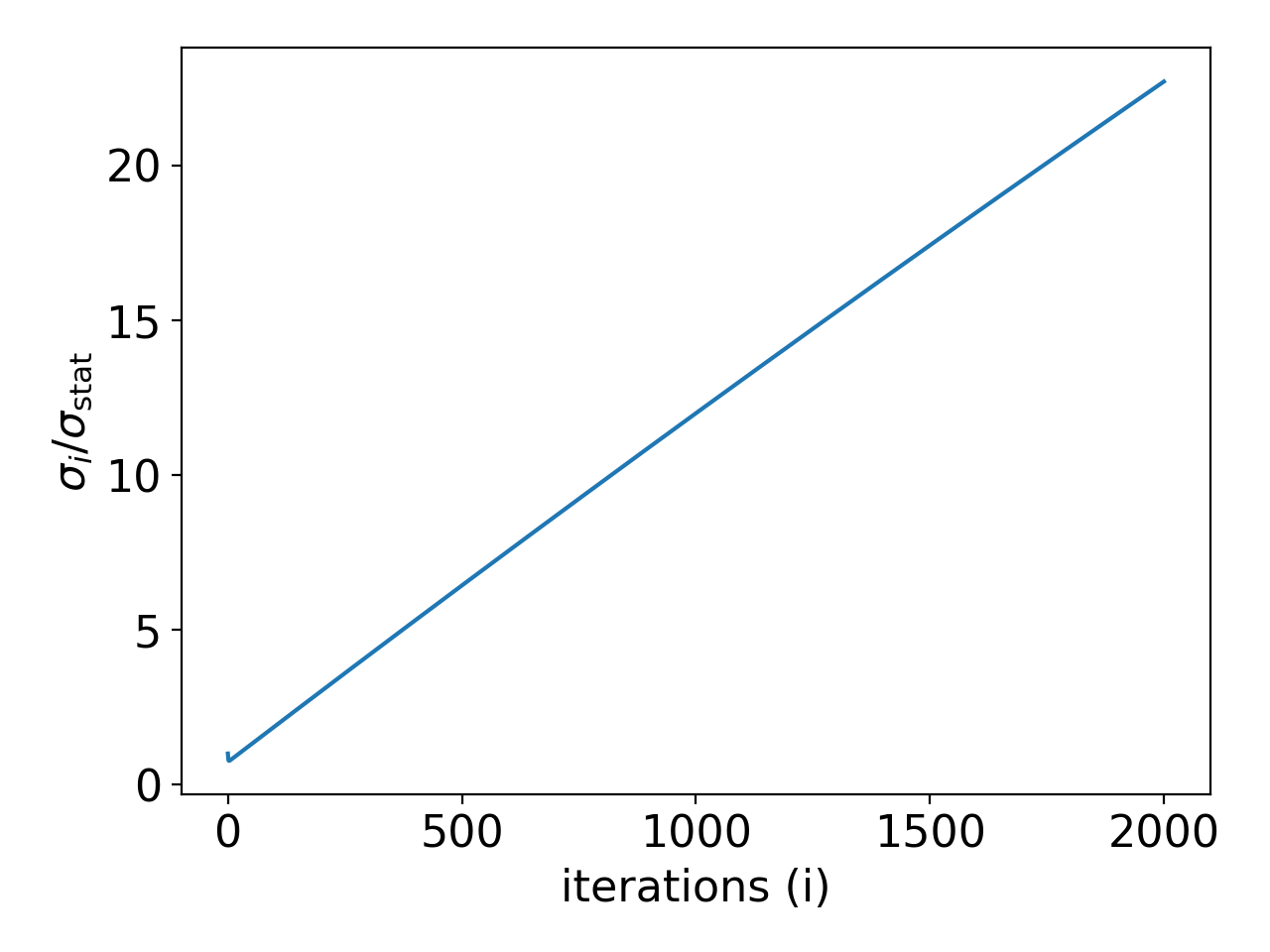}
    \caption{Un-convergence of graphene resistivity, plotted with respect to the statically screened solution $\sigma_0$, as a functions of iterations, if $\tau^{-1}_{\rm ANH}=0$ everywhere, for the VED case.}
    \label{fig:nonconvcoupledved}
\end{figure}

\begin{figure}
    \centering
    \includegraphics[width=\linewidth]{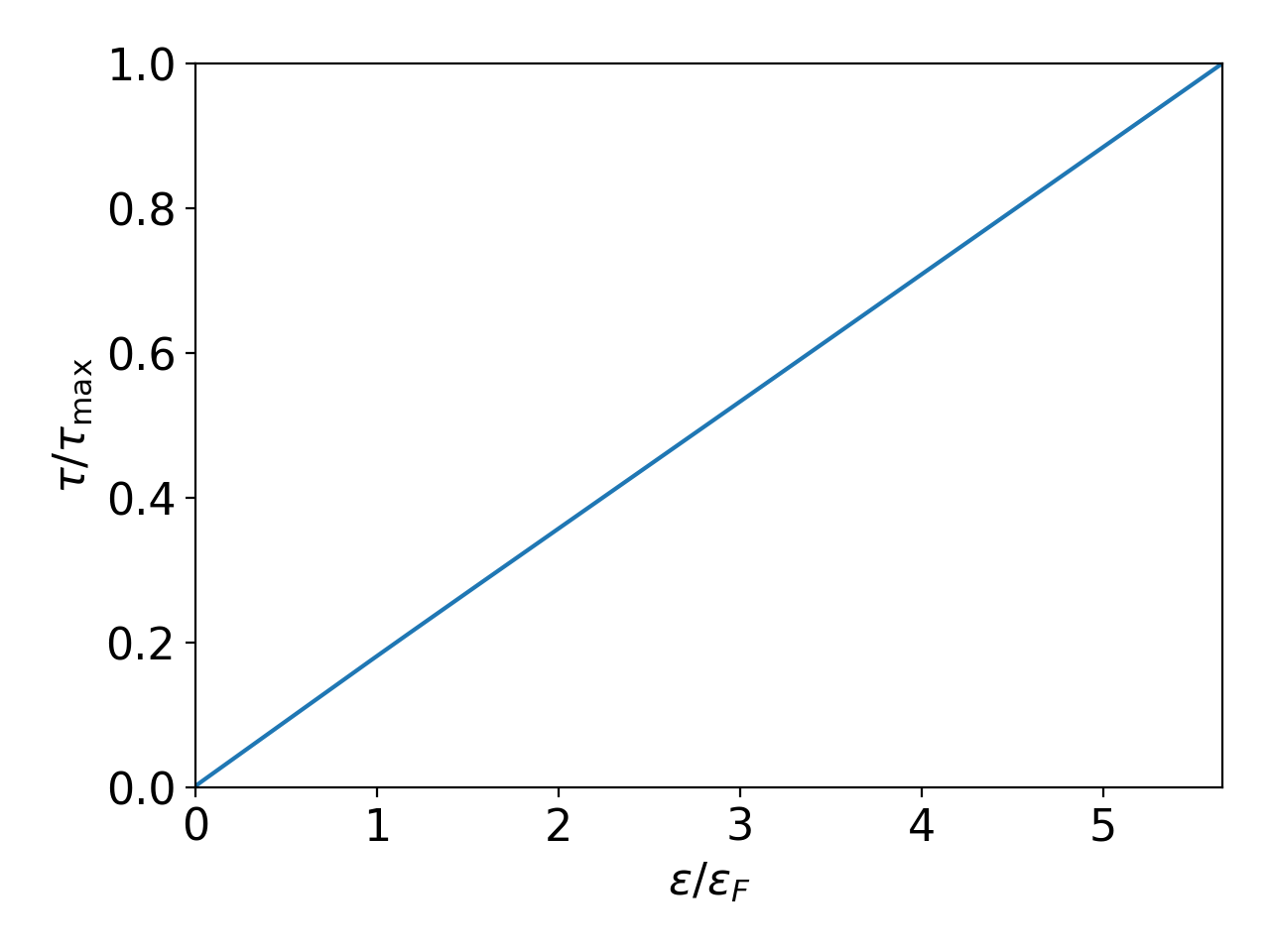}
    \includegraphics[width=\linewidth]{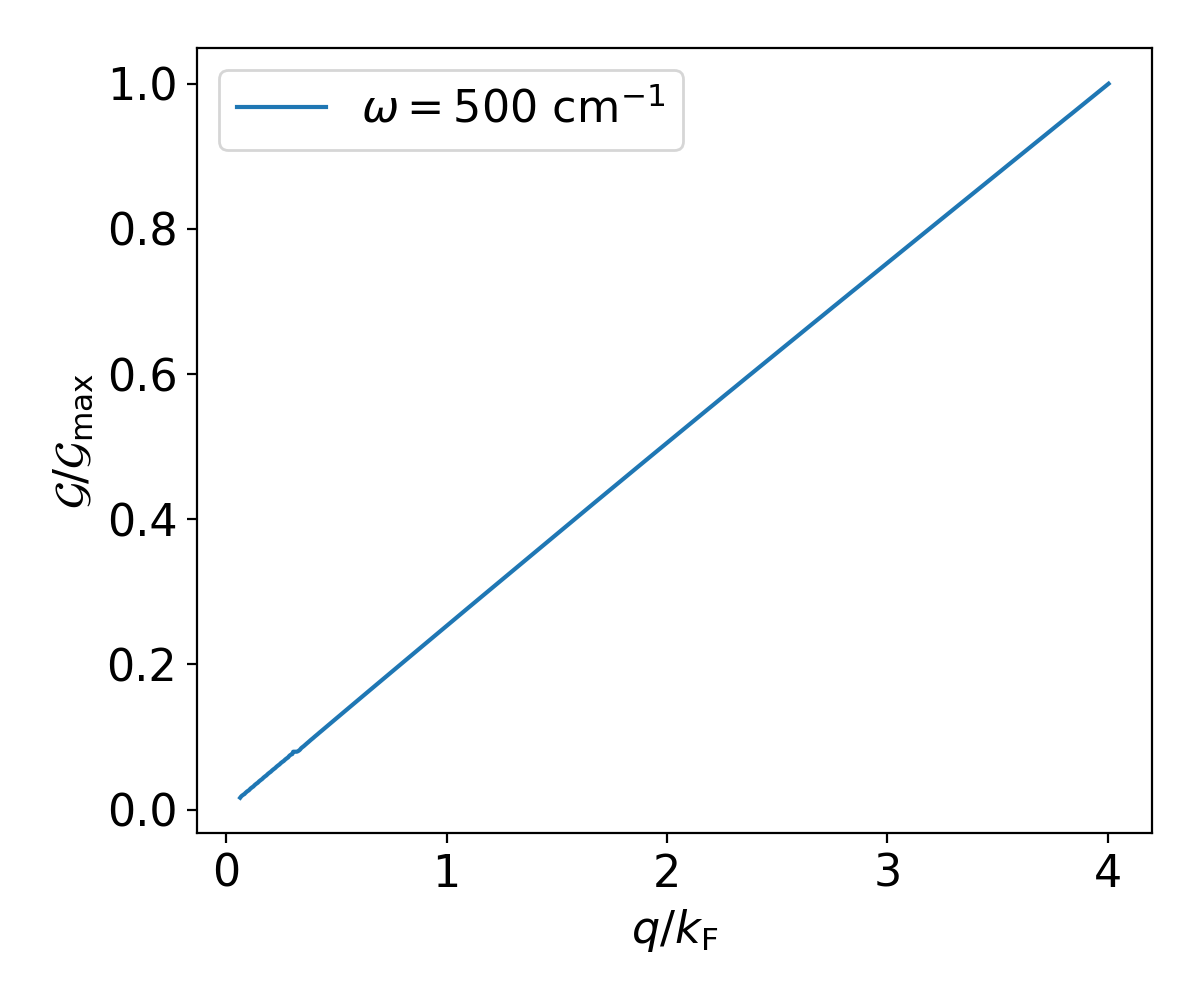}    \caption{$\tau(\varepsilon)/\tau_{\rm max}$ and $\mathcal{G}(q,\omega=500 \textrm {cm}^{-1})/\mathcal{G}_{\rm max}$, where $\tau_{\rm max}$ and $\mathcal{G}_{\rm max}$ are the maximum values contained in the plot, as found from the solution of the coupled BTE with $\tau^{-1}_{\rm ANH}=0$. The linearity of the lifetime and of $\mathcal{G}$ follow from the Peierls' argument applied to the graphene case.}\label{fig:tauGcoupledved}
\end{figure}

\textit{VED case---} The convergence of the VED case, given the different methodology with which is obtained, is by cosntruction very fast and therefore we do not report it. We just plot In Fig. \ref{fig:nonconvcoupledved} $\sigma_i/\sigma_{\rm stat}$ for the VED system with all the interactions turned on, but with $\tau_{\rm ANH}=0$. Conductivity diverges as per the Peierls' argument. We also plot \ref{fig:tauGcoupledved} the electronic lifetime (equal for both bands) and the $G$ population. By inspection, they are linear, as expected from the Peierls' argument. 

\section{Technicalities}
\label{app:real}
\subsection{Change of integration}
Below Eq. \eqref{eq:prePz} we use that $B^{\mathbf{k}}_{nm}(\mathbf{q},\omega')=-[B^{\mathbf{k}}_{nm}(\mathbf{q},-\omega')]^*$. This follows from the fact that
\begin{align}
[B^{\mathbf{k}}_{nm}(\mathbf{q},-\omega')]^* =\int d\mathbf{r} d\mathbf{r'}dz dz'd\bar{z} [F^{nm}_{\mathbf{k}\mathbf{k+q}}(\mathbf{r},z,\mathbf{r'},z')]^* \nonumber \\
\times \frac{2}{A}\textrm{Im}\left[-\epsilon_{\rm tot}^{-1}(\mathbf{q},z,\bar z,-\omega'+i0^+)v(\mathbf{q},\bar z,z')\right]
\label{eq:Bstar}
\end{align}
We now use that, in presence of time reversal, the reciprocity relations are valid (Eq. (3.59) of Ref. \onlinecite{Giuliani2005}) and can be used together with the symmetry relations for the response functions (Eq. (3.56) of Ref. \onlinecite{Giuliani2005}) to have
\begin{align}
\epsilon^{-1}_{\rm tot}(\mathbf{q},z,\bar z,\omega)v(\mathbf{q},\bar z,z')=\epsilon^{-1}_{\rm tot}(\mathbf{q},z',\bar z,-\omega)^*v(\mathbf{q},\bar z,z).
\end{align}
We further notice that
\begin{align}
\int d\mathbf{r}d\mathbf{r'}[F^{nm}_{\mathbf{k}\mathbf{k+q}}(\mathbf{r},z,\mathbf{r'},z')]^*=\int d\mathbf{r}d\mathbf{r'}[F^{nm}_{\mathbf{k}\mathbf{k+q}}(\mathbf{r},z',\mathbf{r'},z)].
\end{align}
Plugging the above equations in Eq. \eqref{eq:Bstar}, the proof is concluded. Notice that a similar proof would hold in the case where the spectral function is expressed in term of retarded and advanced response functions.

\bibliography{biblio}
\end{document}